\documentclass[a4paper,12pt]{article}
\usepackage{epsfig,psfig}

\usepackage{citesort}
 \normalsize
\def\du#1#2{{\left(\delta^u_{#1}\right)_{#2}}}
\def\dd#1#2{{\left(\delta^d_{#1}\right)_{#2}}}

\newcommand{\W}{{\scriptscriptstyle W}}

\newcommand{\bea}{\begin{eqnarray}}
\newcommand{\eea}{\end{eqnarray}}

\begin{document}

%%%%%%%%%%%%%%%%%%%%%%%%%%%%%%%%%%%%%%%%%%%%%%%%%%%%%%%%%%%%%%%%%%%%%%%
\def\lsim{\raise0.3ex\hbox{$\;<$\kern-0.75em\raise-1.1ex\hbox{$\sim\;$}}} 

\def\gsim{\raise0.3ex\hbox{$\;>$\kern-0.75em\raise-1.1ex\hbox{$\sim\;$}}}

\def\Frac#1#2{\frac{\displaystyle{#1}}{\displaystyle{#2}}}
\def\no{\nonumber\\}
%-------------------------------------------------------------------------

%-------------------------------------------------------------------------
\def\dofig#1#2{\centerline{\epsfxsize=#1\epsfig{file=#2, width=10cm,
height=8cm, angle=0}}}
%---------------------------------------------------------------------------
\def\dobox#1#2{\centerline{\epsfxsize=#1\epsfig{file=#2, width=10cm,
height=5.5cm, angle=0}}}
%-------------------------------------------------------------------------%

%-------------------------------------------------------------------------
%\def\dofig#1#2{\centerline{\epsfxsize=#1\epsfig{file=#2, width=6cm, 
%height=7cm, angle=-90}}}
%
\def\dofigs#1#2#3{\centerline{\epsfxsize=#1\epsfig{file=#2, width=6cm, 
height=7.5cm, angle=-90}\hspace{0cm}
\hfil\epsfxsize=#1\epsfig{file=#3,  width=6cm, height=7.5cm, angle=-90}}}
\def\dofourfigs#1#2#3#4#5{\centerline{
\epsfxsize=#1\epsfig{file=#2, width=6cm,height=7.5cm, angle=-90}
\hspace{0cm}
\hfil
\epsfxsize=#1\epsfig{file=#3,  width=6cm, height=7.5cm, angle=-90}}

\vspace{0.5cm}
\centerline{
\epsfxsize=#1\epsfig{file=#4, width=6cm,height=7.5cm, angle=-90}
\hspace{0cm}
\hfil
\epsfxsize=#1\epsfig{file=#5,  width=6cm, height=7.5cm, angle=-90}}
}

\def\dosixfigs#1#2#3#4#5#6#7{\centerline{
\epsfxsize=#1\epsfig{file=#2, width=6cm,height=7cm, angle=-90}
\hspace{-1cm}
\hfil
\epsfxsize=#1\epsfig{file=#3,  width=6cm, height=7cm, angle=-90}}

\centerline{
\epsfxsize=#1\epsfig{file=#4, width=6cm,height=7cm, angle=-90}
\hspace{-1cm}
\hfil
\epsfxsize=#1\epsfig{file=#5,  width=6cm, height=7cm, angle=-90}}

\centerline{
\epsfxsize=#1\epsfig{file=#6, width=6cm,height=7cm, angle=-90}
\hspace{-1cm}
\hfil
\epsfxsize=#1\epsfig{file=#7,  width=6cm, height=7cm, angle=-90}}
}
%--------------------------------------------------------------------------

\def\no{\nonumber\\}
\def\slash#1{\ooalign{\hfil/\hfil\crcr$#1$}}
\def\ep{\eta^{\prime}}
\def\susy{\mbox{\tiny SUSY}}
\def\sm{\mbox{\tiny SM}}
\def\bsg{$b\to s \gamma~$}
\def\bbar{$B-\bar{B}~$}
%%%%%%%%%%%%%%%%%%%%%%%%%%%%%%%%%%%%%%%%%%%%%%%%%%%%
%
\begin{titlepage}
\vspace*{-2cm}
\begin{flushright}
IPPP/04/38\\
DCTP/04/76\\
HIP-2004-33/TH\\
CERN-PH-TH/2004-133
\end{flushright}
\vspace{0.1cm}

{\Large
\begin{center}
{\bf Comparative Study of CP Asymmetries 

in Supersymmetric Models}
\end{center}
}
\vspace{.5cm}

\begin{center}
{E. Gabrielli$^{a,c}$, K. Huitu$^{a,b}$, and S. Khalil$^{d,e}$}
\\[5mm]
{$^{a}$\textit{Helsinki Institute of Physics,
P.O.B. 64, 00014 University of  Helsinki, Finland }}\\[0pt]
{$^{b}$\textit{Div. of HEP, Dept. of Phys.,
P.O.B. 64, 00014 University of  Helsinki, Finland}} \\[0pt]
{$^{c}$\textit{CERN PH-TH, Geneva 23, Switzerland}}\\[0pt]
{$^{d}$\textit{IPPP, University of Durham, South Rd., Durham
DH1 3LE, U.K.}}\\[0.pt]
{$^{e}$\textit{Dept. of Math., German University in Cairo - GUC,
New Cairo, El Tagamoa Al Khames, Egypt.}}\\
[0pt]

\begin{abstract}
We systematically analyze the supersymmetric contributions to the mixing
CP asymmetries and branching ratios of $B\to \phi K_S$ and $B\to 
\eta^{\prime} K_S$ processes. We consider both  gluino and chargino 
exchanges in a model independent way by using the mass insertion 
approximation method. 
While we adopt the QCD factorization 
approach for evaluating the corresponding 
hadronic matrix elements, a critical comparison with 
predictions in naive factorization one is also provided.
We find that pure chargino contributions cannot 
accommodate the current experimental results on CP asymmetries, 
mainly due to $b\to s \gamma$ constraints.
We show that charged Higgs contributions can relax these constraints
making chargino responsible for large asymmetries.
On the other hand, pure gluino exchanges can easily saturate both the 
constraints on  $B\to \phi K_S$ and 
$B\to \eta^{\prime} K_S$ CP asymmetries.
Moreover, we also find that the simultaneous contributions from gluino and 
chargino exchanges could easily account for the present experimental 
results on the mentioned asymmetries. Remarkably, large experimentally
allowed enhancements of 
$B\to \eta^{\prime} K_S$ branching ratio can easily be achieved by 
the contribution of two mass insertions in gluino exchanges.
Finally, we analyze the correlations between the CP asymmetries of these 
processes and the direct CP asymmetry in $b\to s \gamma$ decay.
When all experimental constraints  are satisfied,
supersymmetry favors
large and positive values of $b\to s \gamma$ asymmetry.
\end{abstract}

\end{center}

\end{titlepage}

%%%%%%%%%%%%%%%%%%%%%%%%%%%%%%%%%%%%%%%%%%%%%%%%%%%%%%%%%%%%%%%

\section{Introduction}
The B-factories are producing interesting experimental results with 
continuously increasing integrated luminosities.
Currently they offer one of the most promising routes to
test the Kobayashi-Maskawa  ansatz for the CP violation.
The understanding of the CP violating mechanism is one of
the major open problems in particle physics. 
There are reasons to believe that
the Standard Model (SM) cannot provide a complete description
for the CP violating phenomena in nature. 
For instance, it is established that the SM mechanism of
CP violation cannot account for the observed size of the 
baryon asymmetry in the Universe, and 
additional sources of CP violation beyond the SM are needed
\cite{bau}.

Recently, BaBar and Belle collaborations \cite{Bfact} announced 
large deviations 
from the SM expectations in the CP asymmetry of $B^0\to \phi K_S$ 
and branching 
ratio of $B^0\to \eta^{\prime} K^0$. These discrepancies
have been interpreted as possible consequences of 
new physics (NP) beyond the SM 
\cite{PhiKs_NP,KM,PhiKs_Rparity,PhiKs_gluino,PhiKs_sugra,CFMS,KK,chargino,our,hpenguin}.
For $B^0$ and $\bar{B}^0$ decays to a CP eigenstate $f_{CP}$,
the time dependent CP asymmetries are usually described by rates
$a_{f_{CP}}(t)$,
\begin{eqnarray}
a_{f_{CP}}(t)=\frac{\Gamma (\overline{B}^0(t)\to f_{CP})-\Gamma
(B^0(t)\to f_{CP})} {\Gamma (\overline{B}^0(t)\to f_{CP})+\Gamma (B^0(t)
\to f_{CP})}
=C_{f_{CP}}\cos\Delta M_{B_d}t+S_{f_{CP}}\sin\Delta M_{B_d}t,
\label{CPasym}
\end{eqnarray}
where $C_{f_{CP}}$ and $S_{f_{CP}}$ represent the parameters
of direct  and indirect CP violations respectively, and
$\Delta M_{B_d}$ is the $B^0$ eigenstate mass difference.

In the Standard Model, the angle $\beta$ in the unitary triangle of
Cabibbo-Kobayashi-Maskawa (CKM) matrix \cite{CKMfitter}, can be measured from
$B$ meson decays.
The golden mode $B^0\to J/\psi K_S$ is dominated by tree contribution
and $S_{J/\psi K_S}=\sin 2\beta +{\cal{O}}(\lambda ^3)$, where $\lambda$
is the Cabibbo mixing angle (see e.g. \cite{Fleischer}).
These uncertainties are less than 1\%.

The dominant part of the decay amplitudes for $B^0\to \phi K_S,\;\eta' K_S$
is assumed to come from the gluonic penguin \cite{London1,London2},
but some contribution from the
tree level $b\to u\bar{u} s$ decay is expected.
The $|\phi\rangle$ is almost pure $|s\bar{s}\rangle$ and consequently
this decay mode corresponds also accurately, up to terms of 
orders ${\cal{O}}(\lambda^2 )$,
to $\sin 2\beta$ in the SM \cite{Grossman}.
The $b\to u\bar{u} s$ tree level contribution to $B_d\to \eta' K$ was
estimated in \cite{London2}.
It was found that the tree level amplitude is less than 2\% of the
gluonic penguin amplitude.
Thus also in this mode one measures the angle $\beta$ with
a good precision in the SM.
Therefore, it is expected that NP contributions to the CP asymmetries in 
$B^0\to \phi K_S,\;\eta' K_S$ decays are more significant than in 
$B^0\to J/\psi K_S$ and can compete with the SM one. 

New sample of data have been recently analyzed 
by BaBar and Belle collaboration and the
results with higher statistics
have been now announced in \cite{giorgi,sakai}.
The new experimental value of the indirect CP asymmetry parameter
for $B^0\to J/\psi K_S$ is given by \cite{giorgi,sakai}
\bea
S_{J/\psi K_S} =0.726\pm 0.037,
\label{Spsi}
\eea
which does not differ much from the previous one \cite{teb,exp1}, and 
agrees quite well with the SM prediction $0.715_{-0.045}^{+0.055}$ \cite{AB}.
However, results of Belle on the corresponding $\sin{2\beta}$ 
extracted for $B^0\to \phi K_S$ process has changed dramatically 
\cite{giorgi,sakai}
\bea
S_{\phi K_S}&=&0.50\pm 0.25^{+0.07}_{-0.04}\;\; 
({\rm BaBar}),\nonumber\\
&=&0.06\pm 0.33 \pm 0.09\;\; ({\rm Belle})\, ,
\label{Sphi}
\eea
where the first errors are statistical and the second systematic,
showing now a better agreement 
than before \cite{phi_babar,phieta_belle}.
However, as we can see from Eq.(\ref{Sphi}),
the relative central values are still different. BaBar results 
\cite{giorgi} are more
compatible with SM predictions, while Belle measurements 
\cite{sakai} still show 
a deviation from the $c\bar{c}$ measurements of about $2\sigma$.
Moreover, the average $S_{\phi K_S}=0.34 \pm 0.20$ is quite 
different from the previous one \cite{hfag},
displaying now 1.7$\sigma$ deviation from Eq.(\ref{Spsi}).

Furthermore, the most recent measured CP asymmetry 
in the $B^0\to \eta^{\prime} K_S$ decay is found by BaBar \cite{giorgi}
and Belle \cite{sakai} collaborations as
\bea
S_{\eta^{\prime} K_S}&=&0.27\pm 0.14\pm 0.03 \;\; ({\rm BaBar}) \nonumber\\
&=&0.65\pm 0.18\pm 0.04 \;\; ({\rm Belle}),
\label{Seta}
\eea
with an average $S_{\eta^{\prime} K_S}= 0.41 \pm 0.11$,
which shows a 2.5$\sigma$  discrepancy from Eq.~(\ref{Spsi}). 
For the previous results 
see (BaBar) \cite{etaBa}
and (Belle) \cite{phieta_belle}.

It is interesting to note that the new results on s-penguin modes
from both experiments differ from the value extracted from the
$c\bar{c}$ mode ($J/\psi$), BaBar by
2.7$\sigma$ and Belle by 2.4$\sigma$ \cite{giorgi,sakai}.
At the same time the experiments agree with each other, and even
the central values are quite close:
\bea
0.42\pm0.10\;\; {\rm BaBar},\;\;\; 0.43^{+0.12}_{-0.11}\;\;{\rm
  Belle}.
\nonumber
\eea

On the other hand, the experimental measurements of the branching
ratios of $B^0\to \phi K^0$ and $B^0\to \eta^{\prime} K^0$ at BaBar 
\cite{BR_babar}, Belle \cite{BR_belle},
and CLEO \cite{BR_cleo} lead to the following averaged  results \cite{hfag} :
\begin{eqnarray}
BR(B^0\to \phi K^0) &=& (8.3^{+1.2}_{-1.0}) \times 10^{-6},
\label{BRphi}\\
BR(B^0\to \eta^{\prime} K^0) &=& (65.2^{+6.0}_{-5.9} ) \times 10^{-6}.
\label{BReta}
\end{eqnarray}
From theoretical side, the SM predictions 
for $BR(B\to \phi K)$\footnote{In order to simplify our notation, from now
on everywhere, where the symbols $B$ and $K$ will appear, they will 
generically
indicate neutral $B^0$ ($\bar{B}^0$) and $K^0$ ($\bar{K}^0$) mesons, 
respectively.}
are in good agreement with Eq.(\ref{BRphi}),
while showing a large discrepancy, being experimentally two to five times
larger, for 
$BR(B\to \eta^{\prime} K)$ in Eq.(\ref{BReta}) 
\cite{BR_eta_SM}.
This discrepancy is not new and it 
has created a growing interest in the subject.
However, since it is observed only in $B\to \eta^{\prime} K$ process, 
mechanisms
based on the peculiar structure of $\eta^{\prime}$ meson, such as 
intrinsic charm \cite{CZ} and gluonium \cite{AS} content, have 
been investigated to solve the puzzle.

Supersymmetry (SUSY) is one of the most popular candidates for physics
beyond the SM \cite{susy}. In SUSY models there are new
sources of CP violation besides the CKM phase \cite{GNR}.
The
soft SUSY breaking (SSB) terms contain several parameters that may be complex,
as may also be the SUSY preserving $\mu$ parameter in the Higgs sector.
Then, new CP violating phases can naturally arise in the SSB sector
of scalar- quarks (squarks) and -leptons (sleptons).
These new phases have significant implications for 
the electric dipole moments (EDM) of the neutron, 
electron, and mercury atom \cite{GNR,abel} 
and can be strongly constrained
by the negative search of CP violation in EDM experiments.
Therefore, it remains a challenge for SUSY to provide an explanation for the 
above mentioned discrepancies in $B$-decays, 
while keeping the above EDMs within their experimental ranges.
This can be possible, for instance, in SUSY models
where the CP violating phases entering in $b\to s$ transitions
are independent from the corresponding ones affecting
EDMs. An interesting example is provided by SUSY models 
with flavor dependent CP violating phases 
\cite{abel,bailin,edm_susy}.

In SUSY framework, the main effect on $B\to \phi K$ is usually
assumed to come from the gluino loop contributions to s-penguin diagrams
\cite{oldPhiKs,PhiKs_gluino,PhiKs_sugra,KK,CFMS}, if R parity is 
conserved.\footnote{Recently, 
SUSY effects to $B\to \phi K$ decays have been analyzed in the context of 
R-parity breaking models \cite{PhiKs_Rparity}.
In this case, as well as in models with extra dimensions \cite{KM}, 
effects are induced at tree-level. 
We will restrict our analysis to the R-parity conserving scenarios, where
SUSY corrections to $B\to \phi K$ process always enter at 1-loop.}
However, charginos could also be responsible for such discrepancy, 
as it has been discussed in \cite{chargino,our},
although $b\to s \gamma$ constraints strongly suppress their contribution
\cite{our}.
Similarly to $B\to \phi K$, one would a priori expect large
effects from SUSY corrections to 
$B\to \eta^{\prime} K$ as well, which may contradict the 
experimental results reported in Eq.(\ref{Seta}).
Possible mechanisms to explain such behavior in the supersymmetric
context have been proposed in Refs.\cite{KK,CFMS,phieta}.

In this paper we perform a complete analysis of SUSY
contributions to the CP asymmetries 
and branching ratios of $B\to \phi K$ and $B\to 
\eta^{\prime} K$ processes. Previous analyses on the same issue
have considered either gluino \cite{PhiKs_gluino} or 
chargino exchanges \cite{our,chargino,PhiKs_sugra}.
We think that in the framework of general SUSY models
a complete analysis involving both effects is in fact needed.
Indeed, we find that a SUSY scenario in which both chargino and gluino 
give sizeable contributions to CP violating processes, is an
interesting viable possibility.
For instance, large effects of 
chargino contributions to CP asymmetries that one would expect to be 
excluded by \bsg constraints \cite{our}, 
could be achieved taking into account gluino exchanges. 
This is due to potentially destructive interferences between chargino and 
gluino amplitudes in $b\to s\gamma$ decay that eventually 
relax $b\to s\gamma$ bounds.

In our analysis we consider both the effects of gluino and chargino 
exchanges by using the mass insertion method
(MIA) \cite{mia}. As known, this method is a useful tool 
for analyzing SUSY contributions to flavor changing 
neutral current processes (FCNC) since it allows
to parametrize, in a model independent way, the main sources of flavor
violations in general SUSY models.

We take into account all the relevant operators involved 
in the effective Hamiltonian for $\Delta B=1$ transition, and provide
analytical expression for the corresponding Wilson coefficients.
We analyze the most interesting scenarios in which one or 
two mass insertions are dominant in both gluino and chargino sector.

An important issue in these calculations is the method of evaluating
the hadronic matrix elements for exclusive hadronic final states,
which may play a crucial role in CP asymmetries.
Many studies have been done  with naive 
factorization approach (NF) \cite{naive,ali}, for the computation
of two-body nonleptonic B decays.
Recently, a new approach has been developed, called QCD factorization (QCDF),
\cite{BBNS,BN,BN2}, which 
offers the possibility to include non-factorizable contributions 
and to calculate the strong phases. The drawback in this approach is that it 
includes undetermined parameters $\rho_{H,A}$ and phases $\phi_{H,A}$, 
characterizing the infrared divergences. 
We critically consider both approaches 
and analyze theoretical uncertainties connected with SUSY predictions.
We provide a comparative study of SUSY contributions from
chargino and gluino to $B\rightarrow \phi K$ and
$B\rightarrow \eta^{\prime} K$ processes in NF and QCDF approaches. 
We also analyze the branching ratios of these decays and 
investigate their correlations with CP asymmetries.
Finally, we discuss the correlations between CP asymmetries of these 
processes and the direct CP asymmetry in $b\to s \gamma$ decay \cite{ACPbsg}.

This paper is organized as follows.
In Section 2 we give the basic ingredients needed for calculations,
with many important formulas provided in the Appendices.
In Section 3 we discuss the QCDF method in the evaluation of 
$B$-meson decay amplitudes of $B\to\phi K$ and $B\to \eta^{\prime} K$.
Moreover, we compare SUSY predictions between NF and QCDF approaches.
Section 4 is devoted to the CP asymmetries of these decay modes.
Both effects from gluino and chargino exchanges are discussed.
In section 5 the branching ratios and their correlations to
asymmetries are considered.
Section 6 contains the analysis of SUSY contributions to 
direct $b\to s\gamma$ asymmetry, and correlations to
the mentioned CP asymmetries are also considered.
Finally, section 7 contains our main conclusions.
%%%%%%%%%%%%%%%%%%%%%%%%%%%%%%%%%%%%%%%%%%%%%%%%%%%%%%%%%%
%
\section{SUSY contributions to the effective Hamiltonian of  
$\Delta B=1$ transitions}
%
%%%%%%%%%%%%%%%%%%%%%%%%%%%%%%%%%%%%%%%%%%%%%%%%%%%%%%%%%%
We start our analysis by considering the supersymmetric effect in 
the non-leptonic $\Delta B=1$ processes.
Such an effect could be a probe for any testable 
SUSY implications in CP violating experiments.
The most general effective Hamiltonian $H^{\Delta B=1}_{\rm eff}$
for these processes can be expressed via the Operator 
Product Expansion (OPE) as \cite{bbl}
\bea
H^{\Delta B=1}_{\rm eff}&=& \left\{\frac{G_F}{\sqrt{2}} 
\sum_{p=u,c} \lambda_p ~\left( C_1 Q_1^p + C_2 Q_2^p + 
\sum_{i=3}^{10} C_i Q_i + C_{7\gamma} 
Q_{7\gamma} + C_{8g} Q_{8g} \right) \right\}
\nonumber\\
&+&\left\{Q_i\to \tilde{Q}_i\, ,\, C_i\to \tilde{C}_i\right\}
\;,
\label{Heff}
\eea
where $\lambda_p= V_{pb} V^{\star}_{p s}$, 
with $V_{pb}$ the unitary CKM matrix elements satisfying the unitarity 
triangle relation 
$\lambda_t+\lambda_u+\lambda_c=0$, and $C_i\equiv C_i(\mu_b)$ are 
the Wilson coefficients at low energy scale 
$\mu_b\simeq {\cal O}(m_b)$. 
The basis $Q_i\equiv Q_i(\mu_b)$ 
is given by the relevant local operators
renormalized at the same scale $\mu_b$, namely
\begin{eqnarray}
Q^p_2 &=& (\bar p b )_{V-A}~~ (\bar s p)_{V-A}\;,~~~~~~~~~~~~~~~
Q^p_1 = (\bar p_{\alpha} b_{\beta})_{V-A}~~ (\bar s_{\beta}
p_{\alpha})_{V-A}\;
\nonumber\\
Q_3 &=& (\bar s b )_{V-A}~~ \sum_q (\bar q q)_{V-A}\;,~~~~~~~~~~~
Q_4 = (\bar s_{\alpha} b_{\beta})_{V-A}~~ \sum_q (\bar q_{\beta}
q_{\alpha})_{V-A}\;,
\nonumber\\
Q_5 &=& (\bar s b )_{V-A}~~ \sum_q (\bar q q)_{V+A}\;,~~~~~~~~~~~
Q_6 = (\bar s_{\alpha} b_{\beta} )_{V-A}~~ \sum_q (\bar q_{\beta}
q_{\alpha})_{V+A}\;,
\nonumber\\
Q_7 &=& (\bar s b )_{V-A}~~ \sum_q\frac{3}{2}  e_q (\bar q q)_{V+A}\;,~~~~~~
Q_8 = (\bar s_{\alpha} b_{\beta} )_{V-A}~~
\sum_q \frac{3}{2} e_q (\bar q_{\beta} q_{\alpha})_{V+A}\;,
\nonumber\\
Q_9 &=& (\bar s b )_{V-A}~~
\sum_q \frac{3}{2} e_q (\bar q q)_{V-A}\;,~~~~~~
Q_{10} = (\bar s_{\alpha} b_{\beta} )_{V-A}~~
\sum_q \frac{3}{2} e_q (\bar q_{\beta} q_{\alpha})_{V-A}\;,
\nonumber\\
Q_{7\gamma} &=& \frac{e}{8\pi^2} m_b \bar s \sigma^{\mu\nu}
(1+ \gamma_5) F_{\mu \nu} b\;,~~~~~~~
Q_{8g}= \frac{g_s}{8\pi^2} m_b \bar s_{\alpha} \sigma^{\mu\nu}
(1+ \gamma_5) G^A_{\mu \nu} t^A_{\alpha\beta} b_{\beta}\;.
\label{Qbasis}
\end{eqnarray}
Here $\alpha$ and $\beta$ stand for color indices, and $t^A_{\alpha\beta}$
the $SU(3)_c$ color matrices, 
$\sigma^{\mu\nu}=\frac{1}{2}i[\gamma^{\mu},\gamma^{\nu}]$. Moreover,
$e_q$ are quark electric charges in unity of $e$,
$(\bar q q)_{V \pm A}\equiv \bar q \gamma_\mu
(1 \pm \gamma_5) q$, and $q$ runs over $u$, $d$, $s$, $c$, 
and $b$ quark labels.
In the SM only the first part  
of right hand side of Eq.(\ref{Heff}) (inside first curly brackets)
containing operators $Q_i$ will contribute, where
$Q^p_{1,2}$ refer to the current-current operators, 
$Q_{3-6}$ to the QCD penguin operators,  
and $Q_{7-10}$ to the electroweak penguin operators, while $Q_{7\gamma}$ and
$ Q_{8g}$ are the magnetic and the chromo-magnetic dipole operators,
respectively.
In addition, operators $\tilde{Q}_{i}\equiv \tilde{Q}_{i}(\mu_b)$
are obtained from $Q_{i}$ by the chirality exchange 
$(\bar{q}_1 q_2)_{V\pm A} \to (\bar{q}_1 q_2)_{V\mp A}$. 
Notice that in the SM the coefficients $\tilde{C}_{i}$ identically 
vanish due to the V-A structure of charged weak currents, 
while in MSSM they can 
receive contributions from both chargino and gluino exchanges
\cite{bbmr,ggms}.
\\

Due to the asymptotic freedom of QCD, 
the calculation of hadronic weak decay amplitudes
can be factorized by the product of 
{\it long} and {\it short} distance contributions. 
The first ones, that will be analyzed in the next section,
are related to the evaluation of 
hadronic matrix elements of $Q_i$ and  contain 
the main uncertainty of our predictions. 
On the other hand, the latter are contained in 
the Wilson coefficients $C_i$ and they
can be evaluated in perturbation theory with high precision \cite{bbl}.
For instance, all the relevant contributions of particle spectra above 
the W mass ($m_W$) scale, including SUSY particle exchanges, 
will enter in $C_i(\mu_W)$ at $\mu_W \simeq \mathcal{O}(m_W)$ scale.

The low energy coefficients $C_i(\mu_b)$ can be extrapolated 
from the high energy ones $C_i(\mu_W)$ by solving the renormalization 
group equations for QCD and QED in the SM.
The solution is generally expressed as follows \cite{bbl}
\begin{equation}
C_i(\mu) = \sum_j \hat{U}_{ij}(\mu, \mu_W) \, C_j(\mu_W),
\end{equation}
where $\hat{U}_{ij}(\mu, \mu_W)$ is the evolution matrix which takes into
account the re-summation of the terms proportional to large logs
$(\alpha_s(\mu_W) \log (\mu_W/\mu_b))^n$ (leading),  
$(\alpha_s^{2}(\mu_W) \log(\mu_W/\mu_b))^n$ (next-to-leading), etc., in QCD.

In our analysis we include the next-to-leading order (NLO) 
corrections in QCD and QED
for the Wilson coefficients $C_{i=1-10}$ as given in 
Ref. \cite{bbl}, while for $C_{7\gamma}(\mu)$
and $C_{8g}(\mu)$ we include only 
the leading order (LO) ones.\footnote{
However, in the evaluation of the BR of $b\to s \gamma$ decay, 
the complete NLO corrections in $C_{7\gamma}(\mu_b)$
have been taken into account \cite{bsg_NLO}.} 
The expressions for the evolution matrix $\hat{U}_{ij}(\mu, \mu_W)$
at NLO in QCD and QED can be found in Ref.\cite{bbl}.
The reason for retaining only the LO accuracy in $C_{7\gamma}(\mu)$
$C_{8g}(\mu)$ is that the matrix elements of the dipole operators 
enter the decay amplitudes only at the NLO \cite{BBNS}.
\\

Next we discuss the SUSY contributions to the effective Hamiltonian in 
Eq.(\ref{Heff}).
The modifications caused by supersymmetry appear only 
in the boundary conditions
of the Wilson coefficients at $\mu_W$ scale and they can be computed through 
the appropriate matching with one-loop Feynman diagrams 
%in Fig.~\ref{Fig1},
where Higgs, neutralino, gluino, and chargino are exchanged, see for 
instance Refs.\cite{bbmr,GG,ggms}.
Only chargino and gluino contributions can provide a potential 
source of new CP violating phase in MSSM and could account 
for the observed large deviations in $B\to \phi K_S$ asymmetry. 
In principle, the neutralino exchange diagrams involve the same mass 
insertions as the gluino 
ones, but they are strongly suppressed compared to the latter.
For these reasons we neglect neutralino in our analysis.
The charged Higgs contributions cannot generate any new source 
of CP violation in addition to the SM one, or
any sizeable effect to operators beyond the SM basis $Q_i$. 
However, when charged Higgs contributions are taken into account
together with chargino or gluino exchanges, their effect is relevant.
In particular, as we will show in the next sections,
due to destructive interferences with chargino and gluino amplitudes, 
$b\to s\gamma$ constraints can be relaxed allowing 
sizeable contributions to the the CP asymmetries.

The results for Wilson coefficients at $\mu_W$ scale 
can be expressed as follows
\begin{eqnarray}
C_i(\mu_W) = C_i^{W} +  C_i^{H} + \lambda_t^{-1}\left\{
C_i^{{\chi^{}}} +C_i^{{\tilde g}}\right\}
\end{eqnarray}
where $C_i^{W}$, $C_i^{H}$, 
$C_i^{\chi^{}}$, and $C_i^{\tilde{g}}$ correspond to
the $W$, charged Higgs, chargino, and gluino exchanges respectively.
In our analysis we will impose 
the boundary conditions for $C_i^{\chi^{}}$, and $C_i^{\tilde{g}}$
at the scale $\mu_W=m_W$, 
although they should apply to the energy scale at which 
SUSY particles are integrated out, namely $M_S$.
However, these threshold corrections, originating from the 
mismatch of energy scales, are numerically not significant since the running
of $\alpha_s$ from $M_S$ to $m_W$ is not very steep.

Finally, for NDR renormalization scheme, the electroweak contributions to 
the Wilson coefficients are given by \cite{bbmr,GG}
\begin{eqnarray}
C_1^{W} &=& \frac{14 \alpha_s}{16 \pi},\nonumber\\
C_2^{W} &=& 1-\frac{11}{6} \frac{\alpha_s}{4\pi}, \nonumber\\
C_3^{(W,H,\chi^{})} &=& \frac{\alpha}{6 \pi} \frac{1}{\sin^2\theta_w}
\left(B_d^{(W,H,\chi^{})} + \frac{1}{2} B_u^{(W,H,\chi^{})} + 
C^{(W,H,\chi^{})}\right)- 
\frac{\alpha_s}{24 \pi} E^{(W,H,\chi^{})},\nonumber\\
C_4^{(W,H,\chi^{})}&=& \frac{\alpha_s}{8 \pi} E^{(W,H,\chi^{})},\nonumber\\
C_5^{(W,H,\chi^{})}&=& - \frac{\alpha_s}{24 \pi} E^{(W,H,\chi^{})},\nonumber\\
C_6^{(W,H,\chi^{})}&=& \frac{\alpha_s}{8 \pi} E^{(W,H,\chi^{})},\nonumber\\
C_7^{(W,H,\chi^{})}&=& \frac{\alpha}{6 \pi} \left( 4 C^{(W,H,\chi^{})}+
D^{(W,H,\chi^{})}\right),\nonumber\\
C_8^{(W,H,\chi^{})}&=& 0,\nonumber\\
C_9^{(W,H,\chi^{})}&=& \frac{\alpha}{6 \pi} \left( 4 C^{(W,H,\chi^{})}+
D^{(W,H,\chi^{})}+ \frac{1}{\sin^2\theta_w} \left(- B_d^{(W,H,\chi^{})} 
+ B_u^{(W,H,\chi^{})} - 4 C^{(W,H,\chi^{})}\right)\right),\nonumber\\
C_{10}^{(W,H,\chi^{})}&=& 0,\nonumber\\
C_{7\gamma}^{(W,H,\chi^{})}&=& M_{\gamma}^{(W,H,\chi^{})},\nonumber\\
C_{8g}^{(W,H,\chi^{})}&=& M_{g}^{(W,H,\chi^{})}
\label{C_EW}
\end{eqnarray}
where $\alpha_s$ and $\alpha$ are evaluated at $m_W$ scale.
The functions appearing above, 
include the contributions from photon-penguins ($D$), $Z$-penguins
($C$), gluon-penguins ($E$), boxes with external down quarks 
($B_d$) and up-quarks $(B_u)$, the magnetic- ($M_{\gamma}$), and
the chromo-magnetic-penguins ($M_{g}$).
The corresponding SM results  are \cite{bbl}
$D^{W}\equiv {\rm D}(x_t)$,
$C^{W}\equiv {\rm C}(x_t)$, $E^{W}\equiv {\rm E}(x_t)$,
and $B^{W}\equiv {\rm B}(x_t)$ with  $x_t=m_{t}^2/m_W^2$, 
and analogously for charged Higgs \cite{GG}, 
$D^{H}\equiv {\rm D_H}(x_H)$,
$C^{H}\equiv {\rm C_H}(x_H)$, $E^{H}\equiv {\rm E_H}(x_H)$,
and $B^{H}\equiv {\rm B_H}(x_H)$ with  $x_H=m_{t}^2/m_H^2$, 
where the loop functions ${\rm B,C,D,E}$ and ${\rm B_H,C_H,D_H,E_H}$ and 
are provided in Appendix C. Regarding the SM and charged Higgs contributions
to magnetic-penguins, we have \cite{bbmr}
\bea
M_{\gamma}^W&=&-x_t \left(F_1(x_t)+\frac{3}{2}F_2(x_t)\right),
~~~~~M_{g}^W=-\frac{3}{2} x_t F_1(x_t)
\nonumber\\
M_{\gamma}^H&=&-\frac{x_H}{2} \left(
\left(\frac{2}{3}F_1(x_H)+F_2(x_H)\right)\cot^2{\beta}+
\frac{2}{3}F_3(x_H)+F_4(x_H)\right)
\nonumber\\
M_{g}^H&=&-\frac{x_H}{2}\left(F_1(x_H)\cot^2{\beta}+F_3(x_H)\right)\, ,
\nonumber
\eea
where the functions $F_i(x)$ are reported in Appendix C.
Finally, the gluino and chargino exact contributions to 
the expressions  appearing in
Eq.(\ref{C_EW}) can be found in Refs.\cite{bbmr,GG}, while here
we will provide only the corresponding results 
in the so called mass insertion approximation.

Regarding the SUSY contributions to the opposite chirality operators 
$\tilde{Q}_i$, we stress that while in the SM the Wilson coefficients 
$\tilde{C}_i$ identically vanish, in SUSY models 
chargino and gluino exchanges could sizeably affect these coefficients.
However, in case of charginos,
these effects are quite small being proportional to the Yukawa 
couplings of light quarks \cite{our} and so we will not include 
them in our analysis.
Moreover, we have also neglected the small contributions
to $C_{1,2}^{\chi^{}}$ 
coming from box diagrams, where both chargino and gluino are exchanged
\cite{our,GG}.

As mentioned in the introduction, 
in order to perform a model independent analysis of FCNC processes in 
general SUSY models, it is very convenient to adopt the mass insertion
approximation (MIA) method \cite{mia}.
This method has been applied in many analyses
of FCNC processes in $K$, $D$, and $B$ meson sector 
and leptonic sector, mediated by gluino and neutralino 
exchanges, respectively \cite{mia}.
More recently, it has been extended to the sector of FCNC processes
mediated by chargino exchanges \cite{our,BBcharg,KL}.
In particular, in \cite{our},
the chargino functions appearing in Eq.(\ref{C_EW}) have been 
calculated at the first order in MIA.

Let us briefly recall the main steps of this approximation.
In MIA framework, one chooses
a basis (called super-CKM basis) where the couplings of 
fermions and sfermions to neutral
gaugino fields are flavor diagonal. 
In this basis, the interacting Lagrangian involving charginos is given by
\bea
\mathcal{L}_{q\tilde{q}\tilde{\chi}^+} &=& - g~ 
\sum_k \sum_{a,b}~ \Big(~ V_{k1}~ 
K_{ba}^*~ \bar{d}_L^a~ (\tilde{\chi}^+_k)^*~ \tilde{u}^b_L -  U_{k2}^*~ 
(Y_d^{\mathrm{diag}} . K^+)_{ab}~ \bar{d}_R^a~ (\tilde{\chi}^+_k)^*~ 
\tilde{u}^b_L \nonumber\\ && - V_{k2}~ (K.Y_u^{\mathrm{diag}})_{ab}~
\bar{d}_L^a~ (\tilde{\chi}^+_k)^*~ \tilde{u}^b_R~ \Big),
\label{vertices}
\eea
where $q_{R,L}=\frac{1}{2}(1\pm \gamma_5)q$, and 
contraction of color and Dirac indices is understood.
Here $Y_{u,d}^{\mathrm{diag}}$ are the diagonal Yukawa matrices, and
$K$ stands for the CKM matrix.
The indices $a,b$ and $k$ label
flavor and chargino mass eigenstates, respectively, and
$V$, $U$ are the chargino mixing matrices defined by
\begin{equation}
U^* M_{\tilde{\chi}^+} V^{-1} = \mathrm{diag}(m_{\tilde{\chi}_1^+},
m_{\tilde{\chi}_2^+}),~ \mathrm{and}~ M_{\tilde{\chi}^+} = \left(
\begin{array}{cc}
M_2 & \sqrt{2} m_W \sin \beta \\
\sqrt{2} m_W \cos \beta & \mu
\end{array}\right) \;,
\label{chmatrix}
\end{equation}
where $M_2$ is the weak gaugino mass, $\mu$ is the supersymmetric 
Higgs mixing term, and $\tan\beta$ is the ratio of the 
vacuum expectation value (VEV) of the up-type Higgs to the VEV 
of the down-type Higgs\footnote{This $\tan\beta$ should not be
confused with the angle $\beta$ of the unitarity triangle.} .
As one can see from Eq.(\ref{vertices}),
the higgsino couplings are suppressed by Yukawas of the light quarks, and 
therefore they are negligible, except for the stop--bottom
interaction which is directly enhanced by the top Yukawa ($Y_t$). 
In our analysis we neglect the higgsino contributions 
proportional to the Yukawa couplings of light quarks 
with the exception of the bottom Yukawa $Y_b$, 
since its effect could be enhanced by large $\tan{\beta}$.
However, it is easy to show that this vertex cannot 
affect dimension six operators of the effective Hamiltonian
for $\Delta B=1$ transitions (operators $Q_{i=1-10}$ in
Eq.(\ref{Heff})) 
and only interactions involving left down quarks will contribute.
On the contrary, contributions proportional to bottom Yukawa $Y_b$ enter
in the Wilson coefficients of dipole operators ($C_{7\gamma}$, $C_{8g}$) 
due to the chirality flip of $b\to s \gamma$ and $b\to s g$ transitions.
%------------------------------------------------------------
\begin{figure}[tpb]
\begin{center}
\dofig{3.1in}{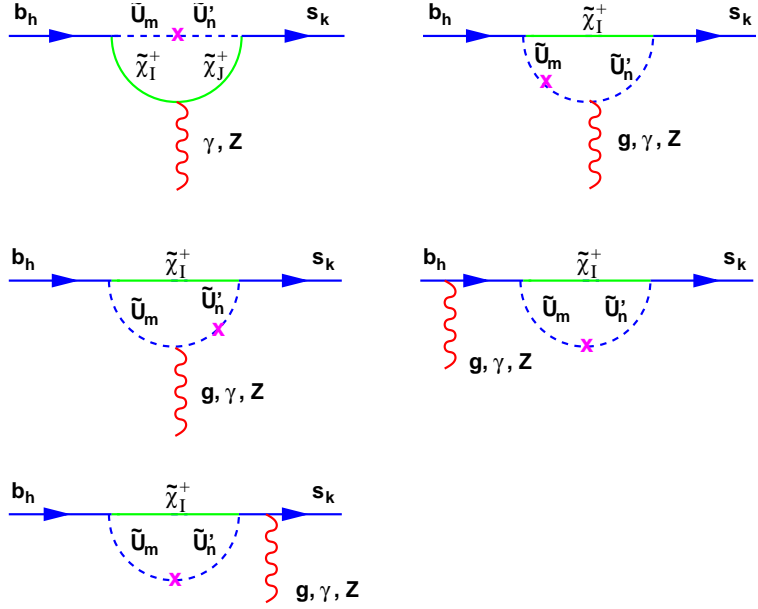}
\end{center}
\caption{\small Penguin diagrams for $\Delta B=1$ transitions with
chargino ($\chi_{\rm I}^+$) exchanges at the first order in mass insertion.
Here $\tilde{U},\tilde{U}^{\prime}=
\{\tilde{u},\tilde{c},\tilde{t}\}$, with indices
$h,k,m,n=\{L,R\}$ and ${\rm I,J}=\{1,2\}$. The cross symbol in the
squark propagator indicates the mass insertion.
The corresponding diagrams at zero order in mass insertion 
are simply obtained by removing the mass insertion in the propagators 
of up-type squarks ($\tilde{U}$). }
\label{penguin_ch}
\end{figure}
%----------------------------------------------------------
%------------------------------------------------------------
\begin{figure}[tpb]
\begin{center}
\dobox{3.1in}{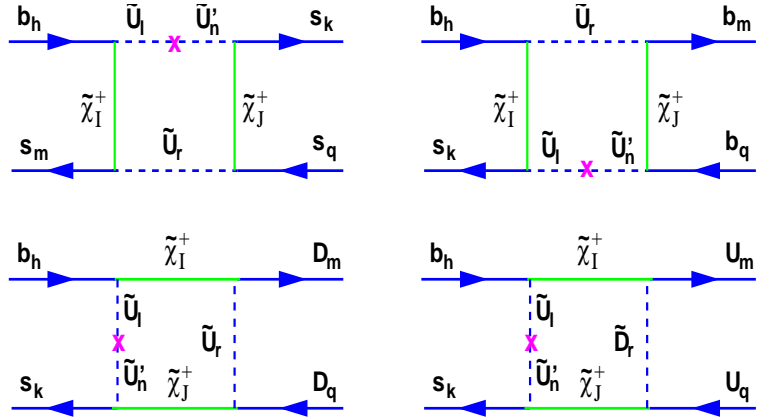}
\end{center}
\caption{\small Box diagrams for $\Delta B=1$ transitions with
chargino exchanges at the first order in mass insertion, where
$\tilde{U},\tilde{U}^{\prime}=\{\tilde{t},\tilde{c},\tilde{u}\}$, $U=\{c,u\}$,
and $D=\{b,s,d\}$, where $h,k,l,n,r,m,q=\{L,R\}$.
}
\label{box_ch}
\end{figure}
%----------------------------------------------------------
%------------------------------------------------------------
\begin{figure}[tpb]
\begin{center}
\dofig{3.1in}{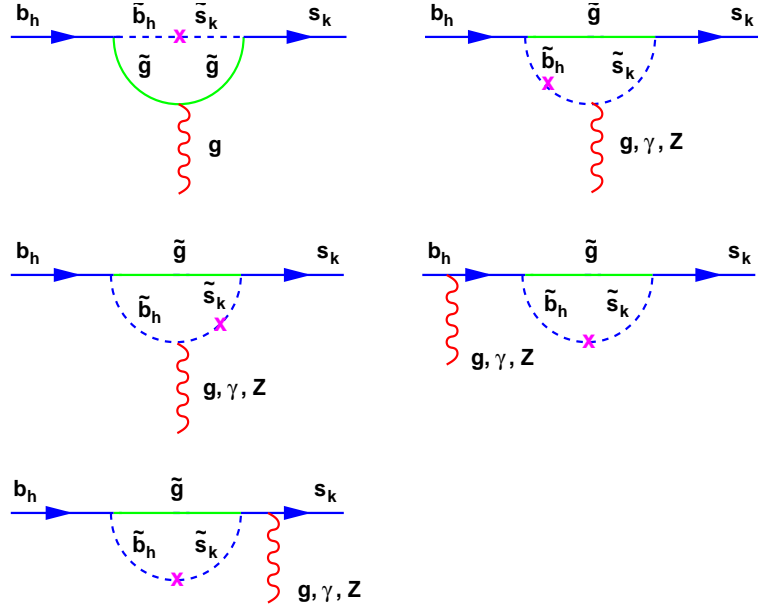}
\end{center}
\caption{\small Penguin diagrams for $\Delta B=1$ transitions 
with gluino exchanges at the first order in mass insertion, 
where $h,k=\{L,R\}$.}
\label{penguin_gl}
\end{figure}
%----------------------------------------------------------
%------------------------------------------------------------
\begin{figure}[tpb]
\begin{center}
\dofig{3.1in}{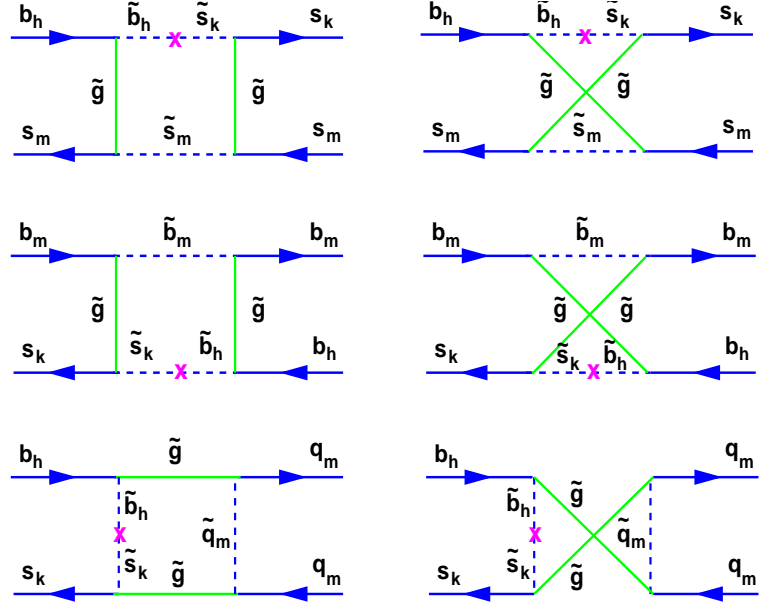}
\end{center}
\caption{\small Box diagrams for $\Delta B=1$ transitions with
gluino exchanges at the first order in mass insertion, where
$q=\{b,c,s,d,u\}$ and $h,k,m=\{L,R\}$.
}
\label{box_gl}
\end{figure}
%----------------------------------------------------------

As mentioned in the case of MIA, the flavor 
mixing is displayed by the non--diagonal entries of the sfermion 
mass matrices. 
Denoting by $\Delta^q_{AB}$ the off--diagonal terms 
in the sfermion $(\tilde{q}=\tilde{u},\tilde{d})$ 
mass matrices for the up and down, respectively, 
where $A,B$ indicate chirality couplings to
fermions $A,B=(L,R)$, the A--B squark propagator can be expanded as 
\begin{equation}
\langle \tilde{q}^a_A \tilde{q}^{b*}_B \rangle = 
i \left(k^2{\bf 1} - \tilde{m}^2 {\bf 1}
- \Delta_{AB}^q\right)^{-1}_{ab} \simeq \frac{i \delta_{ab}}
{k^2 - \tilde{m}^2} + 
~\frac{i (\Delta_{AB}^q)_{ab}}{(k^2 -\tilde{m}^2)^2} + 
{\cal O}(\Delta^2) ,
\end{equation}
where $q=u,d$ selects up or down sector, respectively, 
$a,b=(1,2,3)$ are flavor indices, ${\bf 1}$ 
is the unit matrix, and $\tilde{m}$ is the average squark mass.
As we will see in the following, it is convenient to parametrize 
this expansion in terms of the dimensionless quantity 
$(\delta_{AB}^q)_{ab} \equiv (\Delta^q_{AB})_{ab}/\tilde{m}^2$.
At the first order in MIA, the penguin and box diagrams which contribute to the
$\Delta B=1$ effective Hamiltonian are given in Figs. \ref{penguin_ch}
and \ref{box_ch}, respectively. Evaluating the diagrams in 
Figs. \ref{penguin_ch} and \ref{box_ch} by retaining only terms proportional
to bottom- and top-quark Yukawa couplings and performing the
matching, the chargino contributions 
to the Wilson coefficients
in Eqs.(\ref{C_EW}) can be determined from the following relations \cite{our}
%-----------------------------------------
\begin{eqnarray}
F^{\chi}\!&=\!&\Big[ \sum_{a,b} K^{\star}_{a2} K_{b3}\du{LL}{ba} \Big] 
R_F^{LL}
+ \Big[\sum_a K^{\star}_{a2} K_{33} \du{RL}{3a}\Big] Y_t \, R_F^{RL}
\nonumber \\
&+ &\Big[ \sum_a K^{\star}_{32} K_{a3} \du{LR}{a3} \Big] Y_t\, R_F^{LR}
+ \Big[ K^{\star}_{32} K_{33}  \left(\du{RR}{33} R_F^{RR}
+ R_F^0 \right) \Big] Y_t^2 \, , 
\label{RFch}
\end{eqnarray}
%-----------------------------------------
where the symbol $F \equiv \left\{D,C,E, B_d, B_u,M_{\gamma},M_g\right\}$, 
and the detailed expressions 
for $R_{F}^{LL}$, $R_{F}^{LR}$, $R_{F}^{RL}$, $R_{F}^{RR}$, and
$R_{F}^{0}$ can be found in Appendix A\footnote{The expression 
$R_{F}^{0}$ was missing in \cite{our}.  The contribution is 
${\cal O}(\lambda^{2})$ and does not change the numerical results.}
. 
Notice that 
the last term in Eq.(\ref{RFch}), proportional to $R_F^0$, is independent 
of mass insertions.
This is due to the fact that for chargino exchanges
the super-GIM mechanism is only partially effective,
when only squarks but not quarks are taken to be degenerate. 

Here we will just concentrate on the dominant contributions 
which turn out to be due to the chromo-magnetic ($M_{g}$) penguin and 
$Z$-penguin ($C$) diagrams \cite{our}. 
From the above expressions, it is clear that $LR$ and $RR$ contributions are 
suppressed by order $\lambda^2$ or $\lambda^3$, 
where  $\lambda=\sin{\theta_c}\simeq 0.22$, with $\theta_c$ the Cabibbo angle.
In our analysis we adopt the approximation of
retaining only terms proportional to order $\lambda=\sin{\theta_c}$. 
In this case, Eq.(\ref{RFch}) simplifies as follows \cite{our}
\begin{equation}
F^{\chi}=\xi_{LL} R_F^{LL}+ Y_t\, \xi_{RL} R_F^{RL}\, ,
\end{equation}
where $\xi_{LL}=\du{LL}{32}+\lambda \, \du{LL}{31}$ and 
$\xi_{RL}=\du{RL}{32}+\lambda \, \du{RL}{31}$.

The functions $R^{LL}_F$ and $R^{RL}_F$ depend on
the SUSY parameters through the chargino masses ($m_{\chi_i}$), 
squark masses ($\tilde{m}$) and the entries of
the chargino mass matrix. For instance for $Z$ and 
magnetic (chromo-magnetic) dipole penguins $R_C^{LL,RL}$
and $R_{M_{\gamma (g)}}^{LL,RL}$, respectively, we have 
\begin{eqnarray}
R_C^{LL}&=&
\sum_{i=1,2}|V_{i1}|^2 \, P_C^{(0)}(\bar x_i)
+\sum_{i,j=1,2} \left[ U_{i1}V_{i1}U_{j1}^{\star}V_{j1}^{\star}\, 
P_C^{(2)}(x_i,x_j) \right. 
\nonumber\\
&&+ \left. |V_{i1}|^2 |V_{j1}|^2
\left(\frac{1}{8}-P_C^{(1)}(x_i, x_j)\right)\right],
\nonumber
\\
R_C^{RL}&=&-\frac{1}{2}
\sum_{i=1,2}\, V_{i2}^{\star}V_{i1}\,  P_C^{(0)}(\bar x_i)
- \sum_{i,j=1,2}\, V_{j2}^{\star}V_{i1}\left(
U_{i1}U_{j1}^{\star}\, P_C^{(2)}(x_i, x_j)
\right.\nonumber \\
&&+
\left. 
V_{i1}^{\star} V_{j1}\, P_C^{(1)}(x_i, x_j)\right),
\nonumber
\\
R_{M_{\gamma, g}}^{LL}&=&\sum_i |V_{i1}|^2\,
x_{Wi}\, P_{M_{\gamma,g}}^{LL}(x_i) - Y_b\sum_i V_{i1} U_{i2}\,
x_{Wi}\, \frac{m_{\chi_i}}{m_b} P_{M_{\gamma,g}}^{LR}(x_i),
\nonumber
\\
R_{M_{\gamma, g}}^{RL}&=&
-\sum_i V_{i1}V_{i2}^{\star}\,
x_{Wi}\, P_{M_{\gamma,g}}^{LL}(x_i),
\label{Rterms}
\end{eqnarray}
where $x_{\W i}=m_W^2/m_{\chi_i}^2$, 
$x_{i}=m_{\chi_i}^2/\tilde{m}^2$, $\bar x_i =\tilde{m}^2/m_{\chi_i}^2$, and
$x_{ij}=m_{\chi_i}^2/m_{\chi_j}^2$.
The loop functions $P_C^{(1,2)}(x,y)$, 
$P_{M_{\gamma, g}}^{LL(LR)}(x)$ are provided in Appendix C.
Finally, $U$ and $V$ are the matrices that diagonalize the chargino 
mass matrix, defined in Eq.~(\ref{chmatrix}).

It is worth mentioning that the large effects of chargino contributions
to $C_{7\gamma}$ and $C_{8g}$ come from the terms in 
$R_{M_{\gamma}}^{LL}$ and $R_{M_{g}}^{LL}$, respectively, which are
enhanced by $m_{\chi_i}/m_b$ in Eq.(\ref{Rterms}). 
However, these terms are 
also multiplied by the Yukawa bottom $Y_b$,  
which leads to enhancing the coefficients of the LL 
mass insertion in $C_{7\gamma}$ and  $C_{8 g}$ at large $\tan\beta$. 
As we will see later on, this effect will play a crucial role in chargino
contributions to $B\to \phi(\eta^{\prime}) K$ decays at 
large $\tan\beta$. 

We will also consider the case in which 
the mass of stop-right ($m_{{\tilde t}_R}$)
is lighter than other squarks. In this case
the functional form of the $R_{F}$ remains unchanged, while
only the expressions of $R_F^{RL}$ should be modified in the way described 
in Appendix B.
\\

Now let us turn to the gluino contributions in the $b\to s$ transition.
In the super-CKM basis, the quark-squark-gluino interaction is given by:
\begin{equation}
\mathcal{L}_{d \tilde{d} \tilde g}=
\sqrt{2}\, g_s \, T^A_{\alpha \beta} \left[
\left(\bar d^{\alpha} \,P_L\, \tilde g^A\right) \,\tilde d^{\beta}_R -
\left(\bar d^{\alpha} \,P_R \,\tilde g^A\right) 
\,\tilde d^{\beta}_L + h.c. \right]\,,
\end{equation}
where $\tilde{g}^A$ are the gluino Majorana fields,  
$\tilde{d}^{\beta}_{R,L}$ the squark
fields, $T^A$ are the $SU(3)_c$ generators, and $\alpha,\beta$ 
are color indices.
The dominant gluino contributions are due to the QCD penguin diagrams, and
the magnetic and chromo-magnetic dipole operators.
At the first order in MIA, the penguin and box diagrams are shown in
Figs. \ref{penguin_gl} and \ref{box_gl}, respectively.
Performing the matching, 
the gluino contributions to the corresponding 
Wilson coefficients at SUSY scale 
are given by \cite{ggms}\footnote{Note that the 
Wilson coefficients of Eq.(\ref{Cgluino})
are different from those reported in Ref.\cite{ggms}
by a minus sign and a rescaling factor. This is due to the different
convention for the Wilson coefficients in the effective Hamiltonian of
Eq.(\ref{Heff}).}
\begin{eqnarray}
C^{\tilde{g}}_3  &=& -\frac{\alpha_s^2}{2 \sqrt{2}G_F m_{\tilde{q}}^2}
\dd{LL}{23} \left[ -\frac{1}{9} B_1(x) - \frac{5}{9} B_2(x) -
\frac{1}{18} P_1(x) -\frac{1}{2} P_2(x) \right],\nonumber\\
C^{\tilde{g}}_4  &=& -\frac{\alpha_s^2}{2\sqrt{2} G_F  m_{\tilde{q}}^2}
\dd{LL}{23}
\left[ -\frac{7}{3} B_1(x) + \frac{1}{3} B_2(x) + \frac{1}{6} P_1(x)
+\frac{3}{2} P_2(x) \right],\nonumber\\
C^{\tilde{g}}_5  &=& -\frac{\alpha_s^2}{2\sqrt{2}G_F m_{\tilde{q}}^2}
\dd{LL}{23}
\left[ \frac{10}{9} B_1(x) + \frac{1}{18} B_2(x) - \frac{1}{18} P_1(x)
-\frac{1}{2} P_2(x) \right],\nonumber \\
C^{\tilde{g}}_6  &=& -\frac{\alpha_s^2}{2\sqrt{2} G_F m_{\tilde{q}}^2}
\dd{LL}{23}
\left[ -\frac{2}{3} B_1(x) + \frac{7}{6} B_2(x) + \frac{1}{6} P_1(x)
+\frac{3}{2} P_2(x) \right],\nonumber\\
C^{\tilde{g}}_{7\gamma} \!\! &=&\!\!\frac{8 \alpha_s \pi}
{9\sqrt{2} G_F m_{\tilde{q}}^2}\!\left[
\!\dd{LL}{23} M_3(x)\!+\!
\dd{LR}{23}\frac{m_{\tilde{g}}}{m_b}
 M_1(x)\!\right]\!,
\nonumber\\
C^{\tilde{g}}_{8g} \!\! &=&\!\!\frac{ \alpha_s \pi}
{\sqrt{2}G_F m_{\tilde{q}}^2}\!\left[
\!\dd{LL}{23}\left( \frac{1}{3} M_3(x)\! + \!3 M_4(x)\right)\!+\!
\dd{LR}{23}\frac{m_{\tilde{g}}}{m_b}
\left(\!\frac{1}{3} M_1(x) \!+\! 3 M_3(x)\right)\!\right]\!,
\label{Cgluino}
\end{eqnarray}
where $\tilde{C}_{i,8g}$ are obtained from $C_{i,8g}$ by
exchanging $L \leftrightarrow R$ in $(\delta_{AB}^d)_{23}$.
The functions appearing in these expressions
can be found in appendix C, with $x=m^2_{\tilde{g}}/m^2_{\tilde{q}}$.

Now we would like to comment about the chiral enhancement in
$C^{\tilde{g}}_{7\gamma, 8g}$ due to the $\dd{LR}{23}$ mass insertion. 
As for chargino in Eq.(\ref{Rterms}), the term proportional to 
$\dd{LR}{23}$ in $C^{\tilde{g}}_{7\gamma, 8g}$ in Eq.(\ref{Cgluino}), 
has also the large enhancement factor 
$m_{\tilde{g}}/m_b$ in front. Moreover, contrary to the chargino case,
this term is not suppressed by the bottom Yukawa coupling. 
As we will see later on, this enhancement factor will be responsible
for the dominant gluino effects in $B\to \phi (\eta^{\prime}) K$ decays.

In concluding this section, we emphasize that in the SM the $b \to s$ 
transition process is dominated by the top quark mediated penguin diagram,
which does not include any CP violating phase. Therefore, this process, as 
observed at the B-factories, opens up the possibility to probe virtual effects 
from new sources of flavor structure and CP violation.
The SUSY contributions through gluino and chargino exchanges are 
independent. The ones from gluino depend on the flavor structure of 
the down squark sector, namely $(\delta^d_{AB})_{23}$, while the other ones
from chargino depend on the up squark sector, particularly 
$(\delta^u_{AB})_{32}$ and $(\delta^u_{AB})_{31}$. So, depending on the
constraints imposed on the flavor structure of the down or up sector
(for instance from $b\to s \gamma$ decay, $K-\bar{K}$ and $B-\bar{B}$ mixing),
gluino or chargino exchanges could give sizeable effects. 
As known, in many SUSY scenarios the lighter 
chargino is expected to be one of the lightest supersymmetric particles. 
Thus, it could contribute significantly in the one-loop processes. 
However, even though gluino in most models is expected to be heavier
than chargino, it is a strongly interacting particle, and may give the
dominant effect as well.
%
%%%%%%%%%%%%%%%%%%%%%%%%%%%%%%%%%%%%%%%%%%%%%%%%%%%%%%%%%%%%%%%%%%%%%%
%
\section{$B\to \phi(\eta^{\prime}) K$ in QCD factorization approach}
%
%%%%%%%%%%%%%%%%%%%%%%%%%%%%%%%%%%%%%%%%%%%%%%%%%%%%%%%%%%%%%%%%%%%%%%%
%
The calculation of $B\to \phi(\eta^{\prime}) K$ decays involves the
evaluation of the hadronic
matrix elements of related operators in the effective Hamiltonian, 
which is the most 
uncertain part of this calculation. 
In the limit in which $m_b \gg \Lambda_{QCD}$ and neglecting QCD
corrections in $\alpha_s$, 
the hadronic matrix elements of B meson decays in two mesons 
can be factorized, for example for $B\to M_1 M_2$, in the form
\begin{equation}
\langle M_1 M_2|Q_i|\bar{B}^0\rangle =\langle M_1|j_1|\bar{B}^0\rangle 
\times \langle M_2|j_2|0\rangle
\end{equation}
where $M_{1,2}$ indicates two generic mesons, 
$Q_i$ are local four fermion operators of the effective Hamiltonian in
Eq.(\ref{Heff}), and $j_{1,2}$ represent bilinear quark currents.
Then, the final results can be usually parametrized 
by the product of the decay constants and the transition form factors. 
This approach is known as naive factorization (NF) \cite{naive,ali}. 
Then, the hadronic matrix elements for $B\to \phi K$ are 
given by~\cite{ali}, 
\begin{eqnarray}
\langle \phi\bar{K}^0  | Q_1 |\bar{B}^0\rangle &=& 0,~~~~~~~~~ 
\langle \phi\bar{K}^0 | Q_2 |
\bar{B}^0\rangle = 0, ~~~~~~~~~ 
\langle\phi\bar{K}^0 | Q_3 |\bar{B}^0\rangle = \frac{4}{3}X,
\nonumber\\
\langle \phi\bar{K}^0 | Q_4 |\bar{B}^0\rangle &=& \frac{4}{3}X,~~~~~~
\langle \phi\bar{K}^0 | Q_5 |\bar{B}^0\rangle = X, ~~~~~~~~~
\langle \phi\bar{K}^0 | Q_6 |\bar{B}^0\rangle = \frac{1}{3}X , \nonumber\\
\langle \phi\bar{K}^0 | Q_7 |\bar{B}^0\rangle &=& \frac{3}{2} e_sX,~~~~
\langle \phi\bar{K}^0 | Q_8 |\bar{B}^0\rangle = \frac{1}{2} e_s X,~~~~~
\langle \phi\bar{K}^0 | Q_9 |\bar{B}^0\rangle = 3 e_sX, \nonumber\\
\langle \phi\bar{K}^0 | Q_{10} |\bar{B}^0\rangle &=& e_s X,~~~~~~
\langle \phi\bar{K}^0 | Q_{7\gamma} |\bar{B}^0\rangle = 0,\nonumber\\
\label{ME_NF}
\end{eqnarray}
where for color number $N_c=3$ one gets
\begin{equation}
X=2F_+^{B\to K}(m_{\phi}^2)f_{\phi}m_{\phi}(p_K \cdot \epsilon_{\phi}).
\label{X}
\end{equation}
In Eq.(\ref{X}), $m_{\phi}$ is the $\phi$ meson mass, 
$F_+^{B\to K}(m_{\phi}^2)$ is the transition form factor evaluated at
transfered momentum of the order of $m_{\phi}$ scale, $p_K$ stands for
$K$ momentum, and $\epsilon_{\phi}$ is the $\phi$ polarization vector.
In our analysis, for the parameters above, we will 
use the central values
$m_{\phi} = 1.02$ GeV, $f_{\phi}=0.233$ GeV, and $F_+^{B\to K}= 0.35 \pm 0.05$,
and for the scalar product $ (p_K \cdot \epsilon_{\phi}) =
\frac{m_B}{m_{\phi}} \sqrt{\left[\frac{1}{2m_B}\left(m_B^2 - m_K^2 +
m_{\phi}^2\right)\right]^2  -m_{\phi}^2} \simeq 13$ GeV.

Since the evaluation of the matrix element of $Q_{8g}$ goes beyond the
standard NF approach, we report here its result as given in \cite{ali}
\begin{equation}
\langle \phi\bar{K}^0 | Q_{8g} |\bar{B}^0\rangle = 
-\frac{\alpha_s}{4 \pi} \frac{m_b}
{\sqrt{\langle q^2\rangle }} \left[ \langle Q_4 \rangle + \langle Q_6
\rangle -\frac{1}{3}\left(\langle Q_3 \rangle + 
\langle Q_5 \rangle\right) \right],
\label{Q8_NF}
\end{equation}
where $\langle Q_i\rangle\equiv \langle \phi\bar{K}^0 | Q_i |\bar{B}^0\rangle$.
In the derivation of Eq.(\ref{Q8_NF}), the following assumption was made
\cite{ali}
\begin{equation}
q^{\mu}=\sqrt{\langle q^2\rangle}\frac{p_b^{\mu}}{m_b}\, .
\label{qmu}
\end{equation}
The momentum $q_{\mu}$ appearing in Eq.(\ref{qmu}),
is connected to the virtual gluon
in the $Q_{8g}$ operator, and it is given by
$q^{\mu}=p_b^{\mu}-p_s^{\mu}$, with $p_{b(s)}^{\mu}$ the momentum
of $b(s)$ quark, and $\langle q^2\rangle$ is an averaged value of $q^2$. 
It has been shown that the physical range of $q^2$ in 
$B\to \phi K_S$ is $m_b^2/4< \langle q^2\rangle < m_b^2/2$ \cite{gerard}. 
Notice that the term $1/\sqrt{\langle q^2\rangle}$ in Eq.(\ref{Q8_NF})
has origin from the propagator of the virtual gluon exchange 
between $Q_{8g}$ and external sources.

Here we stress that in the SM,
the large uncertainty on $\langle q^2\rangle$ does not necessarily 
convert in a large uncertainty on the $B\to \phi K$ amplitude, since 
the contribution of $Q_{8g}$ is not the dominant source in the SM.
On the contrary, $\langle q^2\rangle$ could cause a large uncertainty in 
the numerical analysis of SUSY models. Indeed, 
as we will show in the next section, in most relevant
SUSY scenarios, $C_{8g}$ provides the dominant source to 
$B\to \phi K$ amplitude.
Moreover, the NF approach suffices a serious problem, namely,
the decay amplitude in this approximation is not scale independent. 
The hadronic matrix elements cannot 
compensate for the scale dependence of the Wilson coefficients. This might be
a hint for the necessity of including higher order QCD
corrections to the hadronic matrix elements.
Furthermore, due to the approximations used in NF, one
cannot predict the direct CP asymmetries due 
to the assumption of no strong re-scattering in the final state, thus leaving
undetermined the predictions of strong phases.
\\

Recently, in the framework of QCD and heavy quark effective theory, 
a more consistent method for the determination of nonleptonic
B meson decays has been developed \cite{BBNS}. 
In this approach, called QCD factorization (QCDF), 
the hadronic matrix elements can be
computed from the first principles by means of perturbative 
QCD and $\Lambda_{QCD}/m_b$ expansions. 
Final results can be simply expressed in terms of form factors and meson
light-cone distribution amplitudes. Then, the usual NF is recovered only in 
the limit in which $\Lambda_{QCD}/m_b$ and $\alpha_s$ corrections 
are set to zero. A nice feature of QCDF is that the strong phases
of non-leptonic two body decays can be predicted.

In QCDF the hadronic 
matrix element for $B \to M K$ with $M=\phi, \eta^{\prime}$ 
in the heavy quark limit $m_b \gg \Lambda_{QCD}$ can be written as
\cite{BBNS}
\begin{equation}
\langle M K  \vert Q_i \vert B \rangle_{QCDF} = 
\langle M K \vert Q_i \vert B \rangle_{NF}.
\left[ 1+ \sum_n r_n \alpha_S^n + 
{\mathcal O}\left(\frac{\Lambda_{QCD}}{m_b} \right) \right],
\end{equation}
where $\langle M K\vert Q_i \vert B \rangle_{NF}$ denotes 
the NF results. The second 
and third term in the bracket represent the radiative
corrections in $\alpha_S$ and $\Lambda_{QCD}/m_b$.
Notice that, even though at higher order in $\alpha_s$ 
the simple factorization is broken, these corrections can be calculated
systematically in terms of short-distance coefficients and meson light-cone
distribution functions.

Now we briefly recall the main results of this method \cite{BBNS,BN}.
In QCDF the decay amplitudes of $B \to \phi(\eta^{\prime})K$ 
can be expressed as
\begin{equation}
\mathcal{A}\left(B\to \phi(\eta^{\prime})K\right) = 
\mathcal{A}^f\left(B\to \phi(\eta^{\prime}) K\right)
+ \mathcal{A}^a\left(B\to \phi(\eta^{\prime}) K\right),
\end{equation}
where
\begin{equation}
\mathcal{A}^f\left(B\to \phi(\eta^{\prime}) K\right) = 
\frac{G_F}{\sqrt{2}}\sum_{p=u,c} 
\sum_{1=1}^{10} V_{pb}V_{ps}^*~ a_i^{\phi(\eta^{\prime})} \langle \phi(\eta^{\prime}) K 
\vert Q_i \vert B \rangle_{NF},
\label{nf-ampl}
\end{equation}
and
\begin{equation}
\mathcal{A}^a\left(B\to \phi(\eta^{\prime}) K\right) = 
\frac{G_F}{\sqrt{2}}f_B f_{K} f_{\phi}
\sum_{p=u,c} \sum_{i=1}^{10} V_{pb}V_{ps}^*~ b_i^{\phi(\eta^{\prime})}.
\label{ann-ampl}
\end{equation}
The first term $\mathcal{A}^f\left(B\to \phi(\eta^{\prime}) K\right)$ 
includes vertex corrections, penguin corrections and hard spectator 
scattering contributions which are involved in the parameters
$a_i^{\phi(\eta^{\prime})}$. The other term 
$\mathcal{A}^a\left(B\to \phi(\eta^{\prime}) K\right)$ 
includes  the weak annihilation contributions which are absorbed in the 
parameters $b_i^{\phi(\eta^{\prime})}$.
However, these contributions contain infrared divergences, and the
subtractions of these divergences are 
usually parametrized as \cite{BN}
\begin{equation}
\int_0^1 \frac{d x}{x} \to X_{H,A} \equiv
\left(1 + \rho_{H,A} e^{i \phi_{H,A}} \right)
\ln\left(\frac{m_B}{\Lambda_{QCD}}\right),
\label{paramXAH}
\end{equation}
where $\rho_{H,A}$ are free parameters expected to be of order 
of $\rho_{H,A} \simeq {\cal O}(1)$, and $\phi_{H,A} \in [0,2\pi]$. 
As already discussed in Ref.\cite{BN}, if one does not require fine tuning 
of the annihilation phase $\phi_A$, the $\rho_A$ parameter gets an upper bound
from measurements on branching ratios, which is of order of
$\rho_A \lsim 2$. Clearly, large values of $\rho_A$
are still possible, but in this case strong fine 
tuning in the phase $\phi_A$ is required.
However, assumptions of very large values of $\rho_{H,A}$, which 
implicitly means large contributions from
hard scattering and weak annihilation diagrams, seem to be quite unrealistic.

Notice that the annihilation topology contribution due to the effective
operator $Q_{8g}$ has not yet been calculated, and it is expected to 
have as well a logarithmic divergence. This effect should 
increase the theoretical uncertainty substantially specially in
models like supersymmetric ones, where the $Q_{8g}$ plays a crucial
rule in the $B \to \phi K$ decay.

Following the scheme and the notation of Ref.\cite{BN} we write the 
decay amplitude of $B \to \phi K$ as:
\begin{eqnarray}
{\mathcal A} (B \to \phi K) &=& - i A_{K, \phi}
\sum_{p=u,c} \lambda_p \Big[
\alpha_3^p + \alpha_4^p - \frac{1}{2} \alpha_{3,EW}^p  - 
\frac{1}{2} \alpha_{4,EW}^p + 
\beta^p_3 \nonumber\\
& -& \frac{1}{2} \beta_{3,EW}^p + \beta_{S3}^p - 
\frac{1}{2} \beta_{S3,EW}^p \Big],
\label{A_Bphi}
\end{eqnarray}
where $\lambda_p=V_{pb} V^*_{ps}$ and 
$A_{ K, \phi}=\frac{G_F}{\sqrt{2}}m_B^2 f_{\phi} F_+^{B\to K}$.
The quantities $\alpha_i^p\equiv 
\alpha_i^p(K,\phi)$ and $\beta_i^p\equiv \beta_i^p(K,\phi)$ 
depend\footnote{Notice that in the notation of Ref.\cite{BN},
the order of the arguments 
in the functions $\alpha_i^p(M_1,M_2)$ and $\beta_i^p(M_1,M_2)$ is fixed and
it is determined by the order of the arguments ($M_1,M_2$) in the pre-factor 
$A_{M_1,M_2}$, where $M_{1,2}$ labels the final states.
In Eq.(\ref{A_Bphi}) this order corresponds to $M_1=K$, $M_2=\phi$.}, in 
addition to the leading contribution of the $b \to s $ transition, on 
the one loop vertex
corrections, hard spectator interactions and penguin corrections \cite{BN}. 

As mentioned in section 2,  NP effects are parametrized in the Wilson 
coefficients, while all the other functions involved in the definition
of  $\alpha_i^p$ and $\beta_i^p$ depend on some theoretical input
parameters like the QCD scale $\Lambda_{QCD}$, 
the value of the running masses, 
and parameters of vector-meson distribution amplitudes. 
Therefore it would be very useful for future analyses involving
any NP scenarios, to provide a
numerical parametrization of  Eq.(\ref{A_Bphi})
in terms of the Wilson coefficients at low energy.
In this respect it is very convenient to define new 
Wilson coefficients ${\bf C}_{ i}$ and 
${\bf \tilde{C}}_{i}$ according 
to the parametrization of the effective 
Hamiltonian in Eq.(\ref{Heff})
\begin{equation}
H^{\Delta B=1}_{\rm eff}=\frac{G_F}{\sqrt{2}}\sum_i \left\{
{\bf C}_{ i} Q_i \, +\, {\bf \tilde{C}}_{i} \tilde{Q}_i\right\}
\label{Heff_NP}
\end{equation}
where the operators basis $Q_i$ and $\tilde{Q}_i$ are the same ones of
Eq.(\ref{Qbasis}).\footnote{Notice that in the notation of \cite{BN}, 
the operators $Q_{1}$ and $Q_{2}$ correspond to our operators 
$Q_{2}$ and $Q_{1}$ respectively.}
Fixing the experimental and 
SM parameters to their center values as given in Table 1 of 
Ref.\cite{BN}, we can present the explicit dependence of the decay
amplitude on the Wilson coefficients relevant for this process. 
In particular, for
\begin{equation}
A(B\to \phi K ) = - i \frac{G_F}{\sqrt{2}} m_B^2 F_+^{B\to K} f_{\phi} 
\sum_{i=1..10,7\gamma,8g} H_i(\phi) ({\bf C}_i+ {\bf \tilde{C}}_i),
\end{equation}
we obtain
\begin{eqnarray}
H_1(\phi)&\simeq& -0.0002 - 0.0002 i \nonumber\\
H_2(\phi) &\simeq& 0.011 + 0.009 i ,\nonumber\\
H_3(\phi) &\simeq& -1.23 + 0.089 i - 0.005 X_A - 0.0006 X_A^2 - 0.013 X_H ,\nonumber\\
H_4(\phi) &\simeq& -1.17 + 0.13 i - 0.014 X_H ,\nonumber\\
H_5(\phi) &\simeq& -1.03 + 0.053 i + 0.086 X_A - 0.008 X_A^2\nonumber\\
H_6(\phi) &\simeq& -0.29 - 0.022 i + 0.028 X_A - 0.024 X_A^2 +0.014 X_H,
\nonumber\\
H_7(\phi) &\simeq& 0.52 - 0.026 i - 0.006 X_A + 0.004 X_A^2,\nonumber\\
H_8(\phi) &\simeq&  0.18 + 0.037 i - 0.019 X_A + 0.012 X_A^2 - 0.007 X_H ,\nonumber\\
H_9(\phi) &\simeq& 0.62 - 0.037 i + 0.003 X_A + 0.0003 X_A^2 + 0.007 X_H ,\nonumber\\
H_{10}(\phi) &\simeq&  0.62 - 0.037 i + 0.007 X_H ,\nonumber\\
H_{7\gamma}(\phi) &\simeq& -0.0004,\nonumber\\
H_{8g}(\phi) &\simeq& 0.047.
\label{H_phi}
\end{eqnarray}
Notice that here both $H_{7\gamma}$ and $H_{8g}$ do not have any dependence
on $X_{H,A}$, since, as we mentioned, the hard scattering and weak
annihilation contributions to $Q_{7\gamma}$ and $Q_{8g}$ have been
ignored. 
 
As can be seen from expressions in Eq.(\ref{H_phi}), 
terms proportional to $X_{A,H}$ represent always small corrections for 
typical values of $X_{H,A}\simeq {\cal O}(1)$. 
Moreover, if we set $X_{A,H}$ to zero, 
the contribution to the amplitude is (incidentally) quite 
close to the corresponding one in NF for $\langle q^2\rangle=m_b^2/4$. 
Remarkably, the strong phases appearing in the terms 
independent on $X_{H,A}$ in $H_i(\phi)$ are also negligible.
Thus, in the limit $X_{H,A}\to 0$, the NF result
(where strong phases are assumed to vanish) seems to be recovered
for the choice $\langle q^2\rangle=m_b^2/4$. 

Note that largest corrections in $X_{A}$ are contained 
in the $H_6(\phi)$ term.
In particular, for values of $\rho_A=1$ and 
$\phi_A \simeq 0$, we have $X_A\simeq 4$ and so 
the term proportional to $X_A^2$ in $H_6(\phi)$ becomes of the same order
as the other term, which is independent on $X_A$. 
Therefore, it is expected that the effect of the 
weak annihilation parameter would be important 
when contributions to $C_6$ becomes large. 
Clearly, this last consideration applies 
in a NP context, where the new contribution to Wilson coefficients $C_i$
could sizeably differ from the corresponding  SM ones.

Finally we would like to comment about the fact that contributions from
$\tilde{C}_i$ and $C_i$ to the decay amplitude $A(B\to \phi K)$ 
are identically the same (with the same sign). This can be simply
understood by noticing that 
\begin{equation}
\langle \phi \bar{K}^0 \vert Q_i \vert \bar{B}^0 \rangle = 
\langle \phi \bar{K}^0 \vert \tilde{Q}_i \vert \bar{B}^0 \rangle\, .
\label{Qi_phi_ME}
\end{equation}
which is due to the invariance 
of strong interactions under parity transformations, and 
to the fact that initial and final states have same parity.
Indeed only  terms proportional to the chiral structure 
$V\times V$ or $A\times A$ in the four fermion operators of $Q_i$ basis
contribute to the matrix elements, and
consequently property in Eq.(\ref{Qi_phi_ME}) easily follows. Analogous 
considerations apply to $Q_{7\gamma}$ and $Q_{8 g}$ as well.
\\

Now we turn to the $B\to \eta^{\prime} K $ process. As known, 
the physical state of $\eta$ and 
$\eta^{\prime}$ are the mixture of $SU(3)$ 
singlet $\eta_s = \vert s\bar{s} \rangle$ and octet 
$\eta_q = \left(\vert u\bar{u}\rangle +\vert d\bar{d}\rangle\right)/\sqrt{2}$ 
components:
\begin{equation}
\left(
\begin{array}{c}
\vert \eta \rangle \\ 
\vert \eta^{\prime} \rangle
\end{array}
\right) = \left(
\begin{array}{cc}
\cos \phi & - \sin \phi\\
\sin \phi & \cos \phi
\end{array}
\right)~ \left(
\begin{array}{c}
\vert \eta_q \rangle \\ 
\vert \eta_s \rangle
\end{array}
\right).
\end{equation}
Therefore, we have the following decay constants: 
$f^q_{\eta^{\prime}} = f_q \sin\phi$ and
$f^s_{\eta^{\prime}} = f_s \sin\phi$ and from a fit to experimental 
data one finds \cite{BN}:
$f_q = (1.07 \pm 0.02) f_{\pi}$, $f_s = (1.34 \pm 0.06) f_{\pi}$ 
and $\phi=39.3\pm 1.0$.

In the NF approach, the hadronic matrix elements 
of the $B\to \eta^{\prime} K$ process are given by \cite{KK,ali}
\bea
&& \langle \eta^{\prime} \bar K^0|Q_1 | \bar B^0 
\rangle = \frac{1}{3} X_2~~~~~~
\langle \eta^{\prime} \bar K^0|Q_2 | \bar B^0 \rangle =  X_2 \nonumber\\
&& \langle \eta^{\prime} \bar K^0|Q_3 |\bar B^0 \rangle = \frac{1}{3} X_1
+2 X_2 + \frac{4}{3} X_3 \nonumber\\
&& \langle \eta^{\prime} \bar K^0|Q_4 |\bar B^0 \rangle =  X_1
+\frac{2}{3} X_2 + \frac{4}{3} X_3  \nonumber\\
&& \langle \eta^{\prime} \bar K^0|Q_5 | \bar B^0 \rangle = \frac{R_1}{3} X_1
-2 X_2 -\left (1- \frac{R_2}{3}\right )   X_3 \nonumber\\
&& \langle \eta^{\prime} \bar K^0|Q_6 | \bar B^0 \rangle = R_1 X_1
-\frac{2}{3} X_2 -\left ( \frac{1}{3}-R_2\right ) X_3
 \nonumber\\
&& \langle \eta^{\prime} \bar K^0|Q_7 | \bar B^0 \rangle = 
\frac{1}{2}\biggr[-\frac{R_1X_1}{3}- X_2
+\left (1-\frac{R_2}{3} \right )X_3 \biggr]\nonumber\\
&& \langle \eta^{\prime} \bar K^0|Q_8 |\bar B^0 \rangle =
\frac{1}{2}\biggr[-R_1X_1- \frac{X_2}{3}
+\left (\frac{1}{3} -R_2 \right )X_3 \biggr]\nonumber\\
&& \langle \eta^{\prime} \bar K^0|Q_9| \bar B^0 \rangle = 
\frac{1}{2}\biggr[-\frac{X_1}{3}+ X_2
-\frac{4}{3} X_3 \biggr]\nonumber\\
&& \langle \eta^{\prime} \bar K^0|Q_{10} |\bar B^0 \rangle = 
\frac{1}{2}\biggr[-X_1+ \frac{X_2}{3}
-\frac{4}{3}X_3 \biggr]
\eea
with 
\begin{eqnarray}
X_1&=&-(m_B^2-m_{\ep}^2)F_1^{B\to\pi}(m_{K}^2)\frac{X_{\ep}}{\sqrt{2}}f_{K}, 
\nonumber \\
X_2&=&-(m_B^2-m_{K}^2)F_1^{B\to K}(m_{\ep}^2)f_{\pi}\frac{X_{\ep}}{\sqrt{2}}, 
\nonumber \\
X_3&=&-(m_B^2-m_{K}^2)F_1^{B\to K}(m_{\ep}^2)\sqrt{2f_{K}^2-f_{\pi}^2}Y_{\ep}, 
\nonumber \\
R_1&=&\frac{2m_{K}^2}{(m_b-m_s)(m_s+m_d)}, \ \ \ \ \ \
R_2=\frac{(2m_{K}^2-m_{\pi}^2)}{(m_b-m_s)m_s}\nonumber
\end{eqnarray}
where $F_1^{B\to \pi}(q^2)=0.35$ is the $B-\pi$ transition form factor 
evaluated at $q^2$ scale, and $f_{K(\pi )}=0.16(0.13)$ GeV
is the decay constant of $K (\pi)$ meson. $X_{\ep}=0.57$ and $Y_{\ep}=0.82$,
which correspond to $\theta_p=-20^{\circ}$, represent the
rate of the $u\bar{u}+d\bar{d}$ and $s\bar{s}$ component 
in the $\ep$. These matrix elements
show that the $B\to \eta^{\prime} K$ 
process receives a small contribution from color 
suppressed $b\to u \bar{u}s$ tree diagram, in addition to 
the $b\to s \bar{q} q$ ($q=u,d,s$) penguin diagrams.
As in case of $B\to \phi K$ decay, for the matrix 
element of the chromo-magnetic operator we have
\begin{equation}
\langle \eta^{\prime}\bar{K}^0 | Q_{8g} |\bar{B}^0\rangle = 
-\frac{\alpha_s}{4 \pi} \frac{m_b}
{\sqrt{q^2}} \left[ \langle Q_4 \rangle + \langle Q_6
\rangle -\frac{1}{3}\left(\langle Q_3 \rangle + 
\langle Q_5 \rangle\right) \right]
\end{equation}
\\

In the QCDF approach, 
the decay amplitude of $B \to \eta^{\prime} K$ is given by 
\cite{BN,BN2}
\begin{eqnarray}
{\mathcal A} (B \to  \eta^{\prime} K ) &=& 
\frac{i}{\sqrt{2}} A_{K, \eta^{\prime}_q} 
\sum_{p=u,c} \Big\{ \lambda_p \Big[
\delta_{pu} \alpha_2 + 2 \alpha_3^p + 
\frac{1}{2} \alpha_{3,EW}^p  + 2 \beta^p_{S3} -
\beta^p_{S3,EW}\Big] \nonumber\\
& +&  iA_{K,\eta^{\prime}_s} \Big[\alpha_3^p + 
\alpha_4^p -\frac{1}{2}\alpha_{3,EW}^p -
\frac{1}{2}
\alpha_{4,EW}^p+ \beta_{3}^p -\frac{1}{2}\beta_{3,EW}^p + 
\beta^p_{S3} - \frac{1}{2} 
\beta_{S3,EW}^p \Big]\nonumber\\
&+&  i A_{K, \eta^{\prime}_c} \Big[\delta_{pc} \alpha_2 + \alpha_3^p \Big]+ 
\frac{i}{\sqrt{2}} A_{\eta^{\prime}_q, K}\Big[\alpha_4^p - \frac{1}{2} 
\alpha_{4,EW}^p + 
\beta_3^p - \frac{1}{2} \beta^p_{3,EW} \Big] \Big\}\, ,
\label{A_Beta}
\end{eqnarray}
where\footnote{Same notation as in ${\mathcal A} (B \to  \phi K )$ 
has been adopted here 
for $\alpha_i^p$ and $\beta_i^p$, where $\alpha_i^p\equiv \alpha_i^p(M_1,M_2)$,
$\beta_i^p\equiv \alpha_i^p(M_1,M_2)$. 
The expressions for  $\alpha_i^p(K,\eta^{\prime})$, 
$\alpha_i^p(\eta^{\prime},K)$, $\beta_i^p(K,\eta^{\prime})$, and
$\beta_i^p(\eta^{\prime},K)$ can be found in Ref.\cite{BN,BN2}. 
}
\bea
A_{K, \eta^{\prime}_q}&=& 
\frac{G_F}{\sqrt{2}} m_B^2 F^{B\to K} f^q_{\eta^{\prime}} 
~\simeq~ 6.8 \times 10^{-6}~~{\rm GeV}\, ,
\nonumber\\
A_{K, \eta^{\prime}_s}&=& \frac{G_F}{\sqrt{2}} m_B^2 F^{B\to K} 
f^s_{\eta^{\prime}} ~\simeq~ 1.05 \times 10^{-5}~~{\rm GeV}\, ,
\nonumber\\
A_{K, \eta^{\prime}_c}&=& \frac{G_F}{\sqrt{2}} m_B^2 F^{B\to K} 
f^c_{\eta^{\prime}} ~\simeq~ -2.3\times 10^{-7}~~{\rm GeV}\, ,
\nonumber\\
A_{\eta^{\prime}_q, K}&=& \frac{G_F}{\sqrt{2}} 
m_B^2 F^{B\to \eta^{\prime}} f_K ~\simeq~ 7.0 \times 10^{-6} ~~{\rm GeV}\, ,
\eea 
where the third contribution is negative due to the definition of negative 
decay constant $f^c_{\eta^{\prime}}$ in \cite{BN2}.
As in the case of $B\to \phi K$, we will provide below the numerical
parametrization in terms of the Wilson 
coefficients defined in Eq.(\ref{Heff_NP}).
By fixing the hadronic parameters with their 
center values as in Table 1 of Ref.\cite{BN}, we obtain
\begin{equation}
A(B\to \eta^{\prime} K ) = - i 
\frac{G_F}{\sqrt{2}} m_B^2 F_+^{B\to K} f^s_{\eta^{\prime}} 
\sum_{i=1..10,7\gamma,8g} H_i(\eta^{\prime}) ({\bf C}_i - {\bf \tilde{C}}_i),
\label{A_etaparam}
\end{equation}
where
\begin{eqnarray}
H_1(\eta^{\prime})&\simeq & 0.44 + 0.0005 i, \nonumber\\
H_2(\eta^{\prime}) &\simeq& 0.076 - 0.064 i + 0.006 X_H,\nonumber\\
H_3(\eta^{\prime}) &\simeq& 2.23 - 0.15 i+ 0.009 X_A + 0.0008 X_A^2 + 
0.014 X_H ,\nonumber\\
H_4(\eta^{\prime}) &\simeq& 1.76 - 0.29 i + 0.026 X_H ,\nonumber\\
H_5(\eta^{\prime}) &\simeq& - 1.52 + 0.004 X_A + 0.008 X_A^2\nonumber\\
H_6(\eta^{\prime}) &\simeq& 0.54 - 0.29 i + 0.006 X_A + 0.027 X_A^2 + 
0.026 X_H ,\nonumber\\
H_7(\eta^{\prime}) &\simeq& 0.078 + 0.001 X_A - 0.004 X_A^2,\nonumber\\
H_8(\eta^{\prime}) &\simeq& - 0.58 + 0.02 i + 0.004 X_A - 0.014 X_A^2 - 
0.004 X_H ,\nonumber\\
H_9(\eta^{\prime}) &\simeq&  - 0.44 + 0.054 i + 0.005 X_A - 0.0004 X_A^2 - 
0.007 X_H ,\nonumber\\
H_{10}(\eta^{\prime}) &\simeq& - 0.80 + 0.02 i - 0.004 X_H ,\nonumber\\
H_{7\gamma}(\eta^{\prime}) &\simeq& 0.0007,\nonumber\\
H_{8g}(\eta^{\prime}) &\simeq& -0.089.
\label{H_eta}
\end{eqnarray}
The sign difference between ${\bf C}_i$ and ${\bf \tilde{C}}_i$
appearing in Eq.(\ref{A_etaparam}) is due to the fact that, 
contrary to the $B\to \phi K$ transition,
initial and final states have here opposite parity. 
Thus, due to the invariance of strong interactions 
under parity transformations, only 
$V\times A$ or $A\times V$ structures of four-fermion operators 
will contribute to the hadronic matrix elements, and so
\begin{equation}
\langle \eta^{\prime} K  \vert Q_i \vert B \rangle = 
- \langle \eta^{\prime} K  
\vert \tilde{Q}_i 
\vert B \rangle.
\label{Qi_eta_ME}
\end{equation}

The comparison between the coefficients $H_i(\phi)$ and $H_i(\eta^{\prime})$
in Eqs.(\ref{H_eta}) tells us that, apart from the 
sign difference of the Wilson coefficients 
$\tilde{C}_i$ in the amplitudes of $B\to \eta^{\prime} K$ and 
$B\to \phi K$, in  QCDF these amplitudes are 
expected to be different, unlike the case of NF as 
discussed in Ref.\cite{KK}. However, notice that the values of 
$H_i(\phi)$ and $H_i(\eta)$ are quite sensitive to the scale $\mu$ where they
have been evaluated. Our results in Eqs.(\ref{H_phi}),(\ref{H_eta})
correspond to the choice $\mu=m_b^{\rm pole}=4.5$ GeV.

%
%%%%%%%%%%%%%%%%%%%%%%%%%%%%%%%%%%%%%%%%%%%%%%%%%%%%%%%%%%%%%%%%%%%%%%%%%%%%
%
\section{CP asymmetry of $B\to \phi (\eta^{\prime}) K_S$: gluino / chargino}
%
%%%%%%%%%%%%%%%%%%%%%%%%%%%%%%%%%%%%%%%%%%%%%%%%%%%%%%%%%%%%%%%%%%%%%%%%%%%%

Here we analyze the supersymmetric contributions to the time dependent
CP asymmetries in $B\to \phi K_S$ and $B\to \eta^{\prime} K_S$ decays
in the framework of mass insertion approximation, 
in gluino and chargino dominated scenarios.

New physics could in principle affect the B meson decay 
by means of a new source of CP violating phase in the corresponding 
amplitude. In general this phase is different from the corresponding SM one.
If so, then deviations on CP asymmetries from SM
expectations can be sizeable, depending on the relative magnitude of SM
and NP amplitudes. For instance, in the SM the
$B\to \phi K_S$ decay amplitude
is generated at one loop and therefore it is
very sensitive to NP contributions.
In this respect, SUSY models with non minimal flavor structure and new 
CP violating phases in the squark mass matrices, 
can easily generate large deviations in the 
$B\to \phi K_S$  asymmetry.

As mentioned in the introduction, the time dependent CP asymmetry 
for $B\to \phi K_S$ can
be described by 
\begin{eqnarray}
a_{\phi K_S}(t)&=&\frac{\Gamma (\overline{B}^0(t)\to\phi K_S)-\Gamma
(B(t)\to\phi K_S)} {\Gamma (\overline{B}^0(t)\to\phi K_S)+\Gamma (B(t)
\to\phi K_S)} 
%\\&=&
= C_{\phi K_S}\cos\Delta M_{B_d}t+S_{\phi K_S}\sin\Delta M_{B_d}t,\nonumber\\
\label{asym_phi}
\end{eqnarray}
where $C_{\phi K_S}$ and $S_{\phi K_S}$ represent
the direct  and the mixing CP asymmetry, respectively and they are given by
\begin{equation}
C_{\phi K_S}=
\frac{|\overline{\rho}(\phi K_S)|^2-1}{|\overline{\rho}(\phi K_S)|^2+1},
\ \
S_{\phi K_S}=\frac{2Im \left[\frac{q}{p}~\overline{\rho}(\phi K_S)\right]}
{|\overline{\rho}(\phi K_S)|^2+1}.  
\label{S_PHI}
\end{equation}
The parameter $\overline{\rho}(\phi K_S)$ is defined by
\begin{equation}
\overline{\rho} (\phi K_S)=\frac{\overline{A}(\phi K_S)}{A(\phi K_S)}.
\end{equation}
where $\overline{A}(\phi K_S)$ and $A(\phi K_S)$ are 
the decay amplitudes of $\overline{B}^0$ and $B^0$  mesons, respectively.
Here, the mixing parameters $p$ and $q$ are defined by
$|B_1\rangle =p|B\rangle +q|\overline{B}^0\rangle ,
\ \ |B_2\rangle =p|B\rangle -q|\overline{B}^0\rangle$
where $|B_{1(2)}\rangle$ are mass eigenstates of $B$ meson. 
The ratio of the mixing parameters is given by
\begin{equation}
\frac{q}{p}=-e^{-2i\theta_d}\frac{V_{tb}^*V_{td}}{V_{tb}V_{td}^*},
\end{equation}
where $\theta_d$ represent any SUSY contribution 
to the $B-\bar{B}^0$ mixing angle.
Finally, the above amplitudes can be written in terms of the matrix element 
of the $\Delta B=1$ transition as
\begin{equation}
\overline{A}(\phi K_S)=
\langle \phi K_S| H_{eff}^{\Delta B=1}|
\overline{B}^0\rangle,  
\ \ \
A(\phi K_S)=
\langle \phi K_S| \left(H_{eff}^{\Delta B=1}\right)^{\dag}|B^0\rangle.
\end{equation}

In order to simplify our analysis, it is useful to parametrize the SUSY 
effects by introducing the ratio of SM and SUSY amplitudes as follows
\bea
\left(\frac{A^{\susy}}{A^{\sm}}\right)_{\phi K_S}&\equiv&
R_{\phi}~ e^{i\theta_{\phi}}~e^{i\delta_{\phi}}
\\
\label{ratioPHI}
\eea
and analogously for the $\eta^{\prime} K_S$ decay mode
\bea
\left(\frac{A^{\susy}}{A^{\sm}}\right)_{\eta^{\prime} K_S}&\equiv&
R_{\eta^{\prime}}~ e^{i\theta_{\eta^ {\prime}}}~e^{i\delta_{\eta^ {\prime}}}
\label{ratioETA}
\eea
where $R_i$ stands for the corresponding absolute values of 
$|\frac{A^{\susy}}{A^{\sm}}|$, 
the angles $\theta_{\phi,~\eta^{\prime}}$
 are the corresponding SUSY CP violating phase,
and $\delta_{\phi,~\eta^ {\prime}}=\delta^{SM}_{\phi,~\eta^ {\prime}}-
\delta^{SUSY}_{\phi,~\eta^ {\prime}}$ 
parametrize the strong (CP conserving) phases.
In this case, the mixing CP asymmetry $S_{\phi K_S}$ in Eq.(\ref{asym_phi}) 
takes the following form 
\bea
S_{\phi K_S}&=&\Frac{\sin 2 \beta +2 R_{\phi} 
\cos \delta_{\phi} \sin(\theta_{\phi}+2 \beta)+
R_{\phi}^2 \sin (2 \theta_{\phi}+2 \beta)}{1+ 2 R_{\phi} 
\cos \delta_{\phi} \cos\theta_{\phi} +R_{\phi}^2}.
\label{cpmixing_phi}
\eea
\vspace{0.5in}
and analogously for $B\to \eta^{\prime} K_S$
\bea
S_{\eta^{\prime} K_S}&=&\Frac{\sin 2 \beta +2 R_{\eta^{\prime}} 
\cos \delta_{\eta^{\prime}} \sin(\theta_{\eta^{\prime}}+2 \beta)+
R_{\eta^{\prime}}^2 \sin (2 \theta_{\eta^{\prime}}+2 \beta)}
{1+ 2 R_{\eta^{\prime}} 
\cos \delta_{\eta^{\prime}} \cos\theta_{\eta^{\prime}} +
R_{\eta^{\prime}}^2}.
\label{cpmixing_eta}
\eea

Assuming that the SUSY contribution to the amplitude is smaller than the
SM one {\it i.e.} $R_{\phi(\eta^{\prime})} \ll 1$, 
one can simplify the above expressions as:
\bea
S_{\phi(\eta^{\prime}) K_S}= \sin 2\beta +2\cos
 2\beta\sin\theta_{\phi(\eta^{\prime})}
\cos\delta_{\phi(\eta^{\prime})} R_{\phi(\eta^{\prime})}
+{\cal O}(R_{\phi,(\eta^{\prime})}^2)\, .
\eea
It is now clear that in order to reduce $S_{\phi K_S}$ smaller 
than $\sin 2 \beta$, the relative
sign of $\sin\theta_{\phi}$ and $\cos\delta_{\phi}$ has to be negative. 
If one assumes that 
$\sin\theta_{\phi}\cos\delta_{\phi} \simeq -1$, then $R_{\phi}\geq 0.1$ 
is required in 
order to get $S_{\phi K_S}$ within $1\sigma$ of the experimental range.
  
We will work in a minimal scheme of
CP and flavor violation. This means that,
in the framework of MIA, we will assume a dominant effect 
due to one single mass insertion. 
In this case  the CP violating SUSY phase will coincide with the 
argument (module $\pi$) of corresponding mass insertion. 
Moreover, in the following we will generalize this scheme by 
including two (dominant) mass insertions 
simultaneously, but assuming that their CP violating phases are the same.
We will perform this analysis in both gluino and 
chargino scenarios.\footnote{There are corrections to the 
effective Hamiltonian $\Delta B=1$, mediated by both chargino and 
gluino box diagrams in which both up and down mass insertion contribute.
See Ref.\cite{our,GG} for further details. 
Then, chargino or gluino contributions cannot be completely 
disentangled by taking active only one mass insertion per time.
However, these corrections affect only the Wilson coefficients 
$C_1$ and $C_2$ in Eq.(\ref{Heff}) 
and their effect is quite small 
in comparison to the other SUSY contributions. 
In our analysis we can safely neglect them.}

In the following subsections we present and discuss our
results separately for ${B\to \phi K_S}$ and 
${B\to \eta^{\prime} K_S}$ processes.
%---------------------------------------------------------------
\subsection{{\bf CP asymmetry in $B\to {\phi K_S}$}}
As shown in Eq.(\ref {Sphi}), the most recent world average
measurement of the CP asymmetry $S_{\phi K_S}$, indicates
a $1.7\sigma$ deviation from the $c\bar{c}$ measurement. In particular, 
\bea
S_{\phi K_S}=0.34\pm 0.20
\eea
allows also for negative values of CP asymmetry at 2$\sigma$ level.
In the light of these 
results, we will analyze here, in a model independent way, the SUSY scenarios, 
which are more favored to explain such deviations.
Let us start first with gluino contributions to the  CP asymmetry 
in $B\to {\phi K_S}$.

These effects have been analyzed in Refs.\cite{PhiKs_gluino,CFMS,KK}
in the framework of NF or QCDF and by adopting the MIA method.
However, from these works a direct comparison between
the results in NF and QCDF predictions is not always possible, 
mainly because of different approaches
used in analyzing the SUSY contributions.
For this reason we reanalyze here the gluino contributions  
in both NF and QCDF approaches in a unique SUSY framework,
by comparing the predictions for the CP asymmetries 
versus the relevant SUSY CP violating phases.
In QCDF we have included the last updated
results for  hard spectator scattering and weak annihilation diagrams
\cite{BN}, as explained in section 3.

% ************************* PHI  KS ***********************************

%------------------ GLUINO ---------------------

%-----------------------------------------------

\begin{figure}[tpb]
\begin{center}
\dofourfigs{3.1in}{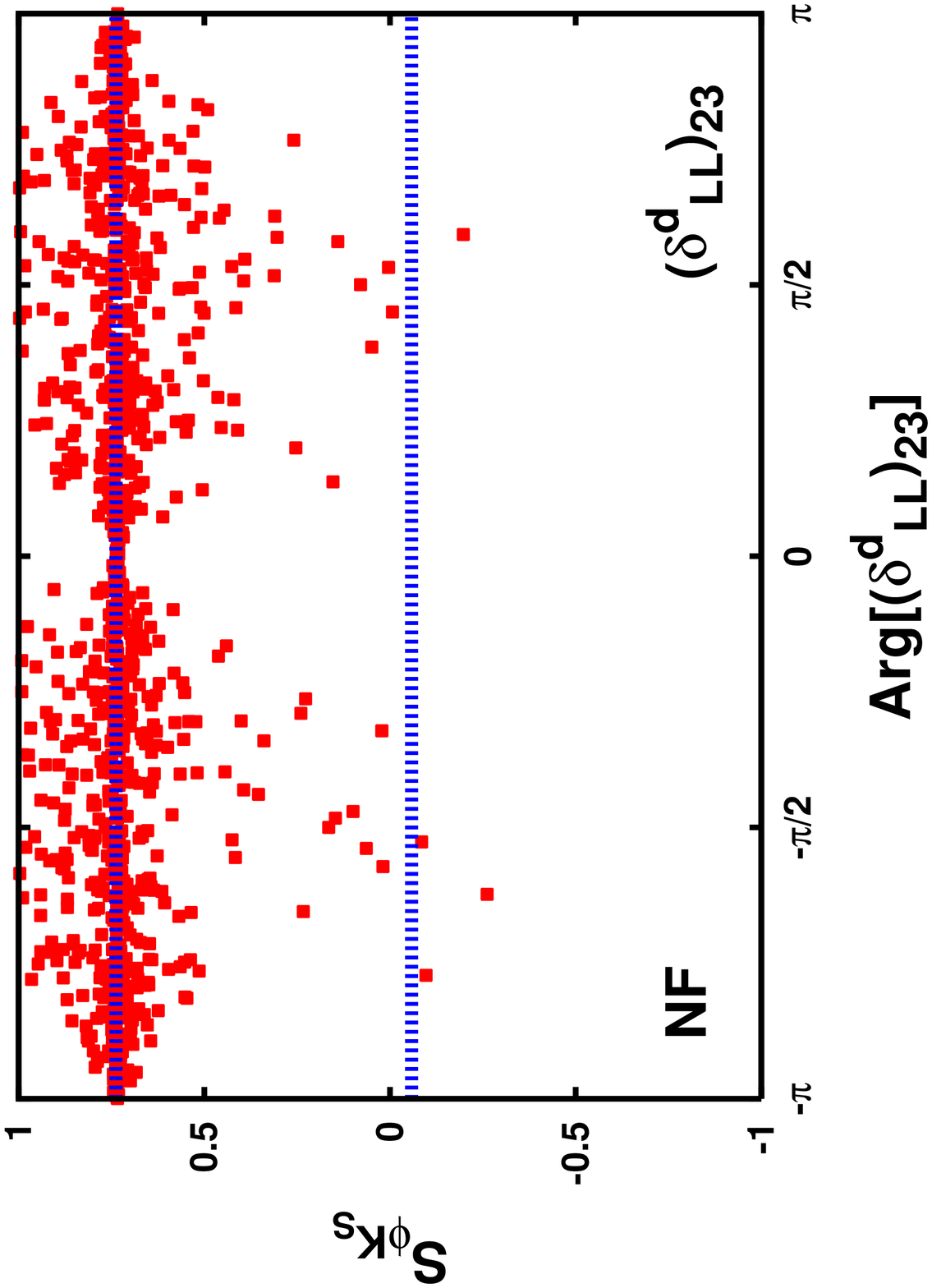}{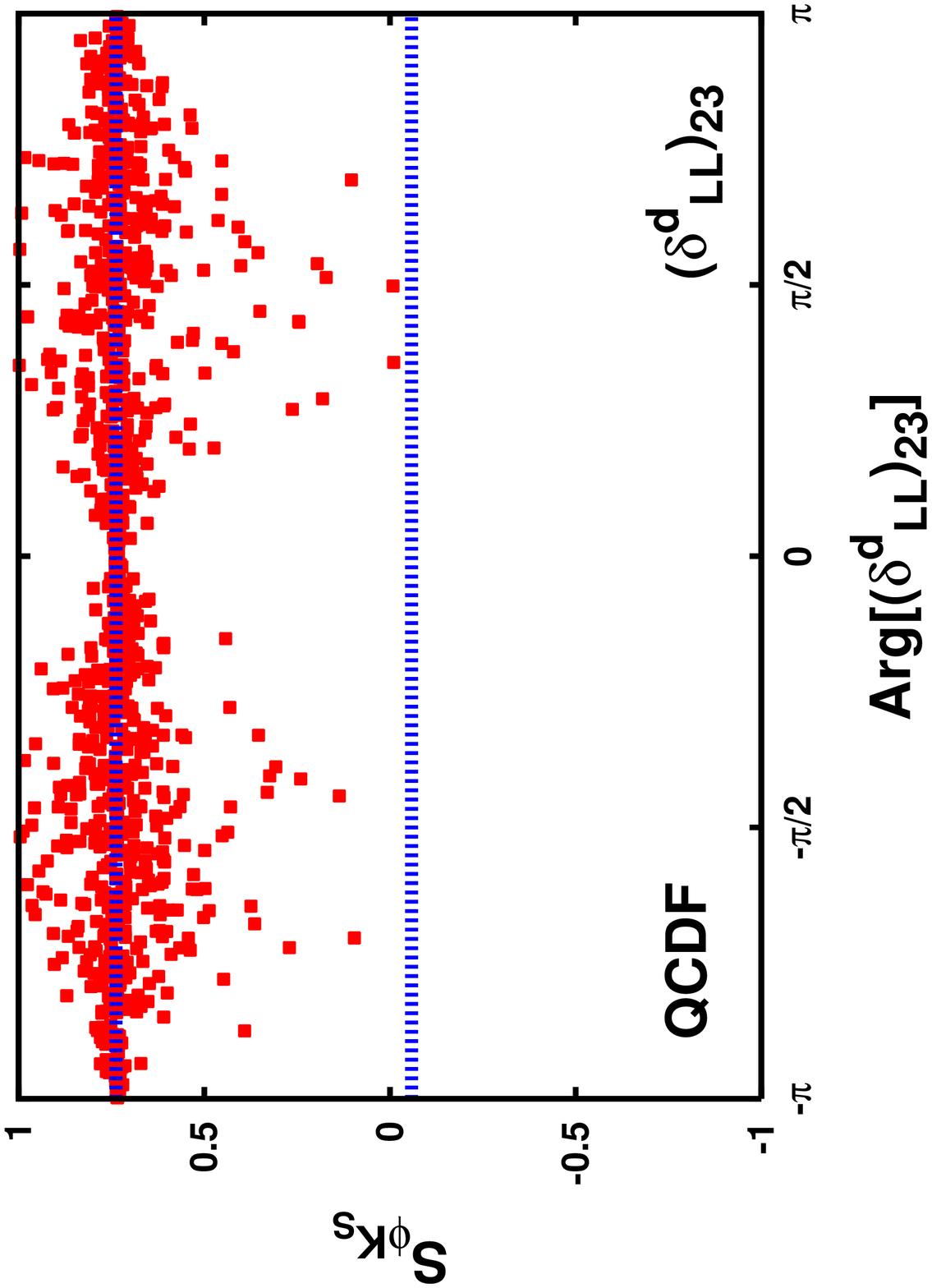}
{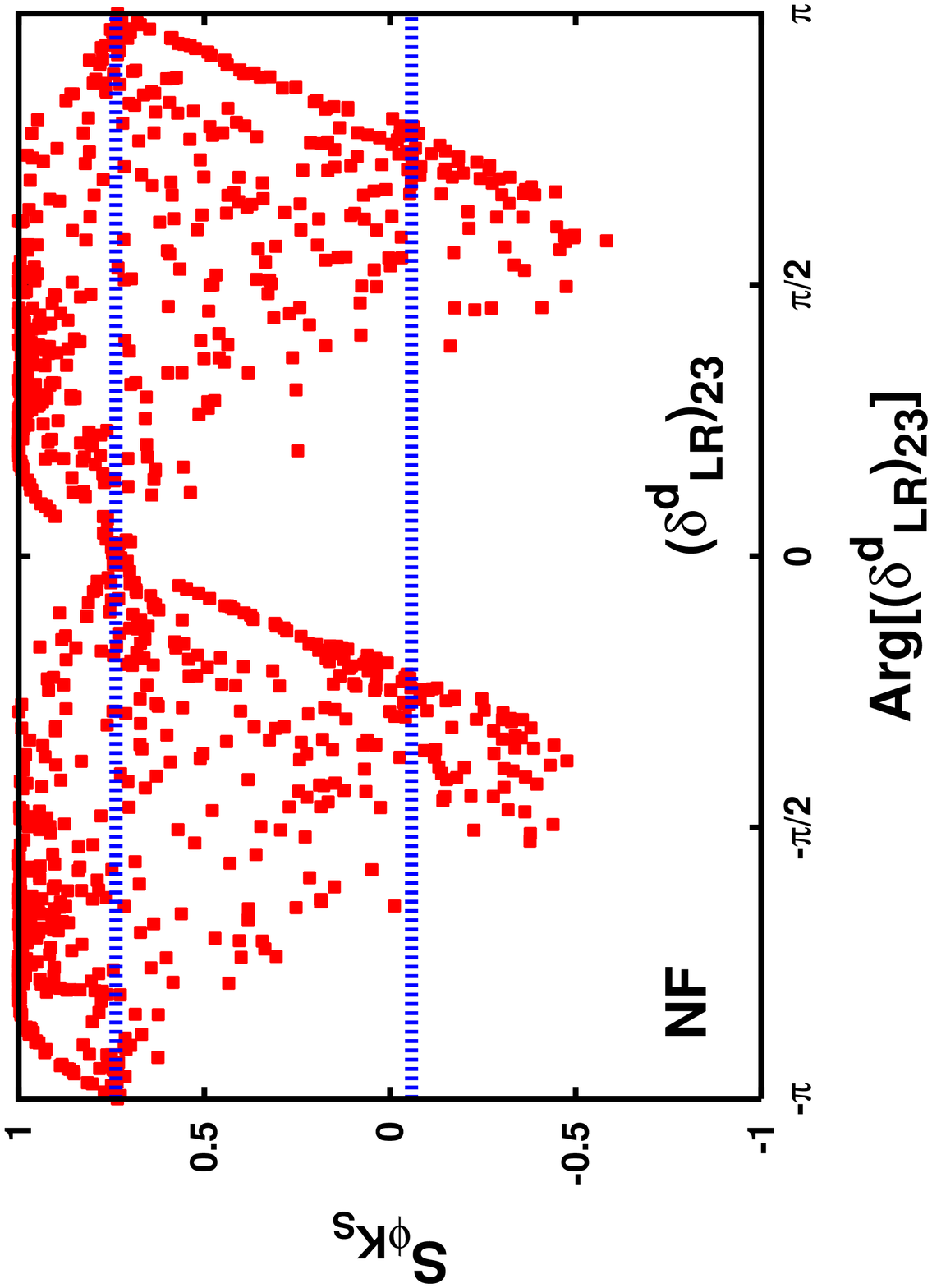}{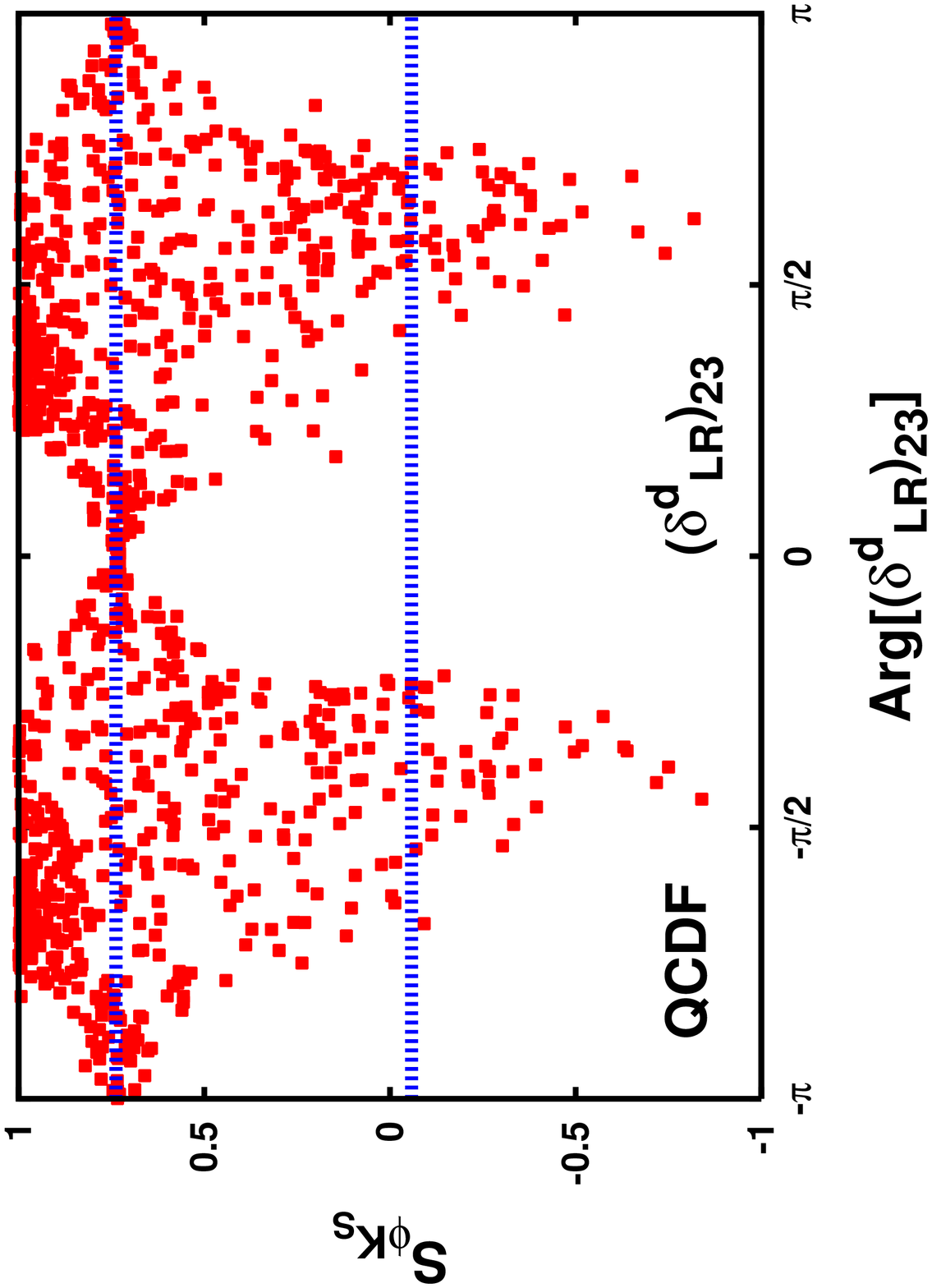}
\end{center}
\caption{\small $S_{\Phi K_S}$ as a function of arg[$\dd{LL}{23}$]
(top) and  arg[$\dd{LR}{23}$] (bottom) with  gluino 
contribution of one mass insertion 
$\dd{LL}{23}$ (top) and $\dd{LR}{23}$ (down). 
Left and right plots correspond to NF and QCDF, respectively. 
The region inside the two horizontal lines 
corresponds to the allowed experimental region at $2\sigma$ level.}
\label{Fig41}
\end{figure}

%------------------ ---------------------

We present our numerical results for the gluino contributions to CP asymmetry
 $S_{\Phi K_S}$  in Figs.~\ref{Fig41}. The left and right plots 
correspond to the evaluation of amplitudes by means of
NF and QCDF methods, respectively. 
In all the plots, regions inside the horizontal lines indicate
the allowed $2\sigma$ experimental range.
In the top and bottom plots only one mass insertion per time is taken active,
in particular this means that we scanned over $|\dd{LL}{23}|<1$ 
and $|\dd{LR}{23}|<1$.
Then, $S_{\Phi K_S}$ is plotted versus $\theta_{\phi}$, which in the
case of one dominant mass insertion should be identified here as
$\theta_{\phi}={\rm arg}[(\delta_{AB}^d)_{ij}]$.

We have scanned over the relevant SUSY parameter space, 
in this case the average squark mass $\tilde{m}$ and gluino
mass $m_{\tilde{g}}$, assuming SM central values \cite{PDG}.
Moreover, we require that the SUSY spectra 
satisfy the present experimental lower mass bounds \cite{PDG}. 
In particular, $m_{\tilde{g}}> 200$ GeV,
$\tilde{m} > 300$ GeV. In addition, we impose that the
branching ratio (BR)\footnote{
The branching ratio (BR) of \bsg is evaluated
at the NLO in QCD, as provided in Ref.\cite{bsg_NLO,KN}. 
However, we have not included
the 2-loop threshold corrections of SUSY contributions at W scale.
For these corrections see Ref.\cite{bsg_SUSY_NLO} for more details.} of \bsg 
and the \bbar mixing are satisfied
at  95\% C.L. \cite{bsgmeas}, namely 
$2\times 10^{-4}\le BR(b\to s\gamma) < 4.5\times 10^{-4}$.
Then, the allowed ranges for $|\dd{LL}{23}|$ 
and $|\dd{LR}{23}|$ are obtained by 
taken into account the above constraints on \bsg and \bbar mixing.

In the plots corresponding to QCDF, we have also scanned over the 
full range of the parameters $\rho_{A,H}$ and $\phi_{A,H}$  in 
$X_A$ and $X_H$, respectively, as defined in  Eq.(\ref{paramXAH}).
We remind here that $X_A$ and $X_H$ 
take into account the (unknown) infrared 
contributions in the hard scattering and annihilation diagrams. 
Regarding the allowed range of $\rho$,
as can be seen from results in Eq.(\ref{H_phi}), the dominant effect is
due to the annihilation contributions proportional to $X_{A}$.
As discussed in section 3, the total width will grow as
$\rho_A^4 \log^4{\left(m_b/\Lambda\right)}$ 
for large $\rho_A$ . Therefore, as we will show in section 5,
by requiring that the SUSY contribution to the 
branching ratio and asymmetries of $B\to \phi K_S$ 
is inside the $2\sigma$ experimental range, 
an upper bound on $\rho_{A}$ of order $\rho_{A}\simeq 2$ is obtained.\footnote{
We would like to stress here that in the literature  
the allowed range of $\rho_A$ has been sometime overestimated.
For instance, in the analysis of 
\cite{CFMS} the experimental upper bound on $BR(B\to \phi K_S)$ 
was not imposed, leaving  the possibility of larger 
values of $\rho_A\;\sim 8$.}

In the corresponding plots evaluated in NF, 
in order to maximize the SUSY contributions to CP asymmetry,
we have fixed the average 
of gluon momenta to its minimum value $\langle q^2\rangle=m_b^2/4$.
We remind here that $\langle q^2 \rangle$ enters as free parameter
in matrix element of the chromo-magnetic operator $Q_{8g}$, see
Eq.(\ref{Q8_NF}).
As we will show later on, the dominant SUSY effect to 
$B\to \phi K_S$ amplitude is given by the SUSY contribution to 
the Wilson coefficient $C_{8g}$. 
Therefore, the smaller the $\langle q^2\rangle$ value is, the larger the SUSY 
contribution to the CP asymmetry can be.

In the framework of NF, the strong phase, which comes from the hadronic
matrix elements, cannot be predicted. For this reason we 
set both SM and SUSY strong phases to zero. 
Therefore, the phase $\delta_{\phi}$ in Eq.(\ref{S_PHI}) can take only the 
values $0$ or $\pi$, corresponding to the relative sign of SM and SUSY 
amplitudes.

By comparing the scatter plots in NF and QCDF in Fig.~\ref{Fig41} we see that 
the predictions are quite similar.
Only the gluino contributions proportional to  $\dd{LR}{23}$
have chances to drive $S_{\Phi K_S}$ toward the region of 
large and negative values, while the pure effect of $\dd{LL}{23}$ 
just approach the negative values region.

This result can be easily understood by noticing that 
the dominant SUSY source to the 
$B\to \phi K_S$ decay amplitude is provided by
the chromo-magnetic operator $Q_{8g}$.
In particular, as already mentioned in section 2, gluino contributions 
to $C_{8g}$, which are proportional to $\dd{LR}{23}$, can be very large
with respect to the SM ones,
being enhanced by terms of order $m_{\tilde{g}}/m_b$.
In addition, large gluino effects in $C_{8g}$ may still escape
\bsg constraints \cite{KCGG}.
This is a remarkable property
and it is due to the fact that in the gluino contributions
to dipole operators $Q_{7\gamma,8g}$, 
the ratio of $|C_{8g}/C_{7\gamma}|$ is
enhanced by color factors with respect to typical contributions of
W or chargino exchanges \cite{bbmr}.

%------------------ CHARGINO ---------------------
%-------------------------------------------------
\begin{figure}[tpb]
\begin{center}
\dofourfigs{3.1in}{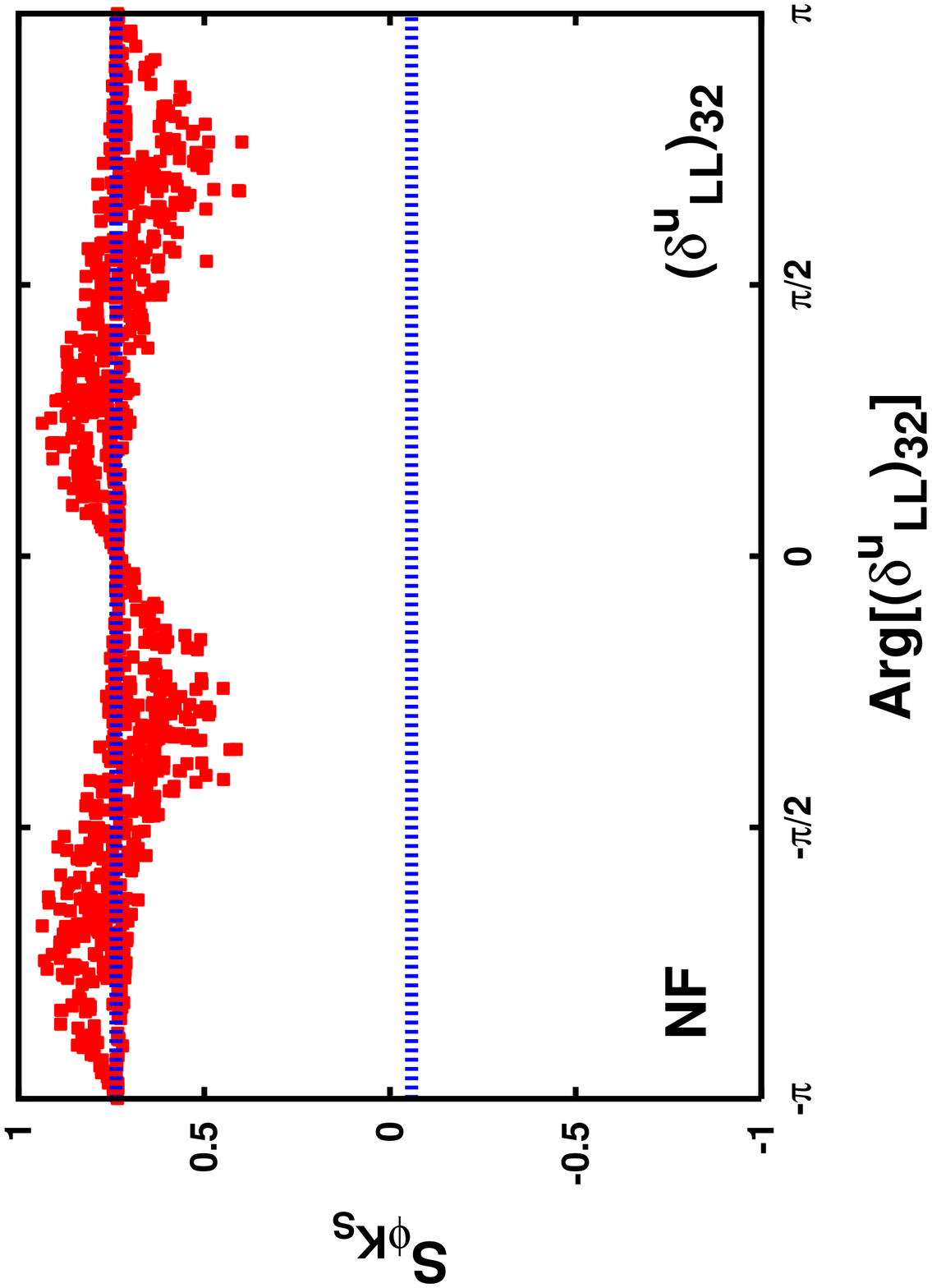}{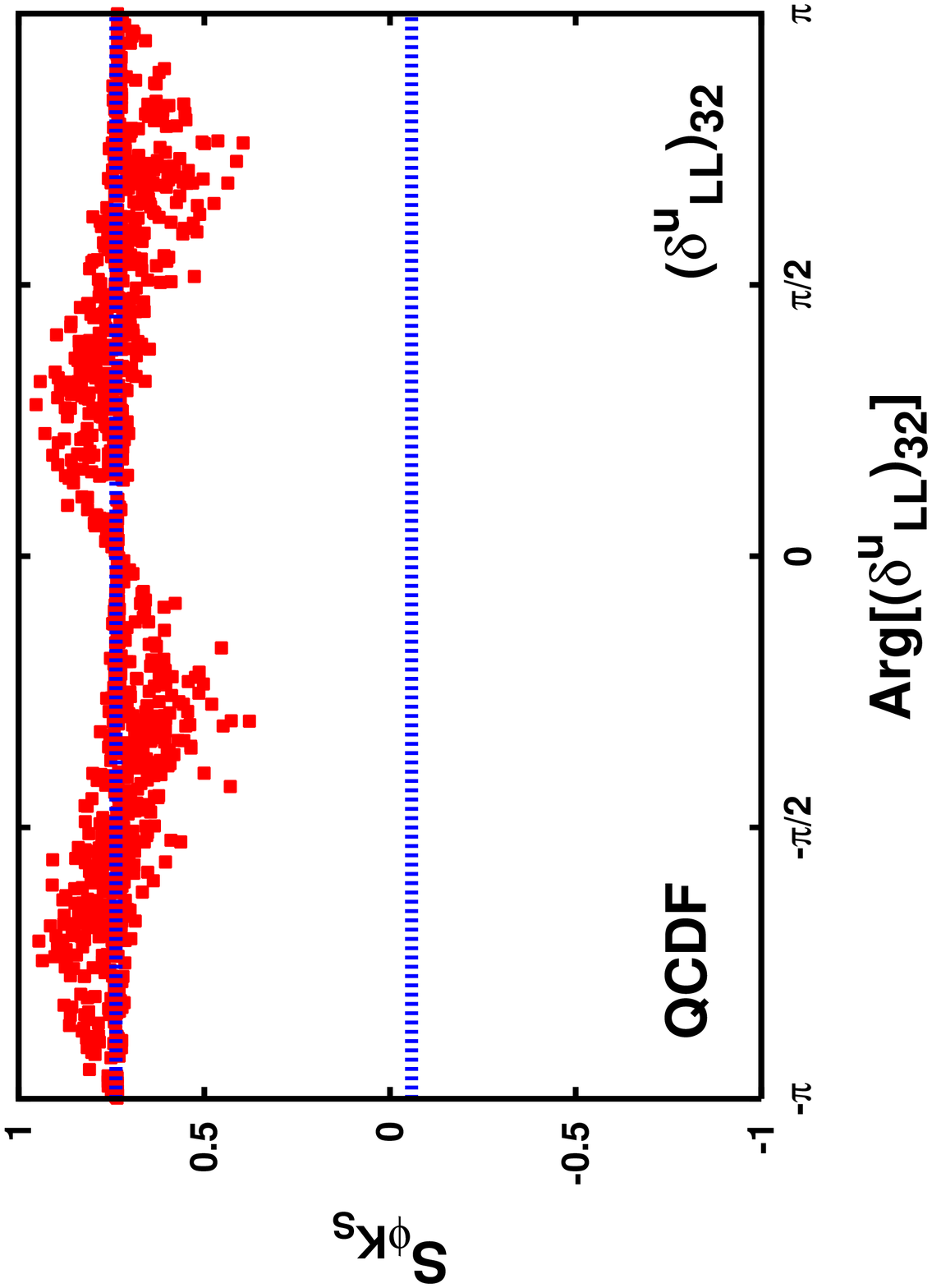}
{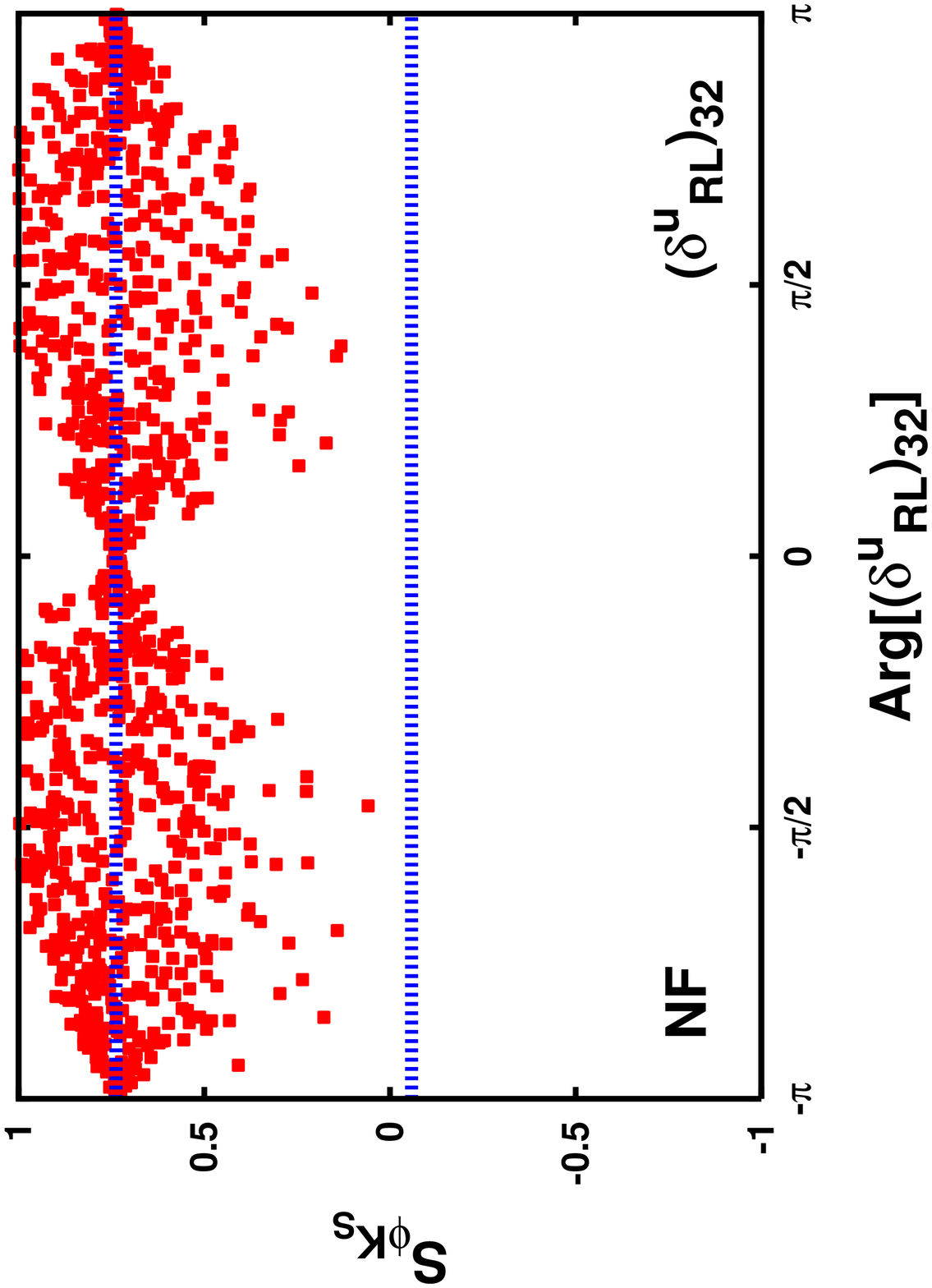}{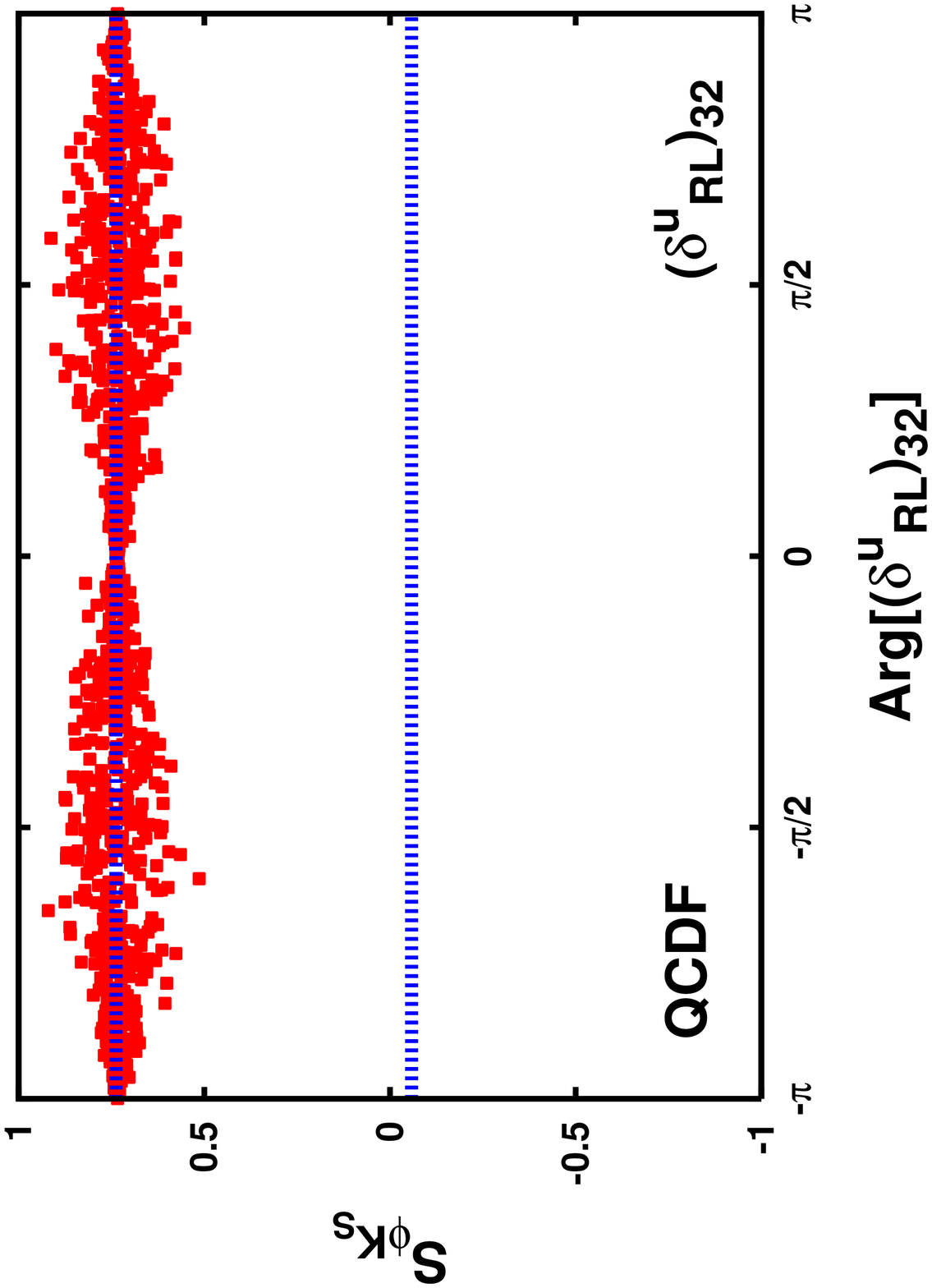}
\end{center}
\caption{\small As in Fig.~\ref{Fig41}, but for
$S_{\Phi K_S}$ as a function of arg[$\du{LL}{32}$]
(top) and  arg[$\du{RL}{32}$] (bottom) with chargino
contribution of one mass insertion 
$\du{LL}{32}$ (top) and $\du{RL}{32}$ (down).}
\label{Fig42}
\end{figure}
%-----------------------------------------------------------------

Now we discuss the chargino effects to $S_{\phi K_S}$, which are
summarized in Fig.~\ref{Fig42}.
These contributions have been first analyzed, in the framework of MIA,
in Ref.\cite{our}, but only using NF approach for evaluating the hadronic
matrix elements.
Here, we extend our previous analysis in \cite{our} 
by including the corresponding predictions in the QCDF approach.
In Fig.~\ref{Fig42}, $S_{\phi K_S}$ is plotted versus the argument of the 
relevant chargino mass insertions, namely 
$\du{LL}{32}$ and $\du{RL}{32}$.
Same conventions as in Fig.~\ref{Fig41} have been adopted 
for left and right plots.

As in the gluino dominated scenario, we have scanned over the 
relevant SUSY parameter space, 
in particular, the average squark mass $\tilde{m}$, 
weak gaugino mass $M_2$, the $\mu$ term, 
and the light right stop mass $\tilde{m}_{\tilde{t}_R}$. 
Also $\tan{\beta}=40$ has been assumed and we take into 
account the present experimental bounds on SUSY spectra, in particular
$\tilde{m} > 300$ GeV, the lightest chargino mass $M_{\chi}>90$ GeV, 
and $\tilde{m}_{\tilde{t}_R} \ge 150$ GeV. As in the gluino case,
we scan over the real and imaginary part of 
the mass insertions $\du{LL}{32}$ and 
$\du{RL}{32}$, by considering the constraints on 
BR(\bsg) and \bbar mixing at 95\% C.L.. 
The \bsg constraints 
impose stringent bounds on $\du{LL}{32}$, specially at large $\tan{\beta}$
\cite{our}.
Finally, as in the other plots, we scanned over the QCDF free parameters
$\rho_{A,H}<2 $ and $0<\phi_{A,H}<2\pi$.

From these results we can see that also for the chargino 
dominated scenario the predictions in  NF and in QCDF are quite close, 
apart from a slight difference in the $\du{RL}{32}$ ones 
that we will discuss below.
The main conclusion in this scenario is that negative 
values of $S_{\phi K_S}$ cannot be achieved neither in NF nor in QCDF.

The reason why extensive regions of negative values of $S_{\phi K_S}$
are excluded here, is only due to the \bsg constraints \cite{our}.
Indeed, as shown in our previous work \cite{our}, 
the inclusion of $\du{LL}{32}$ mass insertion 
can generate large and negative values of $S_{\phi K_S}$,
by means of chargino contributions to chromo-magnetic operator $Q_{8g}$
which are enhanced by terms of order $m_{\chi^{\pm}}/m_b$.
However, contrary to the gluino scenario, the ratio $|C_{8g}/C_{7\gamma}|$ 
is not enhanced by color factors and large contributions to
$C_{8g}$ leave unavoidably to the breaking of \bsg constraints.

On the other hand, the contribution of $\du{RL}{32}$ is independent 
of $\tan{\beta}$ and so large effects in $R_{\phi}$ that 
could drive $S_{\phi K_S}$ toward the region of negative values 
cannot be achieved. 
However, as can be seen from Fig.~\ref{Fig42}, 
while both 
in NF $\du{RL}{32}$ and $\du{LL}{32}$ contributions are within the 
2$\sigma$ experimental range, in NF the $\du{RL}{32}$ contribution
fits better than in QCDF.
This can be explained by the fact that 
$\du{RL}{32}$ mainly gives contribution to the electroweak operators, whose
matrix elements are more sensitive, with respect to the other operators,
to the approach adopted for their evaluation.

As shown in our previous work \cite{our}, by scanning over the two relevant
mass insertions $\du{RL}{32}$ and $\du{LL}{32}$, the
\bsg constraints on $\du{LL}{32}$ are a bit more relaxed, but 
a large amount of fine tuning between SUSY parameters is necessary
if small values of $S_{\phi K_S}$ are required.
\\

In order to understand the behavior of these results, it is useful
to look at the 
numerical parametrization of the ratio of amplitudes
${\cal R}_{\phi}\equiv \frac{A_{SUSY}(B\to \phi K_S)}{A_{SM}(B\to \phi K_S)}$
in terms of the relevant mass insertions.
Below, we present numerical results for ${\cal R}_{\phi}$
in both NF and QCDF. In QCDF 
we set to zero the effect of annihilation and hard scattering diagrams, 
corresponding to the choice of $\rho_{A,H}=0$ and $\phi_{A,H}=\pi$. 
In this case, we expect QCDF predictions to be quite close to the NF ones.
In particular, for a gluino mass and average squark mass of order
$\tilde{m}=m_{\tilde{g}}=500$ GeV, we obtain
\bea
{\cal R}_{\phi}|^{NF}_{\tilde{g}}&\simeq& 
\left\{
-0.08\, \dd{LL}{23}\,-\,
120\, \dd{LR}{23}\right\}
\,+\,
\left\{L\leftrightarrow R\right\},
\nonumber \\
\nonumber \\
{\cal R}_{\phi}|^{QCDF}_{\tilde{g}}&\simeq& 
\left\{-0.14\,\times e^{-i\,0.1}
 \dd{LL}{23}\,-\,
127\, \times e^{-i\,0.08}
\dd{LR}{23}\right\}
\,+\,
\left\{L\leftrightarrow R\right\}
\label{Rgl_Phi},
\eea
while in the case of chargino, by using gaugino mass $M_2=200$ GeV, 
$\mu= 300$ GeV, $\tilde{m}_{\tilde{t}_R}=150$ GeV, and $\tan{\beta}=40$,
we have
\bea
{\cal R}_{\phi}|^{NF}_{\chi}&\simeq& 
1.83\, \du{LL}{32}\,-\,
0.32\, \du{RL}{32}\,+\,
0.41\, \du{LL}{31}\,-\,
0.07\, \du{RL}{31},
\nonumber \\
\nonumber \\
{\cal R}_{\phi}|^{QCDF}_{\chi}&\simeq& 
1.89\times e^{-i\,0.07}
\, \du{LL}{32}\,-\,
0.11\times e^{-i\,0.17}
\, \du{RL}{32}
\nonumber \\
&+&
0.43\times e^{-i\,0.07}
\, \du{LL}{31}\,-\,
0.02\times e^{-i\,0.17}
\, \du{RL}{31}.
\label{Rch_Phi}
\eea
where the first symbol ${\cal R}_{\phi}|_{\tilde{g}}^{NF}$ 
means that the corresponding quantity ${\cal R}_{\phi}$ 
has been calculated in NF approach including only gluino contributions.
Analogously for the other cases of QCDF and for chargino ($\chi$) exchanges.

From results in Eqs.(\ref{Rgl_Phi})--(\ref{Rch_Phi}), 
it is clear that the largest SUSY
effect is provided by the gluino and chargino
contributions to the chromo-magnetic operator which are proportional to
$\dd{LR}{23}$ and $\du{LL}{32}$, respectively.
However, the \bsg constraints play a crucial role in this case.
For the above SUSY configurations, the \bsg decay set the following
(conservative) constraints on gluino and chargino contributions:
$|\dd{LR}{23}|< 0.019$ and $|\du{LL}{32}|<0.20$.
Implementing these bounds in Eqs.(\ref{Rgl_Phi})--(\ref{Rch_Phi}),
we see that gluino contribution
can easily achieve larger values of $R_{\phi}=|{\cal R}_{\phi}|$ (see 
Eqs.(\ref{ratioPHI}),(\ref{cpmixing_phi}))  than chargino one, and this 
is the main reason why extensive regions with negative value of 
$S_{\phi K_S}$ are favored and
disfavored in Figs.~\ref{Fig41} and \ref{Fig42}, respectively.

%%%%%%%%%%%%%%%%%%%% CP ASYMMETRY in B-> Eta(prime) K_S %%%%%%%%%%%%%%
\vspace{0.5cm}
\subsection{{\bf CP asymmetry in $B\to {\eta^{\prime} K_S}$} decay}
%%%%%%%%%%%%%%%%%%%%%%%%%%%%%%%%%%%%%%%%%%%%%%%%%%%%%%%%%%%%%%%%%%%%%%

Recent measurements of CP asymmetry in $B\to {\eta^{\prime} K_S}$
show another discrepancy with SM predictions.
In particular from Eq.(\ref{Seta}), the world average is 
\bea
S_{\eta^{\prime} K_S}&=&0.41\pm 0.11 \, ,
\eea
and which is about $2.5\sigma$ deviation from SM expectations.
From these results we see that, similarly to what happens in 
$B\to \phi  K_S$ decay, large deviations from SM are possible.

Since SUSY contributes to both CP asymmetries with the same 
CP violating source, it is possible that the SUSY effects driving 
$S_{\phi K_S}$ towards negative values, could also sizeably decrease 
$S_{\eta^{\prime} K_S}$.
The main reason for that is because the leading SUSY contributions 
to the amplitudes of $B\to \phi K_S$ and $B\to \eta^{\prime} K_S$
enter through the Wilson coefficient of $C_{8g}$ and the 
operator $Q_{8g}$ has a comparable matrix elements 
in both processes. 
However, since NP corrections enter through the quantity 
$R_{\eta^{\prime}}$, the role of the SM contribution  will be crucial.
Indeed, while the $B\to \phi K_S$ amplitude is purely generated at one-loop
in the SM, the $B\to \eta^{\prime} K_S$ one 
receives tree-level contribution from the SM by means 
of non-vanishing matrix element of $Q_2$. 
Therefore, the increase of SUSY contributions to 
$C_{8g}$ is now compensated in $R_{\eta^{\prime}}$,
by the large SM amplitude contribution.

\begin{figure}[tpb]
\begin{center}
\dofourfigs{3.1in}{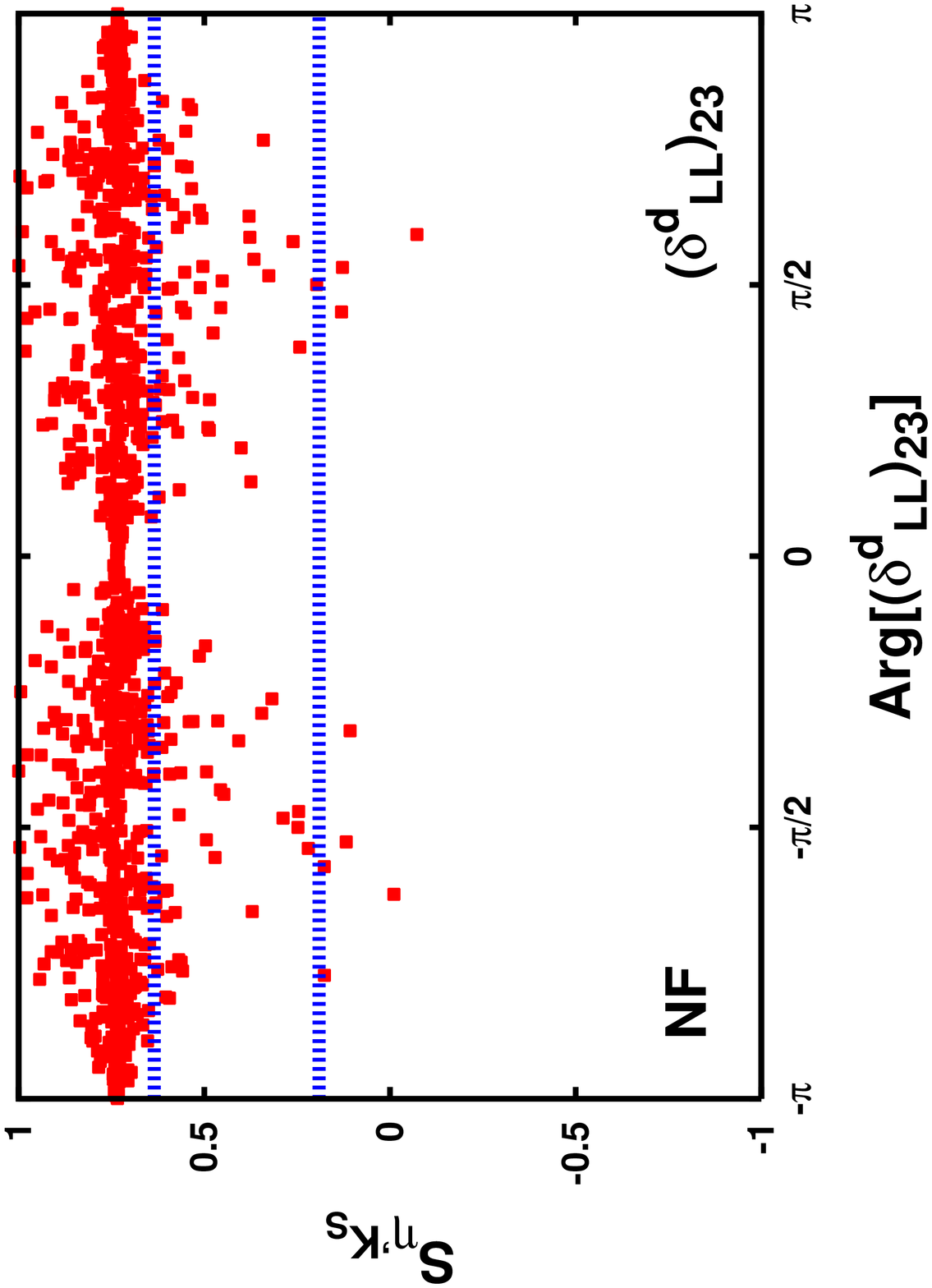}{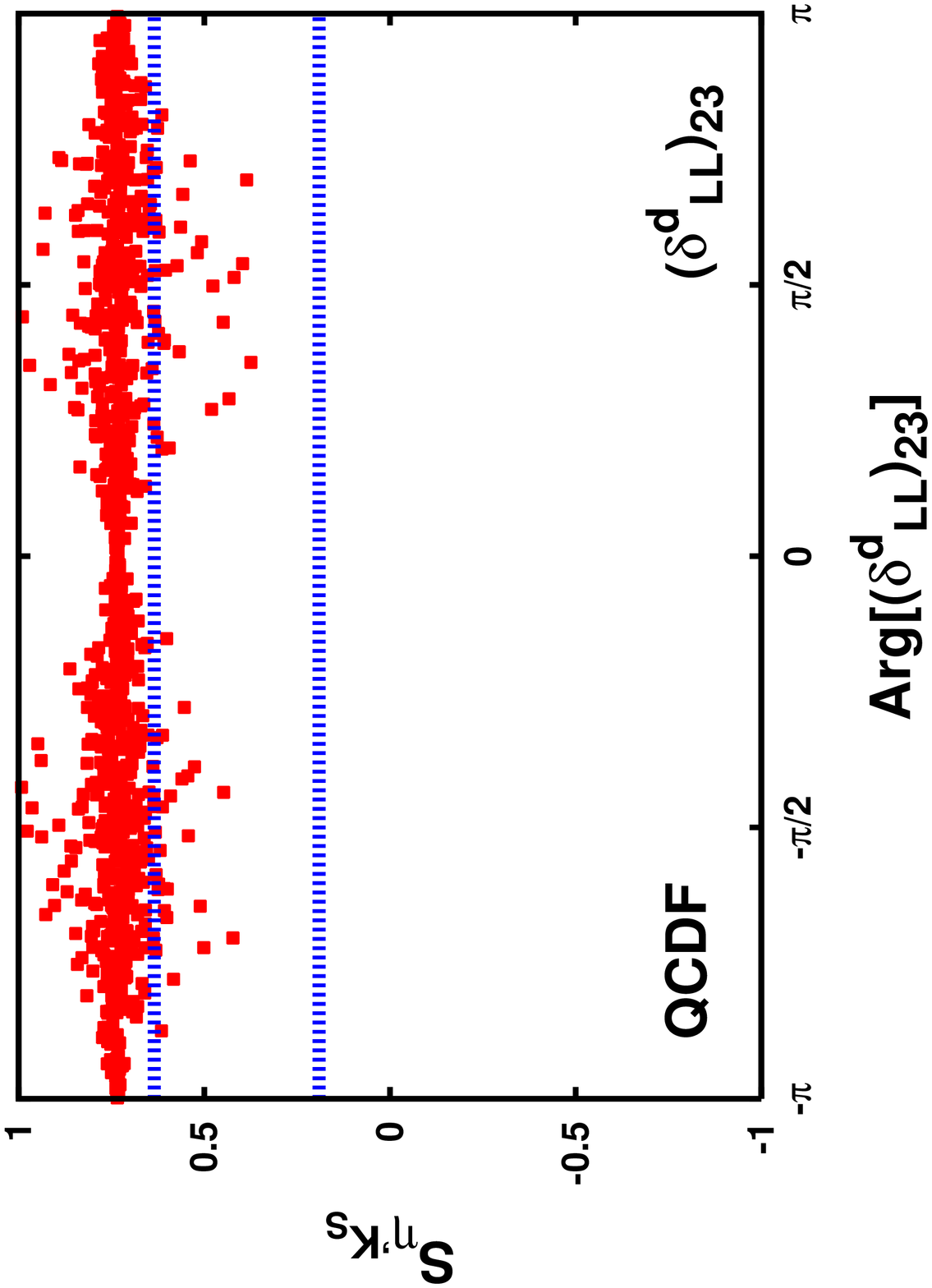}
{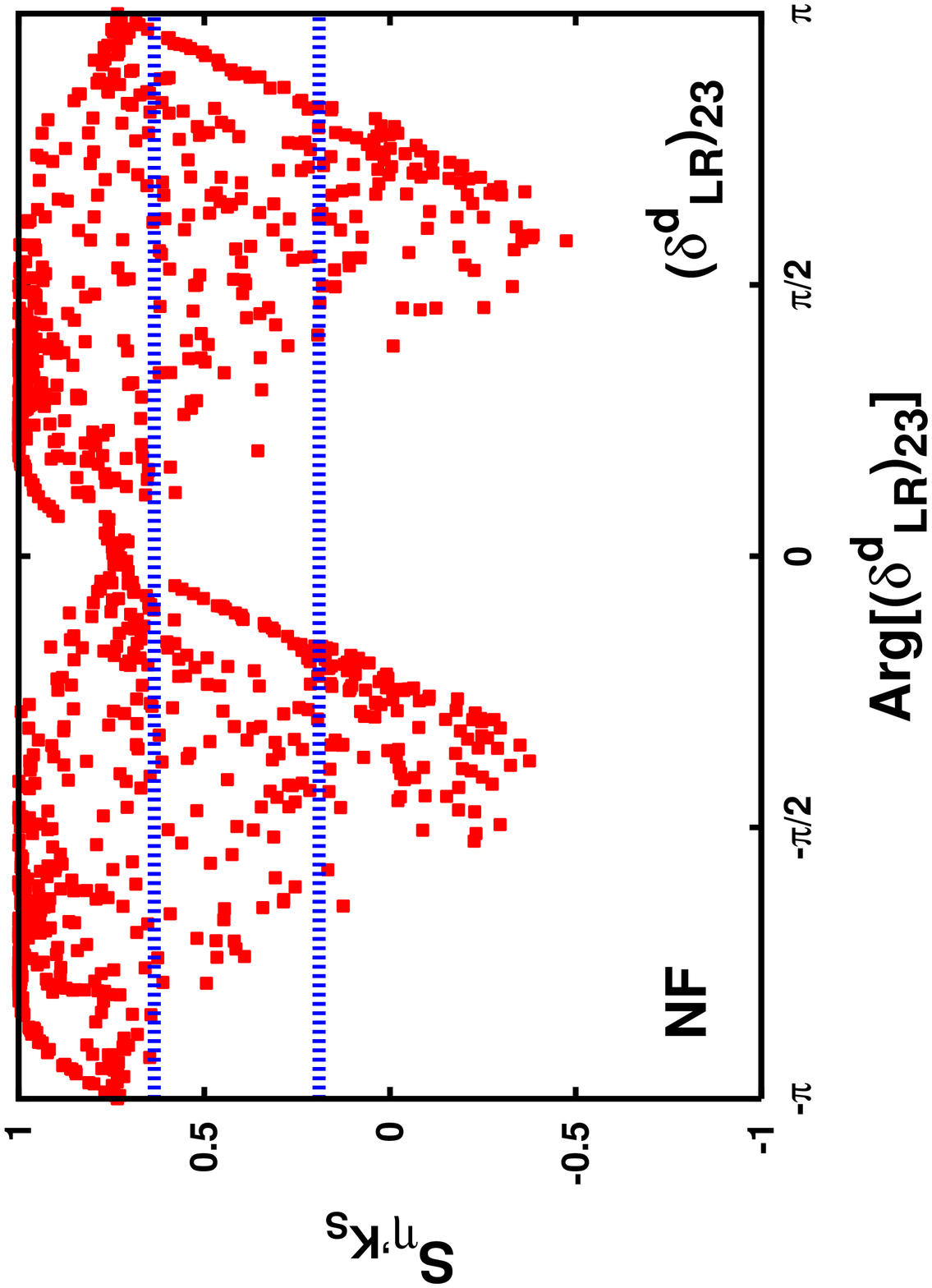}{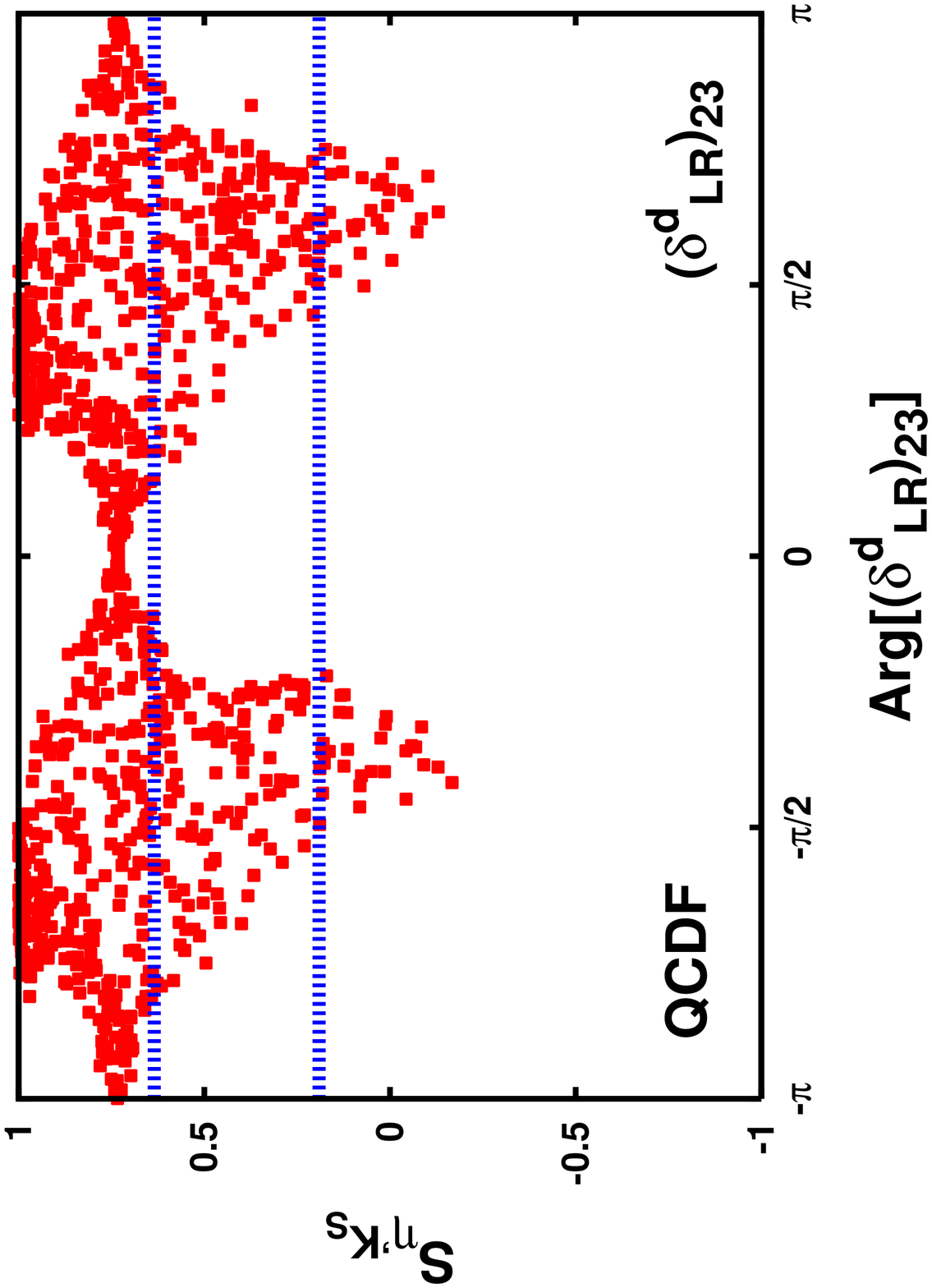}
\end{center}
\caption{\small As in Fig.~\ref{Fig41}, but for
$S_{\eta^{\prime} K_S}$.}
\label{Fig43}
\end{figure}

We show our results for gluino in Fig.~\ref{Fig43}, where we have just 
extended the same analysis of $B\to \phi K_S$.
Same conventions as in figures for $B\to \phi K_S$ have been adopted here.
As we can see from these results, there is a depletion of the 
gluino contribution in  $S_{\eta^{\prime} K_S}$, precisely for the reasons 
explained above. Regions of negative values of $S_{\eta^{\prime} K_S}$
are more disfavored with respect to $S_{\phi K_S}$, 
but a minimum of $S_{\eta^{\prime} K_S}\simeq 0$ can be easily 
achieved. Comparing NF and QCDF, we can also see that 
SUSY predictions in NF and QCDF are very close.

\begin{figure}[tpb]
\begin{center}
\dofourfigs{3.1in}{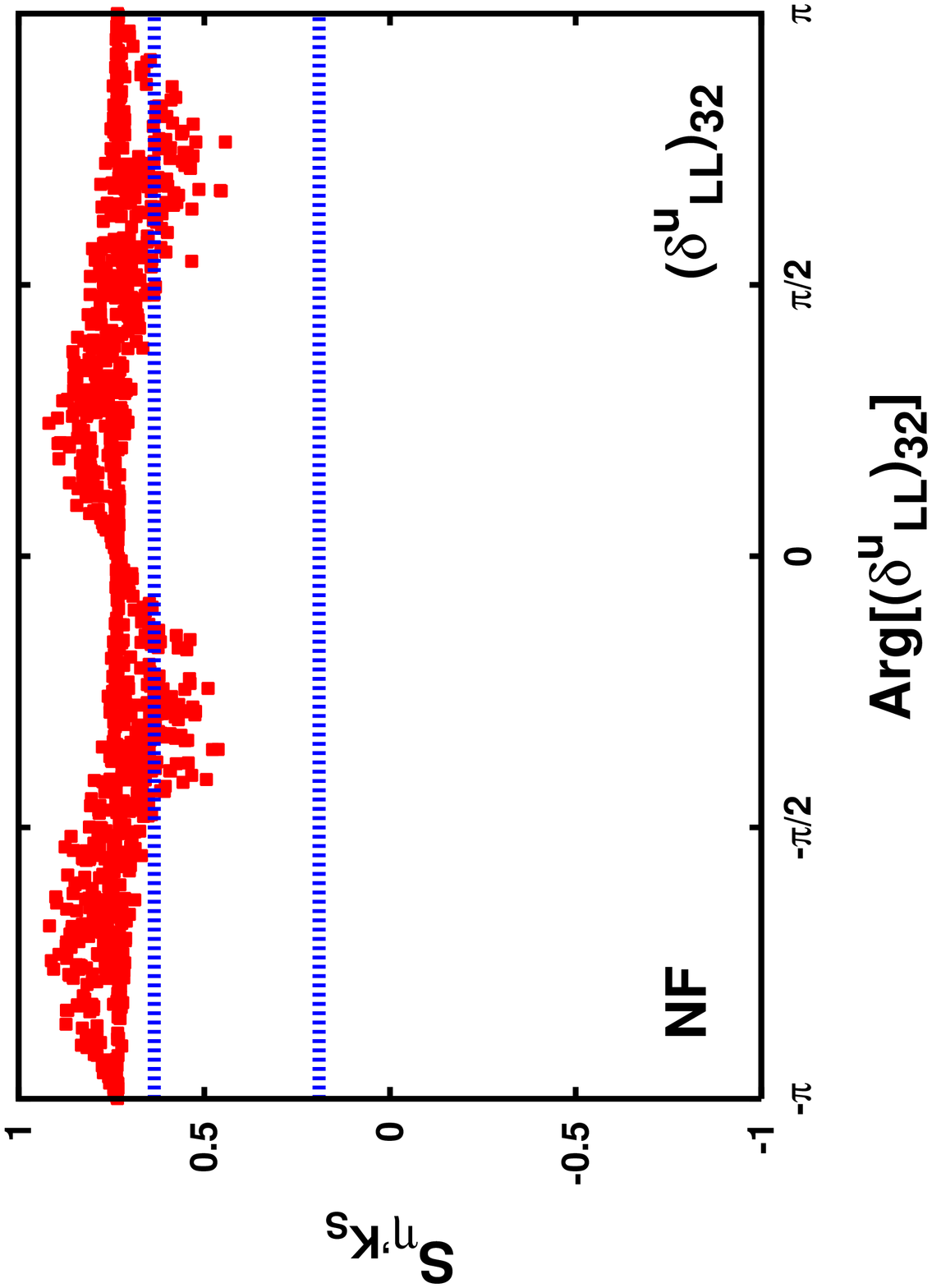}{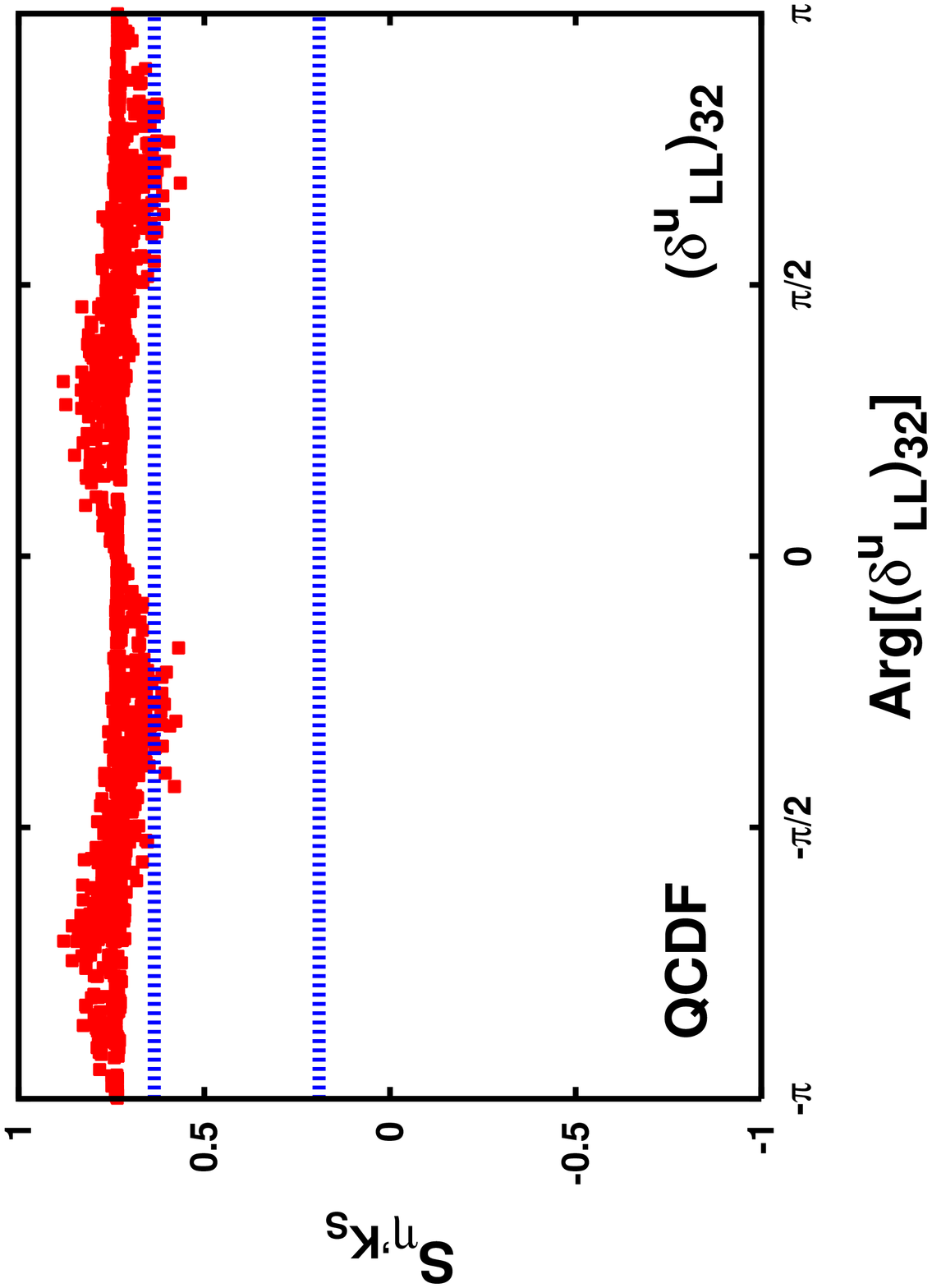}
{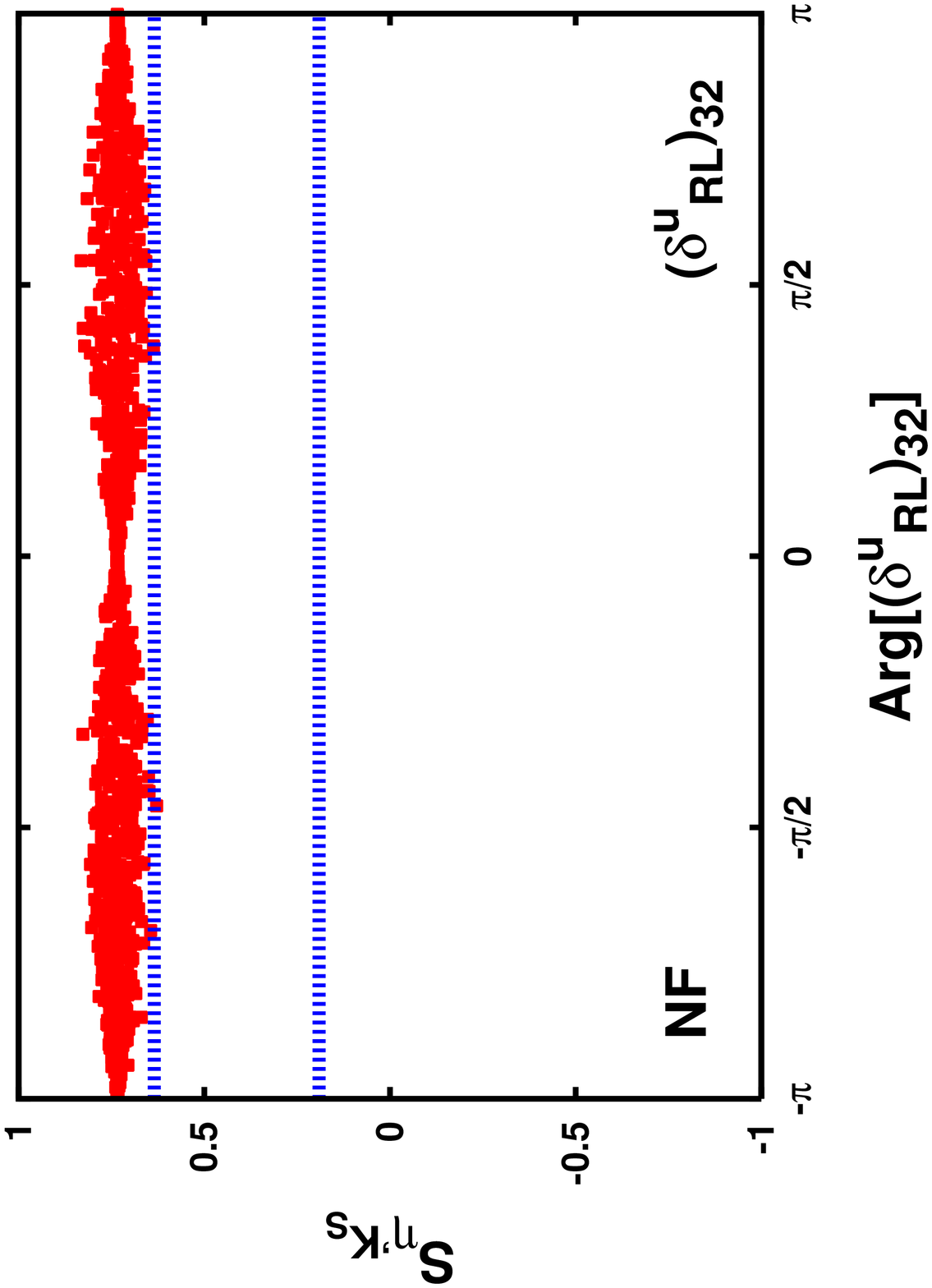}{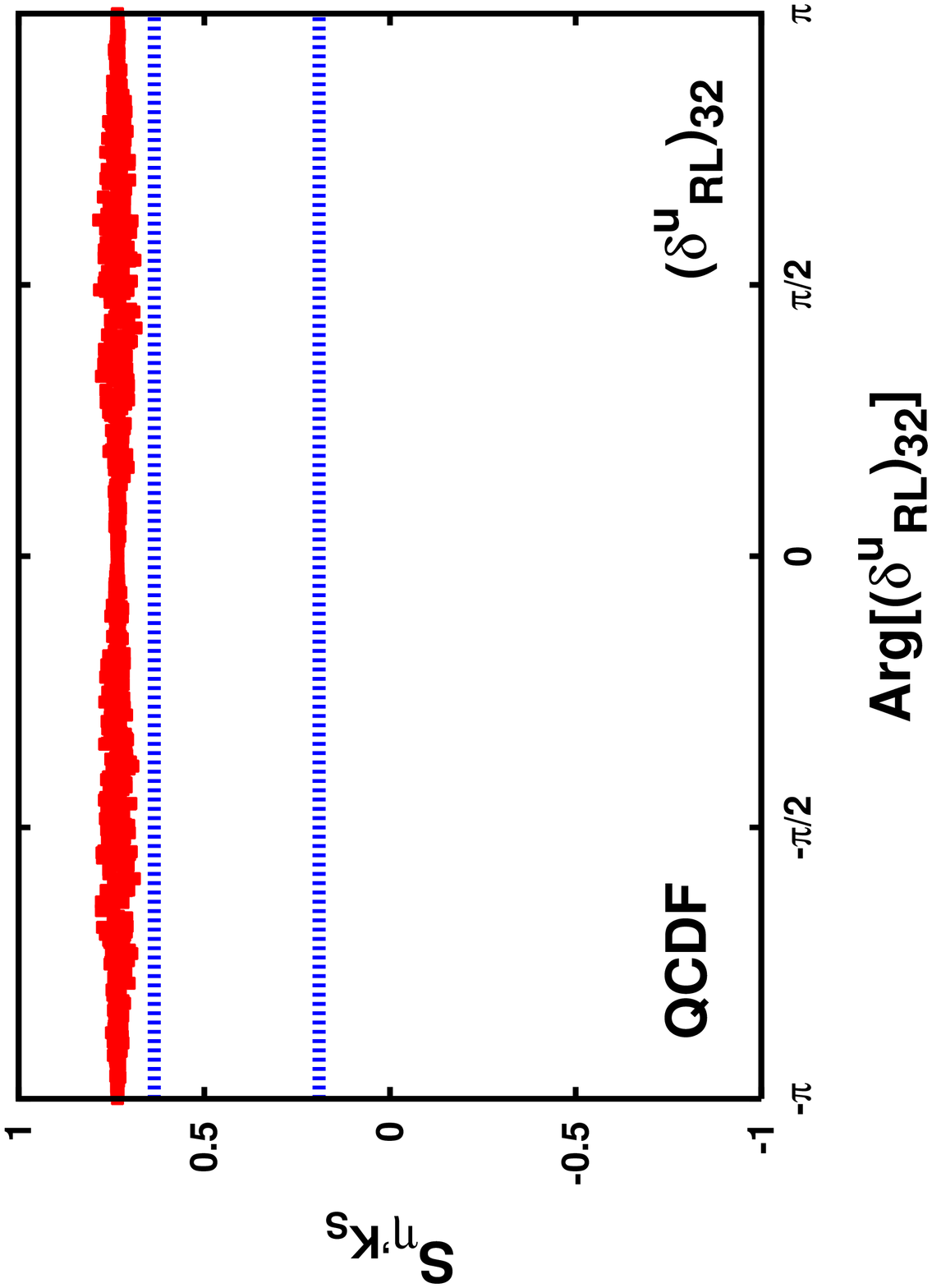}
\end{center}
\caption{\small As in Fig.~\ref{Fig42}, but for
$S_{\eta^{\prime} K_S}$.}
\label{Fig44}
\end{figure}

Finally, in Fig.~\ref{Fig44} we present our results for chargino contributions.
Here we see that charginos can produce at most a
deviation from SM predictions of about $\pm 20$ \%, and the most 
conspicuous effect is achieved by $\du{LL}{32}$.
These results again show the relevant role played by the chromo-magnetic
operator.

% ************************* ETA(PRIME)  KS ************************************
%------------------ GLUINO ---------------------

%------------------------------------------------------------
\vspace{0.1cm}

%------------------ CHARGINO ---------------------

%------------------------------------------------------------
\vspace{0.1cm}

%------------------ CORRELATION OF ASYMMETRIES -------------

%------------------------------------------------------------
\vspace{0.1cm}

\begin{figure}[tpb]
\begin{center}
\dofigs{3.1in}{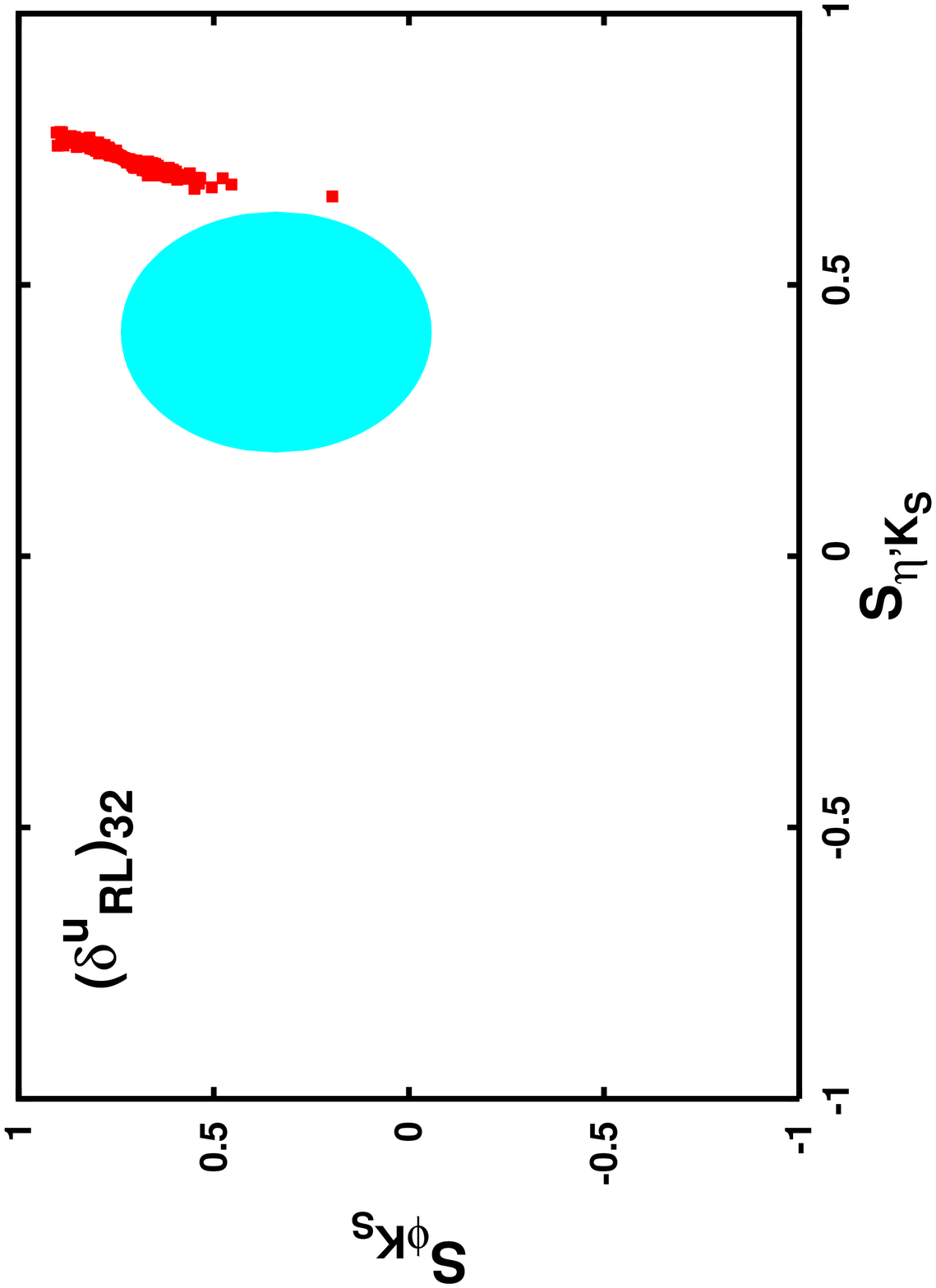}{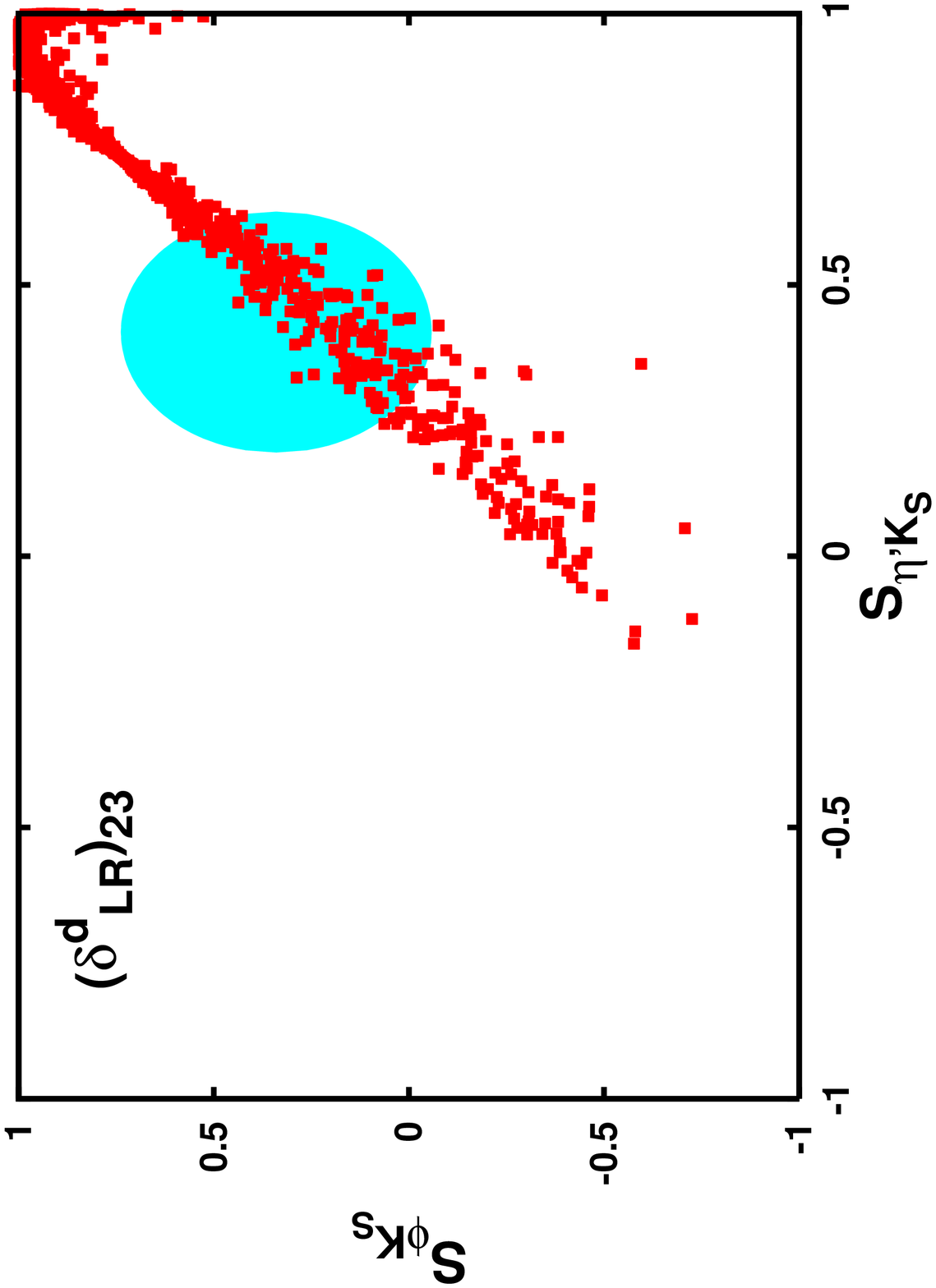}
\end{center}
\caption{\small Correlation of 
asymmetries $S_{\Phi  K_S}$ versus 
$S_{\eta^{\prime} K_S}$ with the contribution of
one mass insertion $\du{RL}{32}$ (left) and 
$\dd{LR}{23}$ (right), for chargino (left) 
and gluino (right) exchanges. 
Region inside the ellipse corresponds to the allowed experimental ranges
at $2\sigma$ level.}
\label{Fig45}
\end{figure}

\begin{figure}[tpb]
\begin{center}
\dofigs{3.1in}{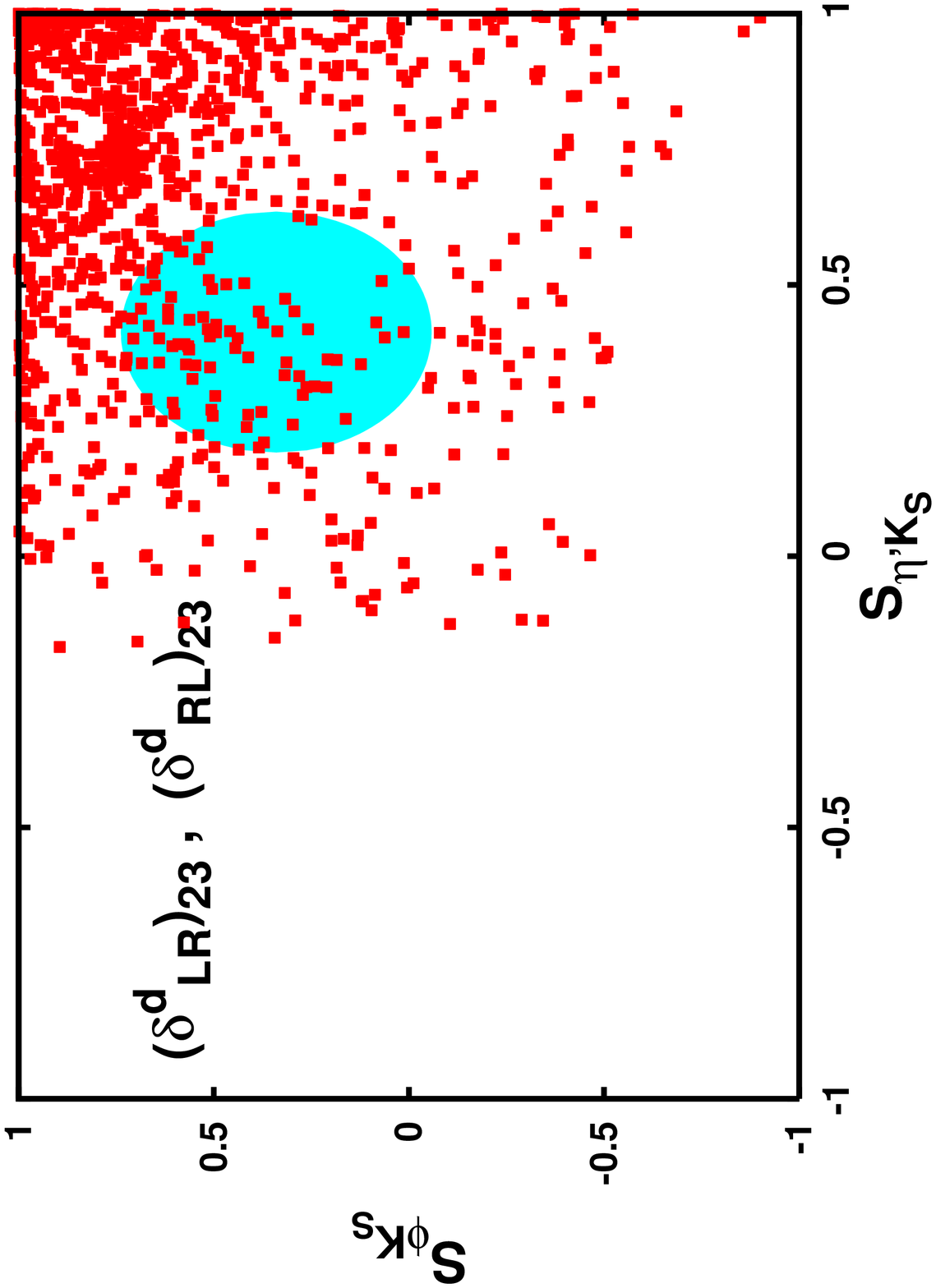}{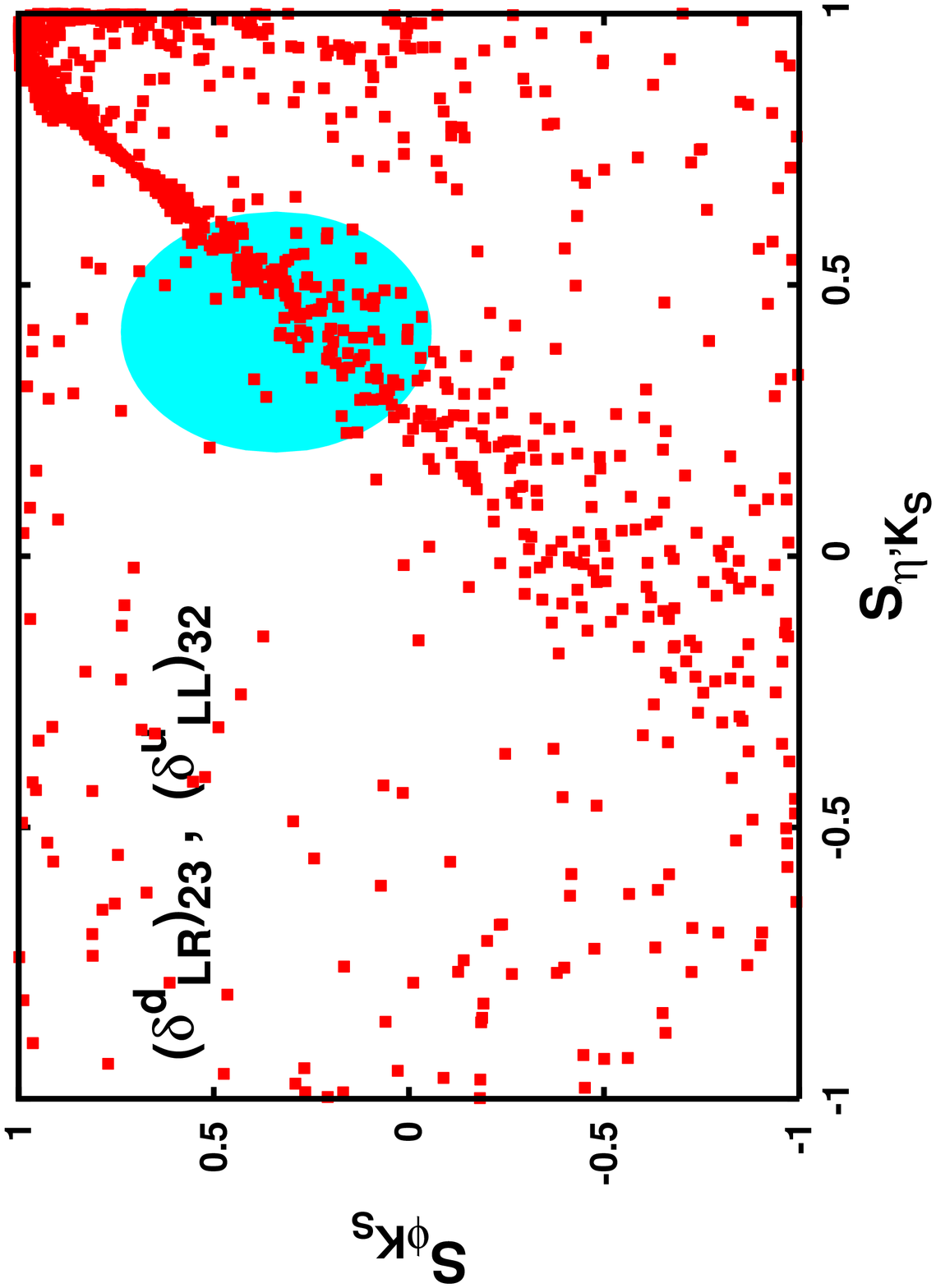}
\end{center}
\caption{\small 
{ As in Fig.~\ref{Fig45}, but for 
arg[$\dd{LR}{23}$]=arg[$\dd{RL}{23}$] (left)
and 
arg[$\dd{LR}{23}$]=arg[$\du{LL}{32}$] (right),
with the contribution of two mass insertions
$\dd{LR}{23}$ \& $\dd{RL}{23}$ (left) for gluino exchanges,
and $\dd{LR}{23}$ \& $\du{LL}{32}$ (right)
for both gluino and chargino exchanges.}}
\label{Fig46}
\end{figure}

As in the case of $B\to \phi K_S$ decay, we present below the parametrization 
of ${\cal R}_{\eta^{\prime}}\equiv 
\frac{A_{SUSY}(B\to \eta^{\prime} K_S)}{
A_{SM}(B\to \eta^{\prime} K_S)}$,
by using the same SUSY inputs 
adopted in Eqs.~(\ref{Rch_Phi}), (\ref{Rgl_Phi}).
For gluino contributions we have
\bea
{\cal R}_{\eta^{\prime}}|^{NF}_{\tilde{g}}&\simeq& 
\left\{
-0.08\, \dd{LL}{23}\,-\,
79\, \dd{LR}{23}\right\}
\,-\,
\left\{L\leftrightarrow R\right\},
\nonumber \\
\nonumber \\
{\cal R}_{\eta^{\prime}}|^{QCDF}_{\tilde{g}}&\simeq& 
\left\{-0.07\,\times e^{i\,0.24}
 \dd{LL}{23}\,-\,
64\dd{LR}{23}\right\}
\,-\,
\left\{L\leftrightarrow R\right\},
\label{Rgl_eta}
\eea
while for chargino exchanges we obtain 
\bea
{\cal R}_{\eta^{\prime}}|^{NF}_{\chi}&\simeq& 
1.2\, \du{LL}{32}\,-\,
0.01\, \du{RL}{32}\,+\,
0.27\, \du{LL}{31}\,-\,
0.002\, \du{RL}{31},
\nonumber \\
\nonumber \\
{\cal R}_{\eta^{\prime}}|^{QCDF}_{\chi}&\simeq& 
0.95\, \du{LL}{32}\,-\,
0.025\times e^{-i\,0.19}
\, \du{RL}{32}
\nonumber \\
&+&
0.21
\, \du{LL}{31}\,-\,
0.006\times e^{-i\,0.19}
\, \du{RL}{31}\, .
\label{Rch_eta}
\eea
Notice that in the second curly brackets in Eq.(\ref{Rgl_eta}) 
there is a minus sign in front.
This takes into account for the minus sign in the matrix elements of
$\tilde{Q}_i$ operators as shown in Eq.(\ref{Qi_eta_ME}).

In both cases of gluino and chargino contributions, we see that 
the coefficient of $\dd{LR}{23}$ and $\du{LL}{32}$
mass insertions is smaller in comparison than the corresponding ones
in $B\to \phi K_S$, see Eqs.(\ref{Rgl_Phi}), (\ref{Rch_Phi}).
As explained before, this depletion comes from the fact that 
the SM amplitude of $A(B\to \eta^{\prime} K_S)$ receives contribution 
from tree-level $W$ exchanges and it is larger than $A(B\to \phi K_S)$.

This general behavior is going in the right direction to explain the 
experimental data. In order to show better this effect, we plotted 
in Figs.~\ref{Fig45} the correlations between 
$S_{\phi K_S}$ versus $S_{\eta^{\prime} K_S}$ for both 
chargino (left plot) and 
gluino (right plot) in QCDF. For illustrative purposes, 
in all figures analyzing correlations,  we colored the area of 
the ellipse corresponding to the allowed experimental range
at $2\sigma$ level.\footnote{ All ellipses here
have axes of length $4\sigma$. As a first approximation, no correlation
between the expectation values of the two observables have been assumed.}
In Fig.~\ref{Fig45} 
we scanned over the real and imaginary part of 
the single relevant mass insertions, namely 
$\du{RL}{32}$ (left plot) and  
$\dd{LR}{23}$ (right plot) for chargino and gluino, respectively.

In conclusion, as can be seen  from the results in Fig.~\ref{Fig45}, 
pure chargino 
exchanges have no chance to fit data at  $2\sigma$ level, while gluino
can fit them quite well.
At this point we want to stress that sizeable chargino contributions 
to the CP asymmetries, in particular from $\du{LL}{32}$ mass insertion,
are ruled out by $b\to s\gamma$ constraints.
Therefore, it would be interesting to see if the effects of
a light charged Higgs exchange could relax these constraints allowing
chargino contributions to fit inside the experimental ranges.
%------------------------------------
\begin{figure}[tpb]
\begin{center}
\dofigs{3.1in}{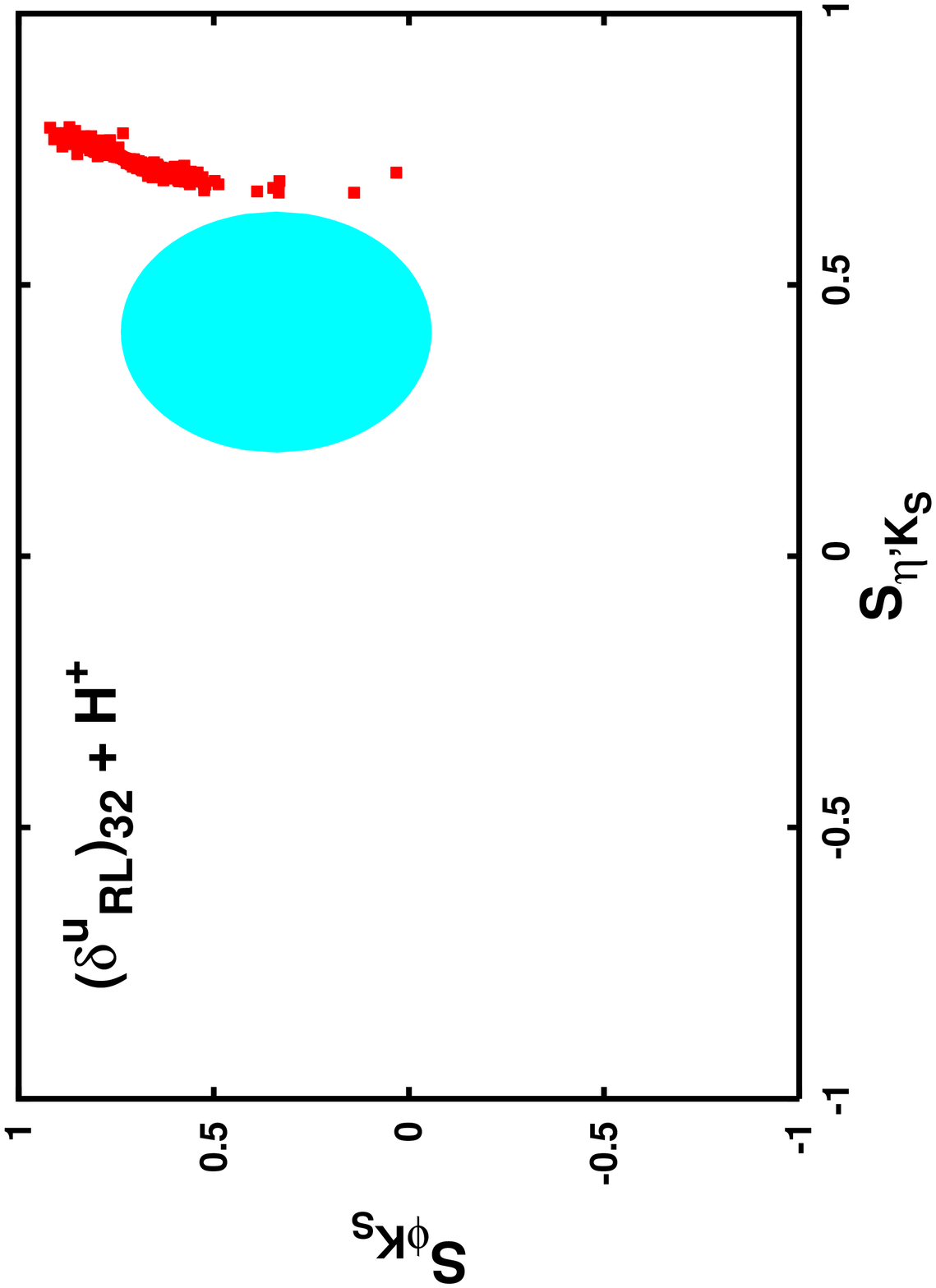}{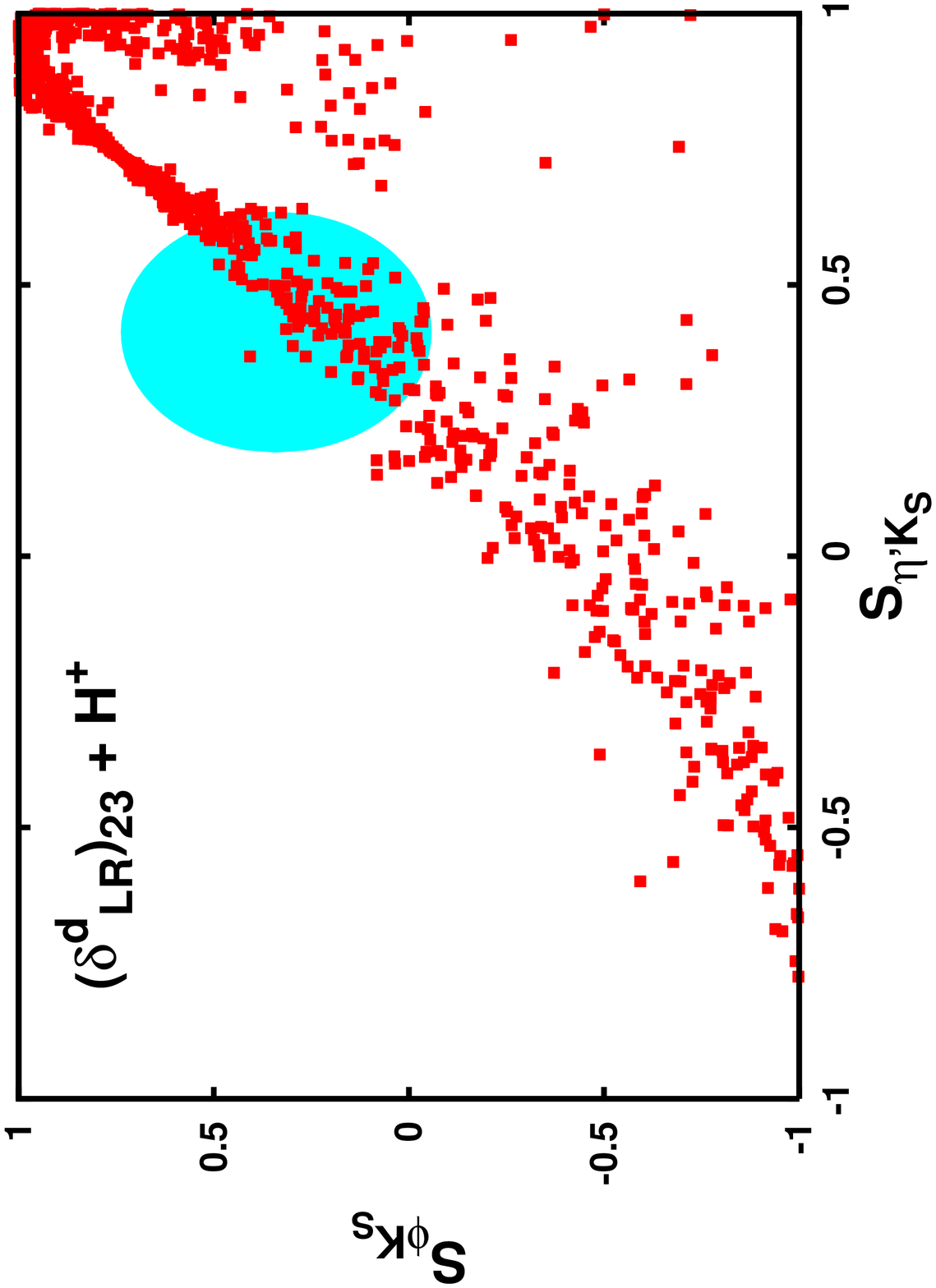}
\end{center}
\caption{\small As in Fig.~\ref{Fig45}, but taking into account
a charged Higgs exchange with mass
$m_H=200$ GeV and $\tan{\beta}=40$.}
\label{FigH1}
\end{figure}
%------------------------------------
\begin{figure}[tpb]
\begin{center}
\dofigs{3.1in}{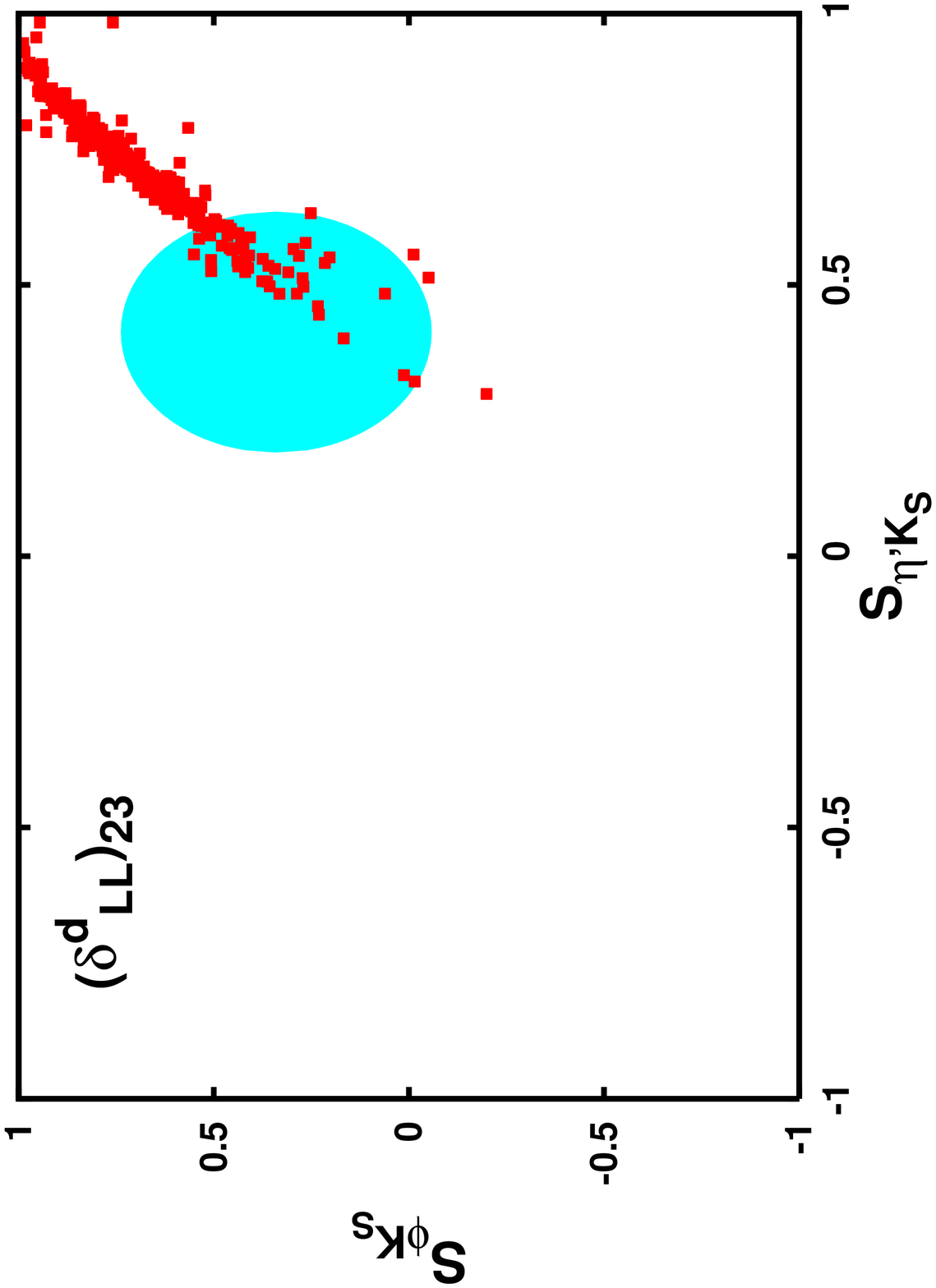}{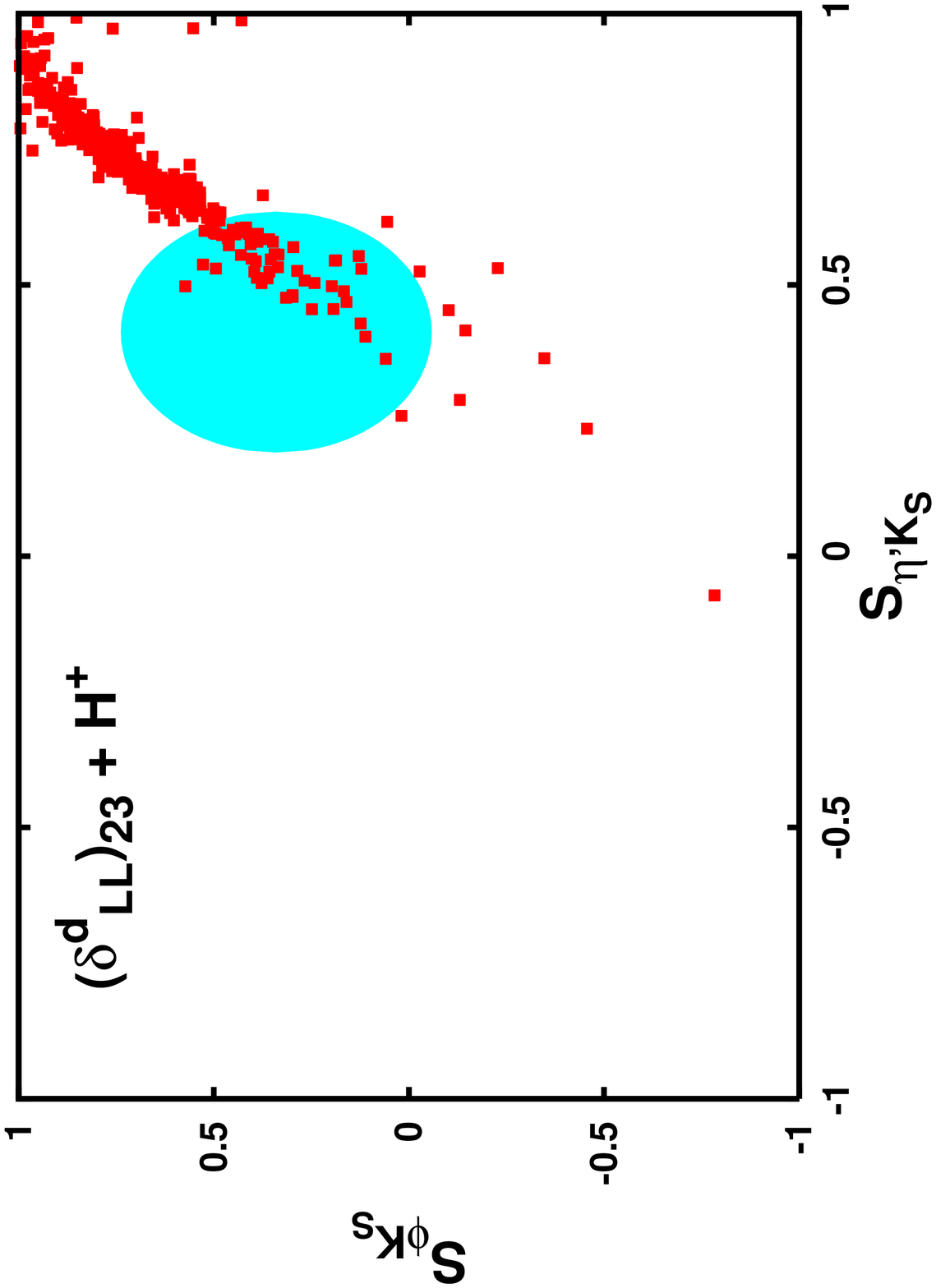}
\end{center}
\caption{\small 
As in Fig.~\ref{Fig45}, but 
for gluino contributions with single mass insertion $\dd{LL}{23}$.
In the right plot the effect of a charged Higgs exchange, with mass
$m_H=200$ GeV and $\tan{\beta}=40$, has been taken into account.}
\label{FigH2}
\end{figure}
%------------------------------------
\begin{figure}[tpb]
\begin{center}
\dofigs{3.1in}{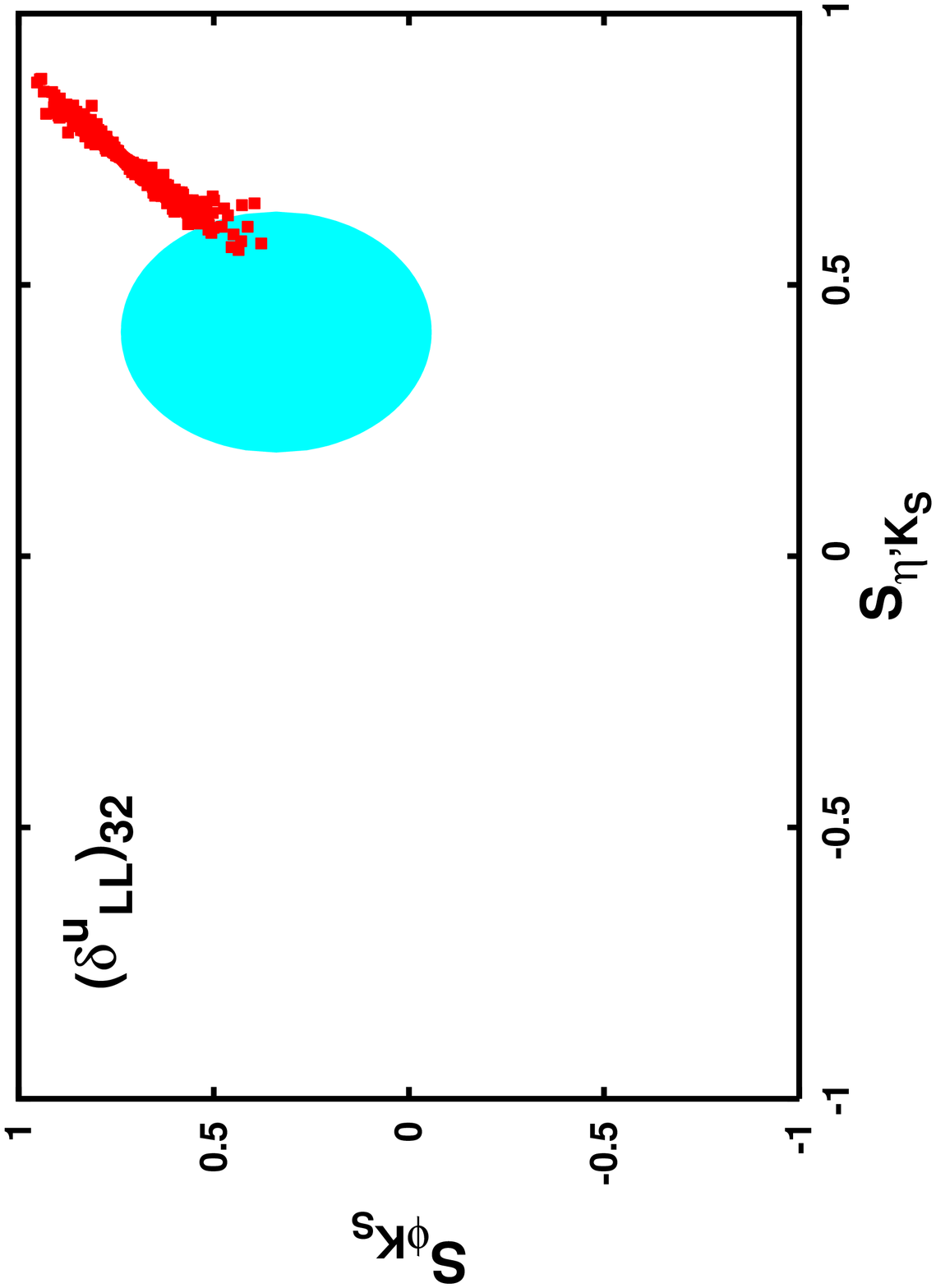}{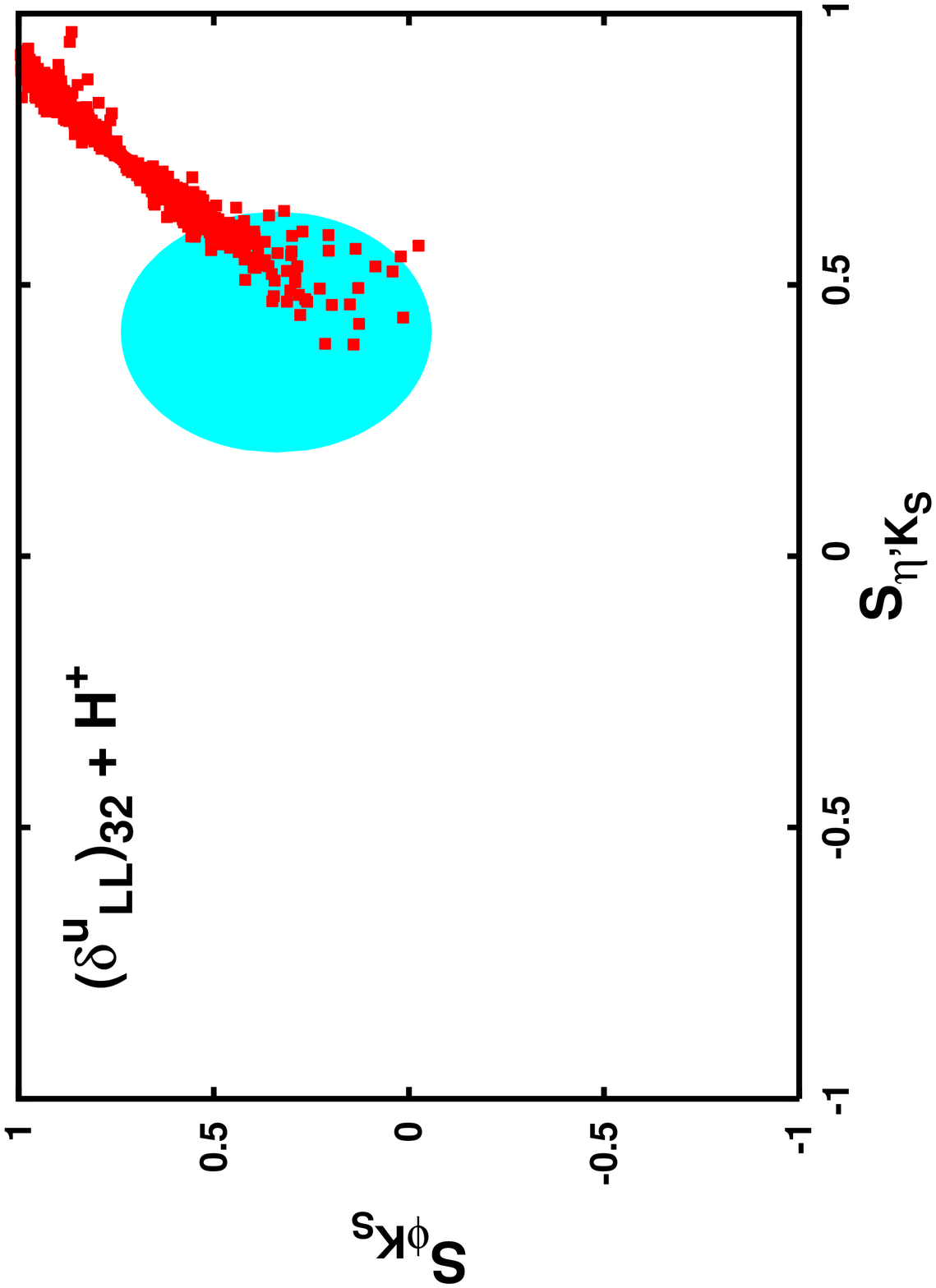}
\end{center}
\caption{\small As in Fig.~\ref{FigH2}, but for chargino contribution with 
single mass insertion $\du{LL}{32}$.}
\label{FigH3}
\end{figure}
%------------------------------------
For this purpose, we present in Figs.~\ref{FigH1}-\ref{FigH3} 
the impact of a light charged Higgs in both chargino and gluino exchanges. 
In particular, plots in Fig.~\ref{Fig45} should be directly compared with the 
corresponding ones in Fig.~\ref{FigH1}, 
where a charged Higgs with mass $m_H=200$ GeV 
and $\tan{\beta}=40$ has been taken into account.
From these results we can see that the effects of charged Higgs exchange in 
the case of $\du{RL}{32}$ mass insertion are negligible,
as we expect from the fact that terms proportional to 
$\du{RL}{32}$ in $b\to s \gamma$ and $b\to s g$ amplitudes are not enhanced by 
$\tan{\beta}$.
On the other hand, in gluino exchanges with $\dd{LR}{23}$ or $\dd{LL}{23}$
(see Figs.~\ref{FigH1},\ref{FigH2}), the most conspicuous 
effect of charged Higgs contribution is in populating the area outside 
the allowed experimental region. This
is clearly due to a destructive interference with $b\to s\gamma$ amplitude, 
which relaxes the $b\to s\gamma$ constraints.
The most relevant effect of a charged Higgs exchange is in the 
scenario of chargino exchanges with $\du{LL}{32}$ mass insertion. In this case,
as can be seen from Fig.~\ref{FigH3}, a strong destructive interference with 
$b\to s\gamma$ amplitude can relax the $b\to s\gamma$ constraints in the 
right direction, allowing chargino predictions 
to fit inside the experimental region.
Moreover, we have checked that, for $\tan{\beta}=40$, 
charged Higgs heavier than approximately 600 GeV cannot affect the CP
asymmetries significantly.

In a recent paper \cite{KK}, it has been 
shown that there exists a particular scenario
in which gluino contribution can sizeably decrease $S_{\phi K_S}$ 
providing on the same time a very modest effect in $S_{\eta^{\prime} K_S}$.
This can be achieved if one assume that both $\dd{LR}{23}$ and 
$\dd{RL}{23}$ mass insertions are of the same order, including
their CP violating phase. In this case gluino contributes with the same weight 
to both $Q_i$ and $\tilde{Q}_i$ operators in Eq.(\ref{Heff}), 
where $\tilde{Q}_i$ are the operators with opposite chirality.
Due to the different parity 
of the final states, as explained in section 3, 
the contribution of the corresponding 
Wilson coefficients to the amplitude of 
$B\to \phi K_S$ and $B\to \eta^{\prime} K_S$ enter in combination as
$C_i + \tilde{C}_i$ and  $C_i - \tilde{C}_i$, respectively.

However, the analysis of Ref. \cite{KK} was based on the 
NF approach, and one may wonder, if their results still hold in QCDF.
For this reason we repeat here the same analysis, but in QCDF
and scanning over the 
strong CP phases $\phi_{A,H}$ and $\rho_{A,H}<2$.
Results of this scenario are shown in Fig.~\ref{Fig46} (left plot),
where we scanned over two mass insertions simultaneously,
namely $\dd{LR}{23}$ and $\dd{RL}{23}$,
by assuming that their CP violating phases are the same.
From these results we can see that
a large number of  SUSY configuration fitting inside the 
$2\sigma$ experimental region, can be obtained.
\\

We have also considered another scenario in which both chargino and gluino 
exchanges are assumed to contribute simultaneously 
with relevant mass insertions, namely $\dd{LR}{23}$ and 
$\du{LL}{32}$.
We plot the corresponding results for 
the correlations between $S_{\phi K_S}$ versus $S_{\eta^{\prime} K_S}$
in the right plot of Fig.~\ref{Fig46}. As in the case of Fig.~\ref{Fig45},
we assume a common SUSY CP violating phase between the two mass insertions.
From these results we can see that also in this case a large number of 
configurations can fit inside the $2\sigma$ experimental regions.
This result shows a remarkable fact. 
The stringent bounds on $\du{LL}{32}$ 
from the experimental limits on $BR(B\to X_s \gamma)$
are relaxed when one considers both gluino and chargino contributions, 
which come with different sign. 
This does not happen, if two chargino mass insertions are
considered.
Then some configurations with large 
$\tan\beta$ are allowed and therefore chargino can contribute 
significantly to the CP asymmetries $S_{\phi K_S}$ and 
$S_{\eta^{\prime} K_S}$.   
\\

Finally, we would like to comment about the stability of
SUSY predictions for $S_{\phi K_S}$ against the low renormalization scale
$\mu_b$, where Wilson coefficients are evaluated.
In NF the scale dependence in the Wilson coefficients
is not compensated by the corresponding one in the matrix elements, so a large
uncertainty is expected.
However, we have noticed that also in QCDF this uncertainty still persists.
In particular the coefficients in the parametrizations in QCDF
in Eq.~(\ref{Rgl_Phi})--(\ref{Rch_Phi})
can vary up to 30-40 \% when the $\mu_b$ scale is changed from 
$\mu_b=m_b/2$ to $\mu_b=m_b$. All the 
numerical results in this paper correspond to the choice
$\mu_b=m_b^{\rm pole}=4.5$ GeV. 
In both NF and QCDF, this residual scale dependence in $R_{\phi}$
is mainly due to the scale dependence in the SM amplitude 
and in the $C_{8g}(\mu_b)$ contribution to the SUSY  amplitude.
However, we noticed that 
the main sensitivity to the scale 
in $R_{\phi}$ comes from the SM contribution.
On the other hand, the main source of sensitivity in $R_{\eta^{\prime}}$
comes from $C_{8g}(\mu_b)$ in the SUSY amplitude, 
since the SM one, receiving tree-level 
contributions, is less sensitive to the renormalization scale.

In the next section we are going to consider the correlations of $S_{\phi K_S}$
and $S_{\eta^{\prime} K_S}$ asymmetries versus the corresponding 
branching ratios of $B\to \phi K_S$ and $B\to \eta^{\prime} K_S$.

%%%%%%%%%%%%%%%%%%%%%%%%%%%%%%%%%%%%%%%%%%%%%%%%%%%%%%%%%%%%%%%%%%%%%%%%
%
\section{SUSY contributions to $BR(B\to \phi(\eta^{\prime})K)$}
%
%%%%%%%%%%%%%%%%%%%%%%%%%%%%%%%%%%%%%%%%%%%%%%%%%%%%%%%%%%%%%%%%%%%%%%%%
In this section we discuss the impact of large SUSY contribution, 
which is required to explain the deviations from SM in
CP asymmetries of $S_{\phi(\eta^{\prime})K_S}$, on 
the branching ratios (BR) of  
$B\to \phi K$ and $B\to \eta^{\prime} K$. 
As shown in Eqs.(\ref{BRphi}),(\ref{BReta}), the experimental measurements of 
these BRs at BaBar, Belle, and CLEO lead to the following averaged results
\cite{{hfag}}:
\begin{eqnarray}
BR(B\to \phi K) &=& (8.3^{+1.2}_{-1.0}) \times 10^{-6},
\label{phiresult}\\
BR(B\to \eta^{\prime} K) &=& (65.2^{+6.0}_{-5.9} ) \times 10^{-6}.
\label{eta'result}
\end{eqnarray}
which mean that $BR(B\to \phi K)$ is in good agreement with SM predictions,
while the experimental value of $BR(B\to \eta^{\prime} K)$ is two 
to five times larger than the SM one. 
Since these two processes are highly correlated and 
both are based on the $b \to s$ transition, it seems a challenge for 
the SUSY contribution
to enhance $BR(B\to \eta^{\prime} K)$ by a factor of two or more, while 
leaving other BRs and asymmetries inside their experimental ranges.

The branching ratio of $B\to M K $, with $M=\phi$ or $\eta^{\prime}$,
can be written in terms of the corresponding amplitude ${\cal M}_M$ as 
\begin{equation}
BR(B\to M K) = \frac{1}{8\pi} \frac{\vert P \vert}{M_B^2} 
\vert \mathcal{M}_M \vert^2 \frac{1}{\Gamma_{tot}},
\label{BR}
\end{equation}
where $\mathcal{M}_M = \langle M \bar{K}^0 \vert H_{eff}^{\Delta B=1} 
\vert \bar{B}^0 \rangle$ and
\begin{equation}
\vert P \vert = \frac{\left[\left(M_B^2 -(m_K+m_M)^2\right)
\left(M_B^2 -(m_K- m_M)^2\right)\right]^{1/2}}{2 M_B}.
\end{equation}

The inclusion of SUSY corrections modifies the BR as
\begin{equation}
BR^{\mbox{\tiny total}}(B\to M K) =BR^{\mbox{\tiny SM}}(B\to M K)\times
\left[1+2\cos(\theta_{M}-\delta_M) R_M+ R_M^2 \right]
\label{BRtotal}
\end{equation}
where $R_M=\left|\frac{A^{\mathrm{SUSY}}(B\to M K )}{A^{\mathrm{SM}}
(B\to M K)}
\right|$,
$\theta_M=arg\left[\frac{A^{\mathrm{SUSY}}(B\to M K)}
{A^{\mathrm{SM}}(B\to M K)}
\right]$  and $\delta_M$ is the corresponding strong phase.
The input parameters that we have used in the previous section 
with $\rho_{A,H}=1$ and
$\phi_{A,H}= \pi$ {\it i.e.} $X_{A,H}=0$ lead to 
$BR^{\mbox{\tiny SM}}(B\to \phi K)
= 2.76\times 10^{-6}$ and $BR^{\mbox{\tiny SM}}(B\to \eta^{\prime}K)
= 11.1\times 10^{-6}$. 

It is remarkable that in order to produce the experimental 
result of $S_{\phi K_S}$ 
by SUSY contribution, the phase $\theta_{\phi}= 
arg\left[\frac{A^{\mathrm{SUSY}}
(B\to \phi K_S)}{A^{\mathrm{SM}}(B\to \phi K_S)}\right]$ 
has to be around $\pm \pi/2$ 
and the strong phase $\delta_{\phi}$ should be of order $\pi$ or $0$, as
shown in section 4. These particular values for the phases suppress 
the leading term in Eq.~(\ref{BRtotal}). Therefore, for typical scenarios 
in which $R_{\phi}\leq 1$, the SUSY contributions 
do not enhance $BR(B\to \phi K)$ much. At most 
a total branching ratio of order $3 \times BR^{\mbox{\tiny SM}}
(B\to \phi K) \simeq 8.3 \times 10^{-6}$ can be achieved, 
as we will explain below,
which is still compatible with the experimental  measurements. 
Clearly, if $R_{\phi} \gg 1$ 
then the total branching ratio would exceed the 
experimental bound in Eq.(\ref{phiresult}). However, 
as shown in section 3, the inclusion of 
$b \to s \gamma$ constraints moderates possible large 
SUSY contributions to $R_{\phi}$. 
\\

Next we briefly discuss the dependence of the branching ratio
on the QCDF free parameters $\rho_{A,H}$, related to annihilation and
hard scattering diagrams. The only relevant
parameter here is the $\rho_A$ one, while regarding 
$\rho_H$, the BR shows a moderate dependence \cite{BN}.
By using dimensional analysis, one can see that for 
large values of $\rho_A$,
$BR^{\mbox{\tiny SM}}(B\to \phi K)$ scales like 
$\rho_A^4$. This result suggests that one can set very strong upper bounds
on $\rho_A$ by requiring that $BR^{\mbox{\tiny SM}}(B\to \phi K)$
does not exceed experimental range.
Clearly, when new corrections beyond the SM ones 
are included, upper bounds on $\rho_A$ could be
relaxed due to possible negative interferences 
between SM and new physics amplitudes.

In order to understand the impact of the annihilation diagrams on 
$BR(B\to \phi K)$, we show below the explicit dependence of the SM 
amplitude on the parameters $\rho_A$ and $\phi_A$:
\begin{equation}
i\, A^{SM}(B\to \phi K_S)  \simeq -1.8 \times 10^{-8} 
+7.9 \times 10^{-10}X_A  -6.1 \times 10^{-10}X_A^2  +2.2 \times 10^{-10}X_H  
\label{Amp_phi}
\end{equation}   
where numbers in the right hand side of Eq.(\ref{Amp_phi}) are 
in unity of GeV, 
and $X_{A,H}$ are defined in Eqs.(\ref{paramXAH}).
We have used SM central values as in Table 1 of \cite{BN}, and 
omitted contributions from imaginary parts which are quite small.

\begin{figure}[tpb]
\begin{center}
\dofigs{3.1in}{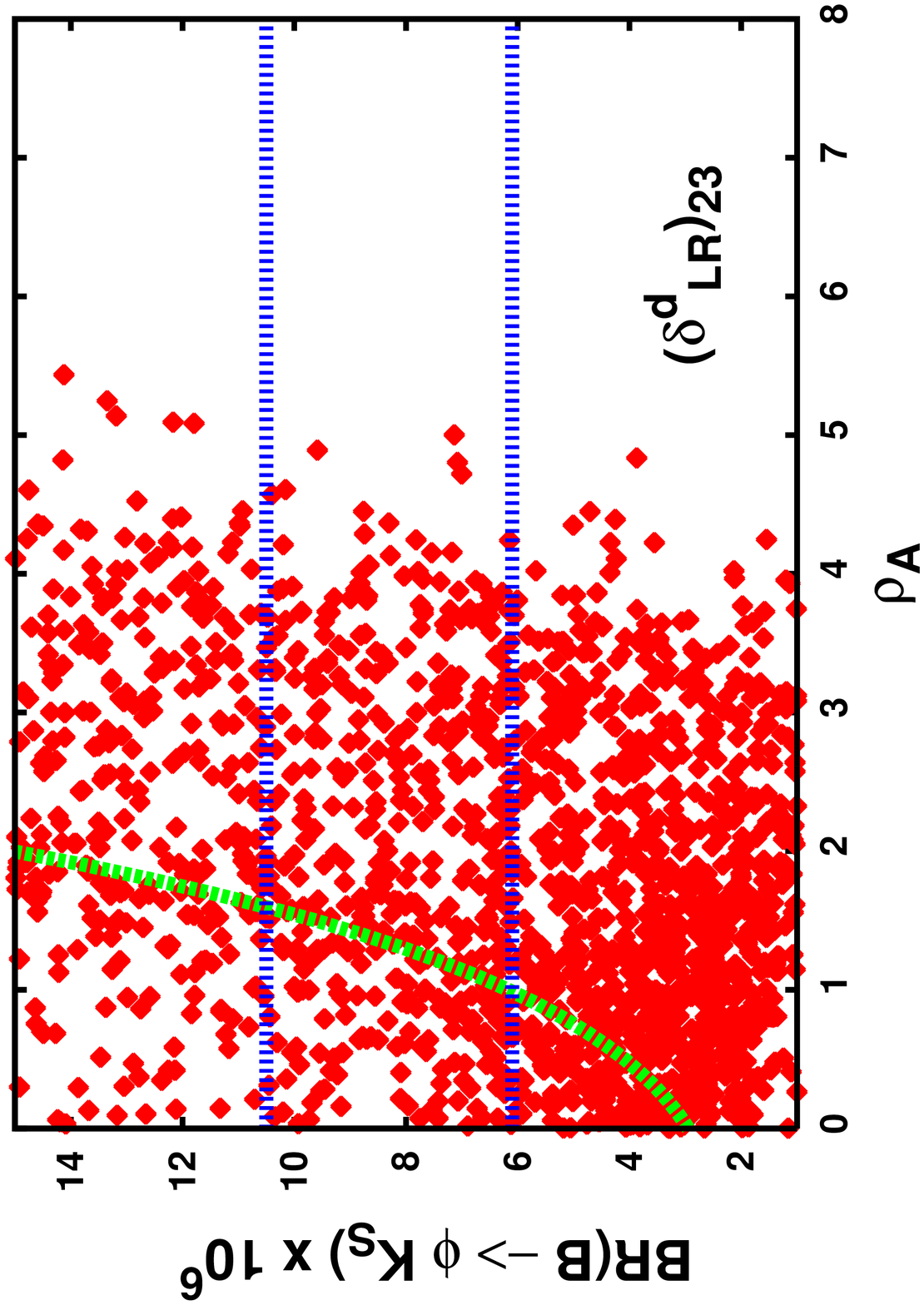}{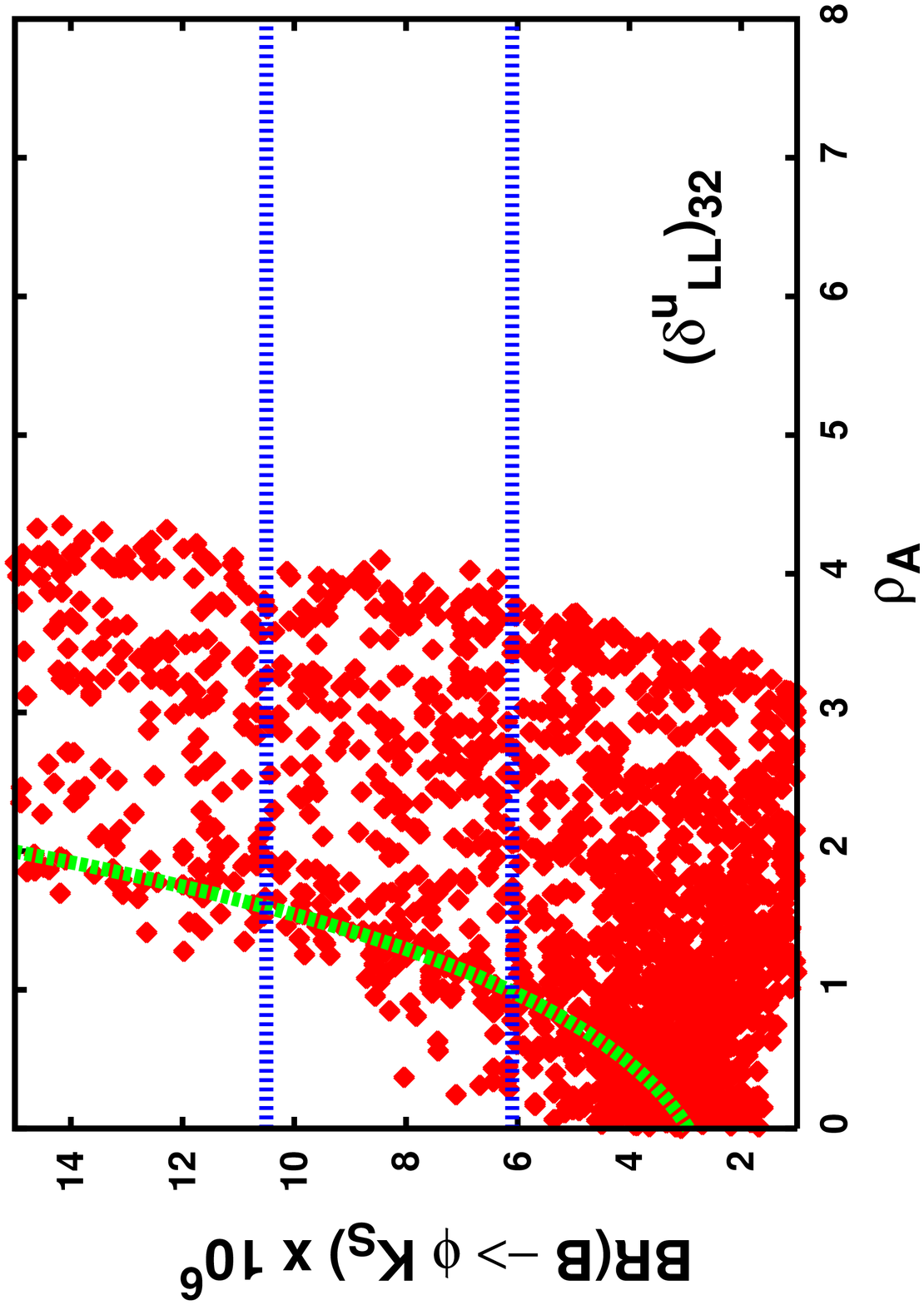}
\end{center}
\caption{\small $BR(B\to \phi K)$ as function of the 
annihilation parameter $\rho_A$ and for $\rho_H=1$, $\phi_{A,H}=0$,
for one mass insertion contribution of gluino $\dd{RL}{23}$ (left) 
and chargino $\du{LL}{32}$ (right).
Lighter dashed line corresponds to the SM case. The region
inside the two horizontal lines corresponds to the allowed 
experimental range at $2\sigma$ level.}
\label{BRrho}
\end{figure}

As can be seen from Eqs.(\ref{Amp_phi}) and (\ref{paramXAH}), 
for $\rho_A \gg 1$
the SM amplitude could be doubled and hence the SM branching ratio would be
about four times the result with $\rho_A <1$, therefore exceeding 
the present experimental range. As shown in Ref.\cite{BN}, and in 
Fig.~\ref{BRrho} (see lighter dashed line), the constraint on $\rho_A$
is already obtained for moderate values of $\rho_A \lsim 2$.

It would be interesting to see how much the upper bounds on $\rho_A$
could be modified by the inclusion of SUSY corrections,
while satisfying all the experimental constraints.
%------------------------------------------------------------
%----------------------------------------------------------
For this purpose, we plot in Fig.~\ref{BRrho} the total branching ratio of 
$B\to K \phi$ as a function of $\rho_A$ for gluino and 
chargino contribution.
We present the gluino results for the case of LR dominant scenario, and 
the chargino ones with the LL dominant case. 
The values of these mass insertions
are varied in the allowed range as done for the other plots.
Also we scanned over the different values of 
the strong phases $\phi_{A,H}$, squark masses, gluino and chargino masses.
In these plots, the dashed lines correspond to 
the SM predictions for $BR(B\to K \phi)$.
As can be seen from this figure, by applying the $2 \sigma$ constraints on
$BR^{SM}(B\to K \phi)$ one can set
very stringent upper bounds on $\rho_A$, namely $\rho \lsim 2$.
On the other hand, this bound can be relaxed to $\rho \lsim 4$ 
and $\rho \lsim 5$,
in the case of  LL and LR mass insertion contributions in up- and down-squark, 
respectively.
However, it is important to note that most of 
the configurations in both gluino and chargino cases that 
allow for $\rho > 2$ lead to CP asymmetry 
$S_{\phi K_S}$ outside the $2\sigma$ range of the experimental results. 
Therefore, as mentioned above, we 
consider  $\rho < 2$ as a conservative bound. 

Regarding the $BR(B\to \eta^{\prime} K)$, 
we find that it is less sensitive to 
$\rho_A$ than  $BR(B\to \phi K)$. This can be easily observed from the 
dependence of the amplitude of $B\to \eta^{\prime} K$ on $X_A$.
Analogously to Eq.(\ref{Amp_phi}), we get
\begin{equation}
i\, A^{SM}(B\to \eta^{\prime}K)  \simeq 3.6 \times 10^{-8} + 0.19 
\times 10^{-10} X_A + 4.2 \times 10^{-10} X_A^2 + 7.7\times 10^{-10} X_H \, ,
\label{Amp_eta}
\end{equation}
where, as in  Eq.(\ref{Amp_phi}), numbers in the r.h.s. of 
Eq.(\ref{Amp_eta}) are expressed in units of GeV.
Thus, comparing Eqs.(\ref{Amp_phi}) and (\ref{Amp_eta}), we see that
the effect of $\rho_A$ in $A_{SM}(B\to \eta^{\prime}K)$ is 
suppressed by two orders of magnitude with respect to the same one in
$A_{SM}(B\to \phi K)$. Hence the strongest bound one 
can obtain 
on $\rho_A$ will come only from $BR(B\to \phi K)$. Therefore,
in all plots of the present work,
including the QCDF ones in section 4, we scanned over $\rho_A$ 
by requiring $\rho_A \leq 2$.
\\

As mentioned before, the experimental measurements of  
$BR(B\to \eta^{\prime} K)$
represent another large discrepancy with the SM prediction. 
There have been various efforts to explain the large observed branching 
ratio in the $B \to \eta^{\prime} K$ process, based on the peculiarity 
of $\eta^{\prime}$ meson. For instance, intrinsic charm \cite{CZ} or 
gluonium contents 
of $\eta^{\prime}$ \cite{AS}, have been investigated as possible new sources
for such an enhancement. Clearly, NP contributions 
could also be responsible of such discrepancy in $BR(B\to \eta^{\prime} K_s)$, 
or at least for a part of it.
In this respect, we are going to analyze next the maximum effect 
that one can obtain from SUSY contributions to $BR(B\to \eta^{\prime} K)$,
by taking into account the experimental constraints on the 
$BR(B\to \phi K)$, $S_{\phi K_S}$, and $S_{\eta^{\prime} K_S}$. 

%------------------------------------------------------------
\begin{figure}[tpb]
\begin{center}
\dofigs{3.1in}{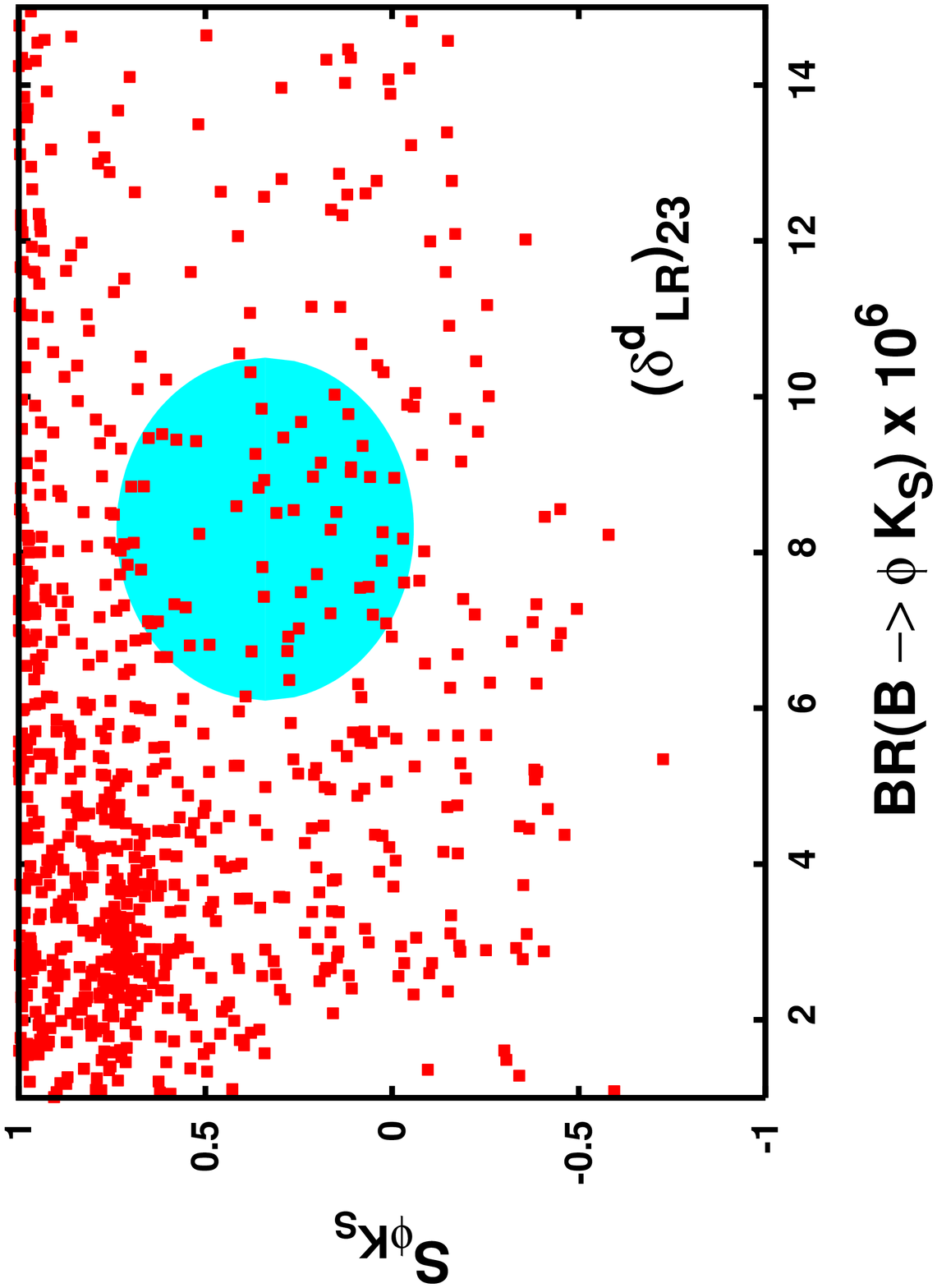}{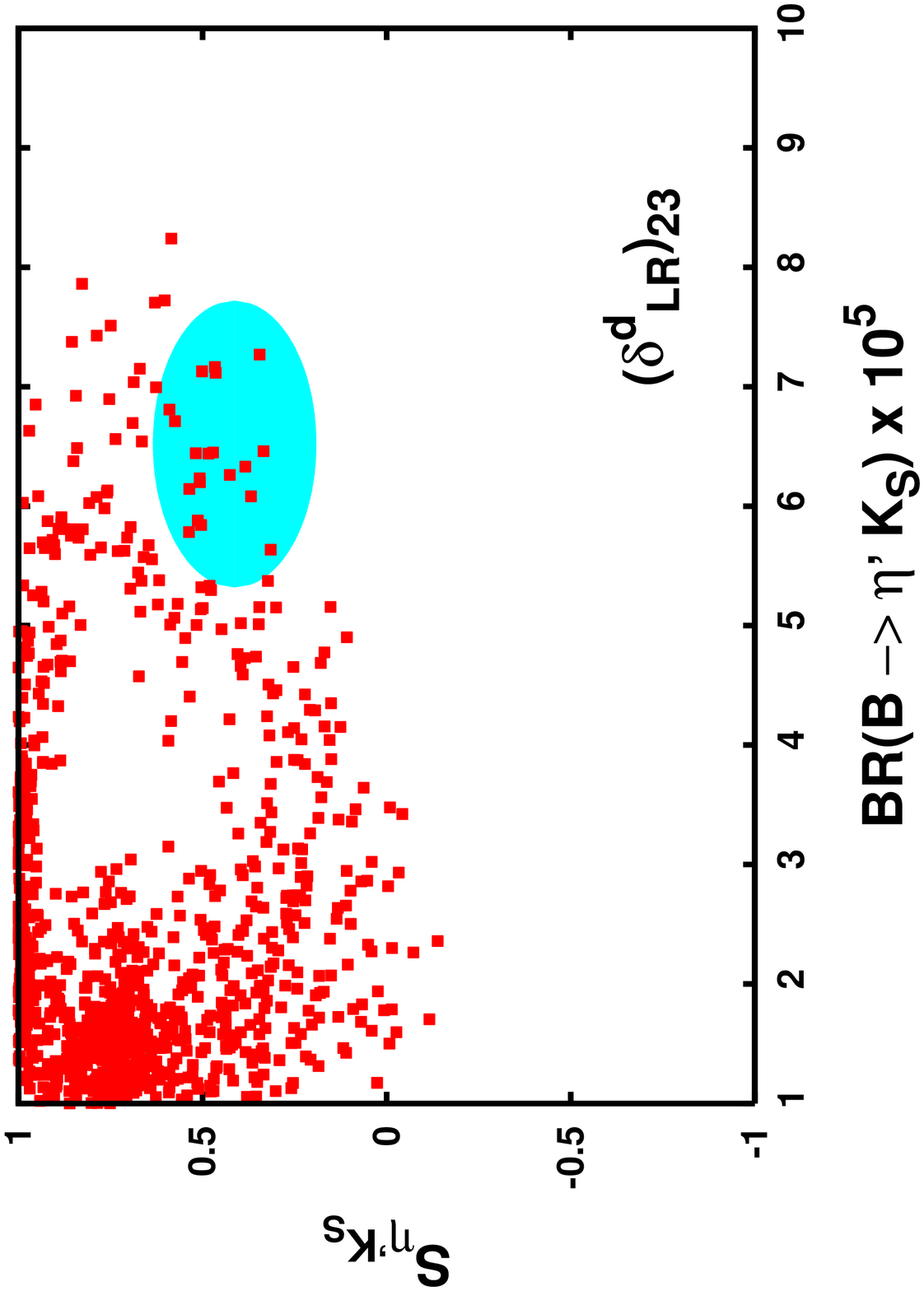}
\end{center}
\caption{\small 
As in Fig.\ref{Fig45}, but for 
correlations of $S_{\phi K_S}$ versus $BR(B\to \phi K )$ (left) and
$S_{\eta^{\prime} K_S}$ versus $BR(B\to \eta^{\prime} K )$ 
(right). Gluino contributions with one single 
mass insertion $\dd{LR}{23}$ has been considered.
}
\label{BRS_DLR}
\end{figure}
%----------------------------------------------------------

In Fig.~\ref{BRS_DLR} we plot the CP asymmetries $S_{\phi K_S}$  versus 
$BR(B\to \phi K)$ and $S_{\eta^{\prime} K_S}$ versus 
$BR(B\to \eta^{\prime} K)$, where 
the area of 
colored ellipse corresponds as usual to the allowed experimental region
for the correlations at $2\sigma$ level.
We consider here the dominant gluino contribution 
due to  $\dd{LR}{23}$ and scan over the other parameters as before. 
One can see from this figure that, gluino contribution can fit quite well 
inside the $2\sigma$ ellipse in both left and right plots.
It is worth noticing that, gluino effects could also rise 
$BR(B\to \phi K)$ to very large values ($\simeq 1.5-2\times 10^{-5}$), 
which are outside the allowed region. 

In the correlation between $S_{\eta^{\prime} K_S}$ and 
$BR(B\to \eta^{\prime} K)$,  
with just one mass insertion, SUSY contributions 
can explain the measured large $BR(B\to \eta^{\prime} K)$ and at
the same time fit the 2$\sigma$ range for $S_{\eta^{\prime}K_S}$.
As can be seen from this figure, for 
$S_{\eta^{\prime}K_S}$ within even 1$\sigma$ region, the
$BR(B\to \eta^{\prime} K)$ can be large as $7\times 10^{-5}$.
However, in this case some
fine tuning between SUSY parameters is necessary.

%------------------------------------------------------------
\begin{figure}[tpb]
\begin{center}
\dofigs{3.1in}{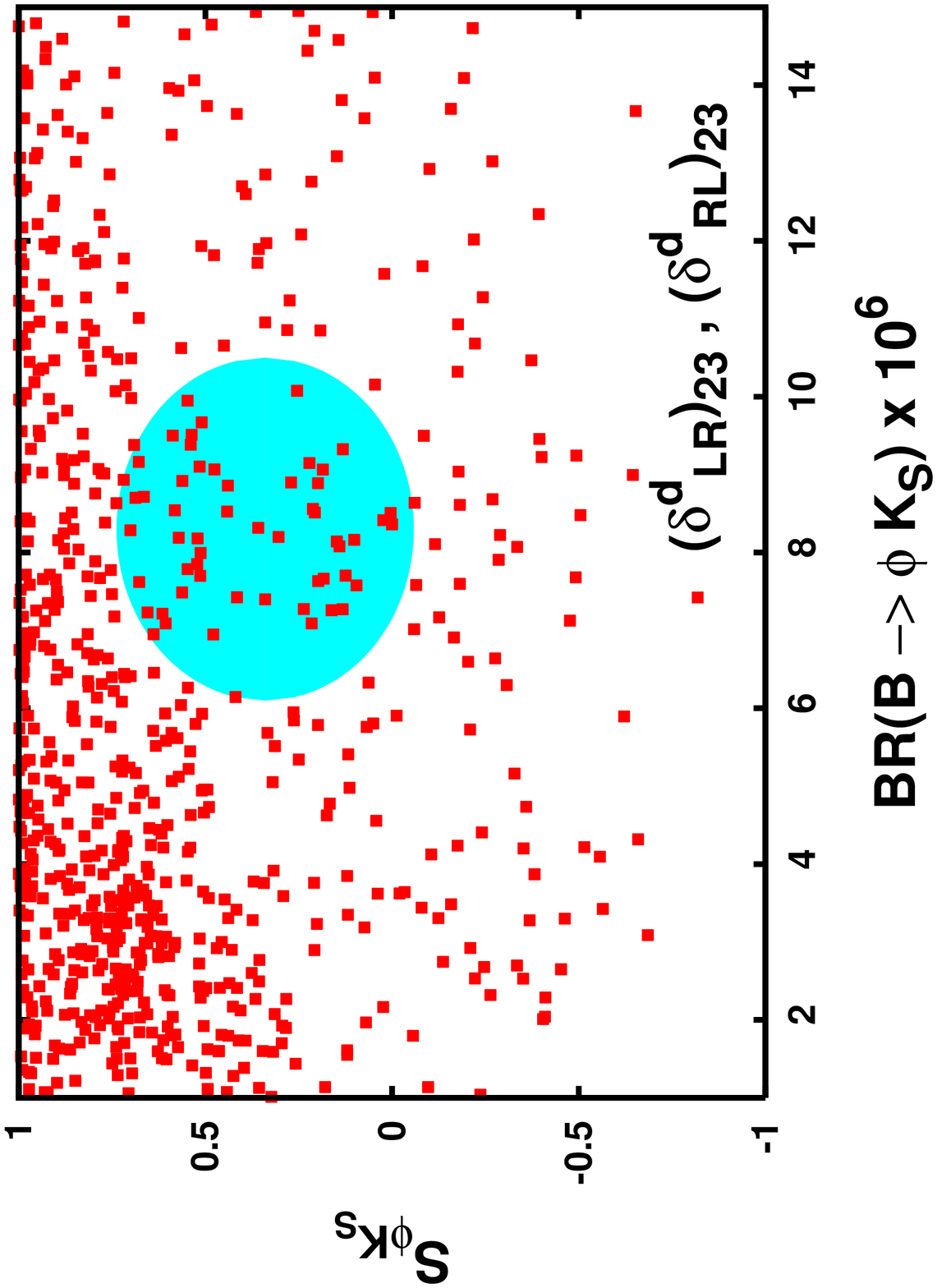}{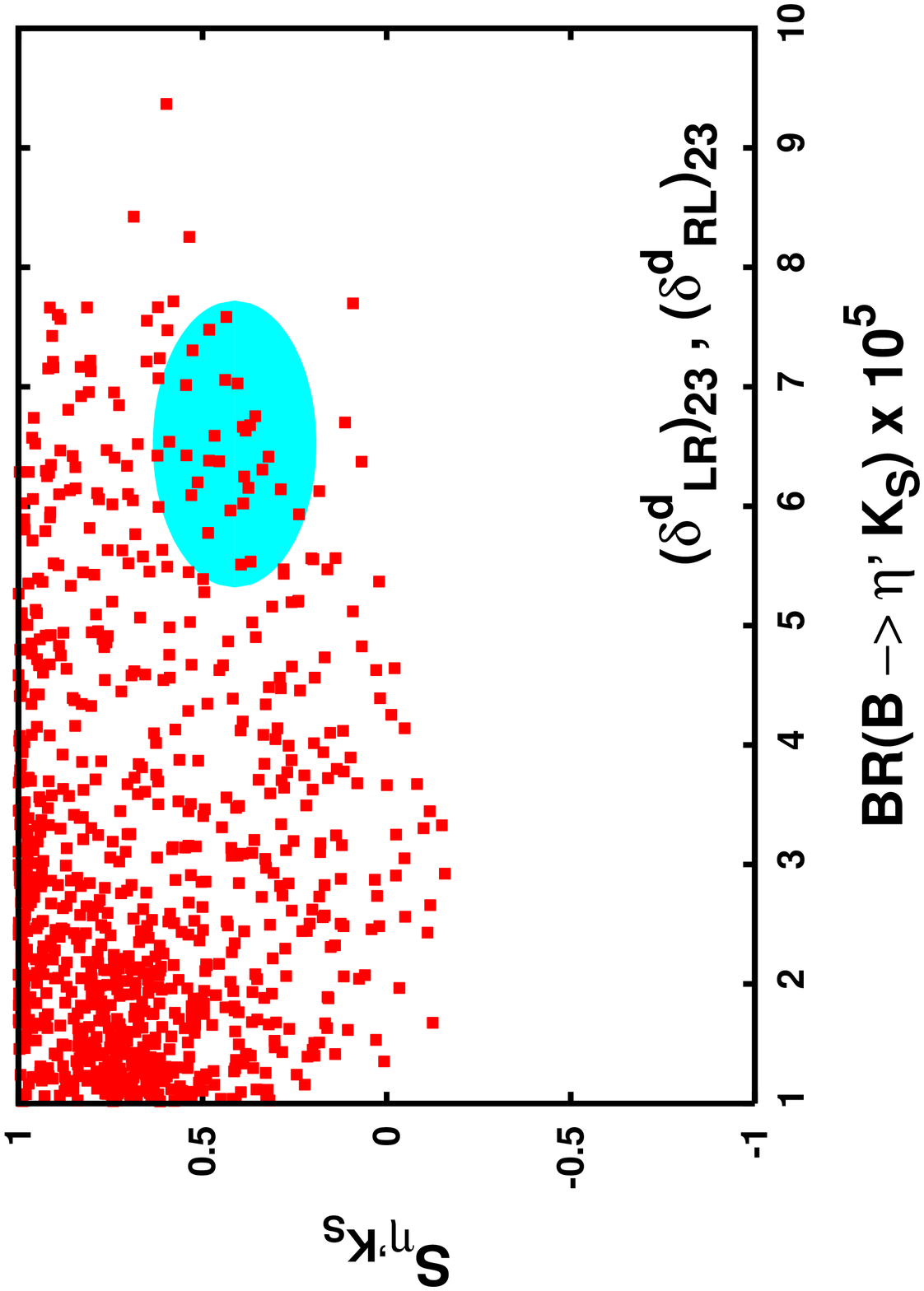}
\end{center}
\caption{\small As in Fig.~\ref{BRS_DLR}, but
for gluino contributions with two 
mass insertions $\dd{LR}{23}$ and $\dd{RL}{23}$, with the assumption of 
${\rm arg}[\dd{LR}{23}]= {\rm arg}[\dd{RL}{23}]$.
}
\label{BRS_DLRRL}
\end{figure}
%----------------------------------------------------------

In Fig.~\ref{BRS_DLRRL}, we present the same correlations as in Fig. 
\ref{BRS_DLR}, but with two simultaneous contributions of mass insertions 
$\dd{LR}{23}$ and $\dd{RL}{23}$ in gluino sector.
In this case, we can easily see that gluino scenario
can saturate simultaneously both $S_{\phi K_S}$ and $BR(B\to \phi K)$ within 
their experimental ranges, covering also larger areas with respect to the 
ones in Fig.~\ref{BRS_DLR}.
%------------------------------------------------------------
\begin{figure}[tpb]
\begin{center}
\dofigs{3.1in}{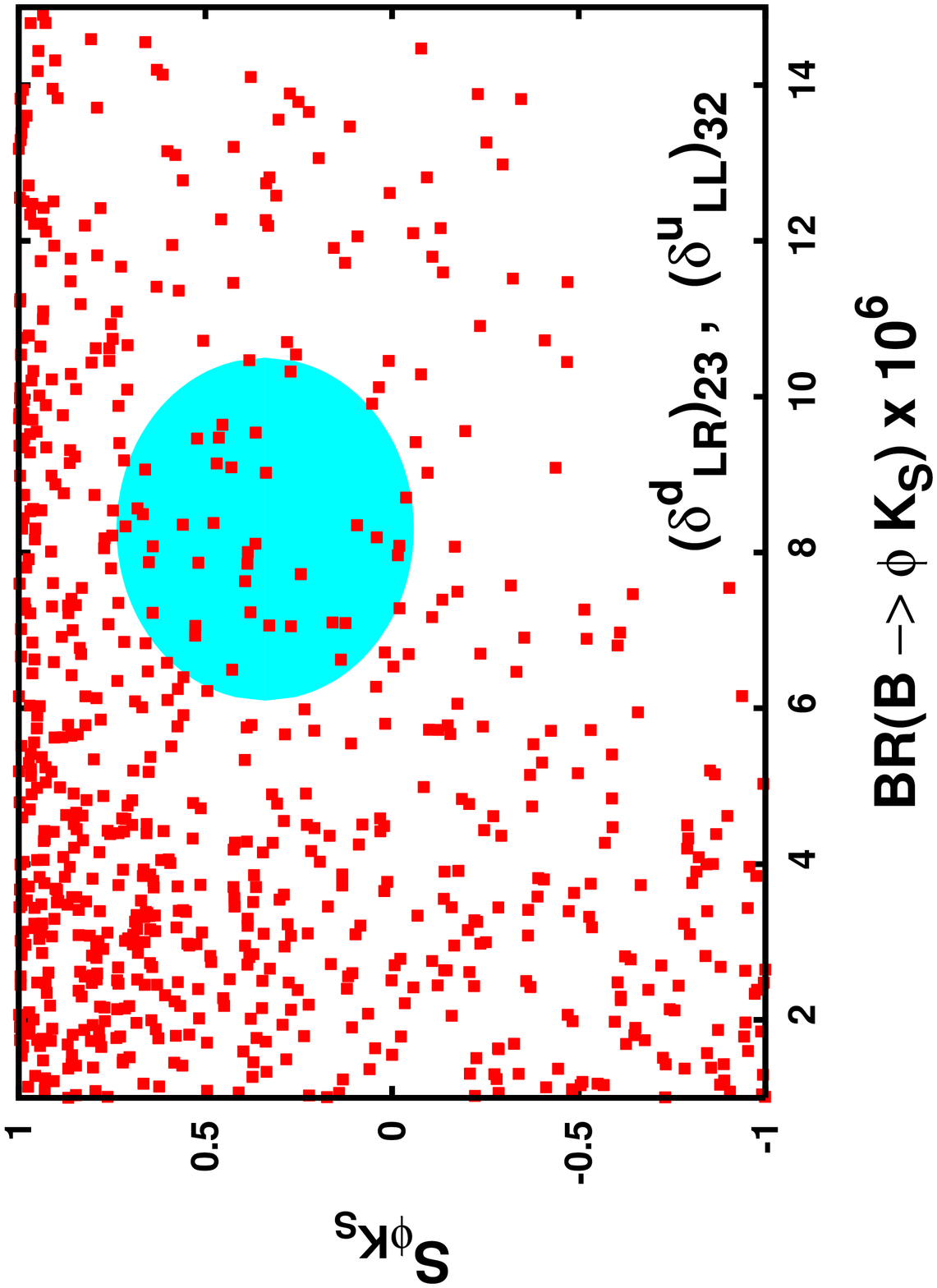}{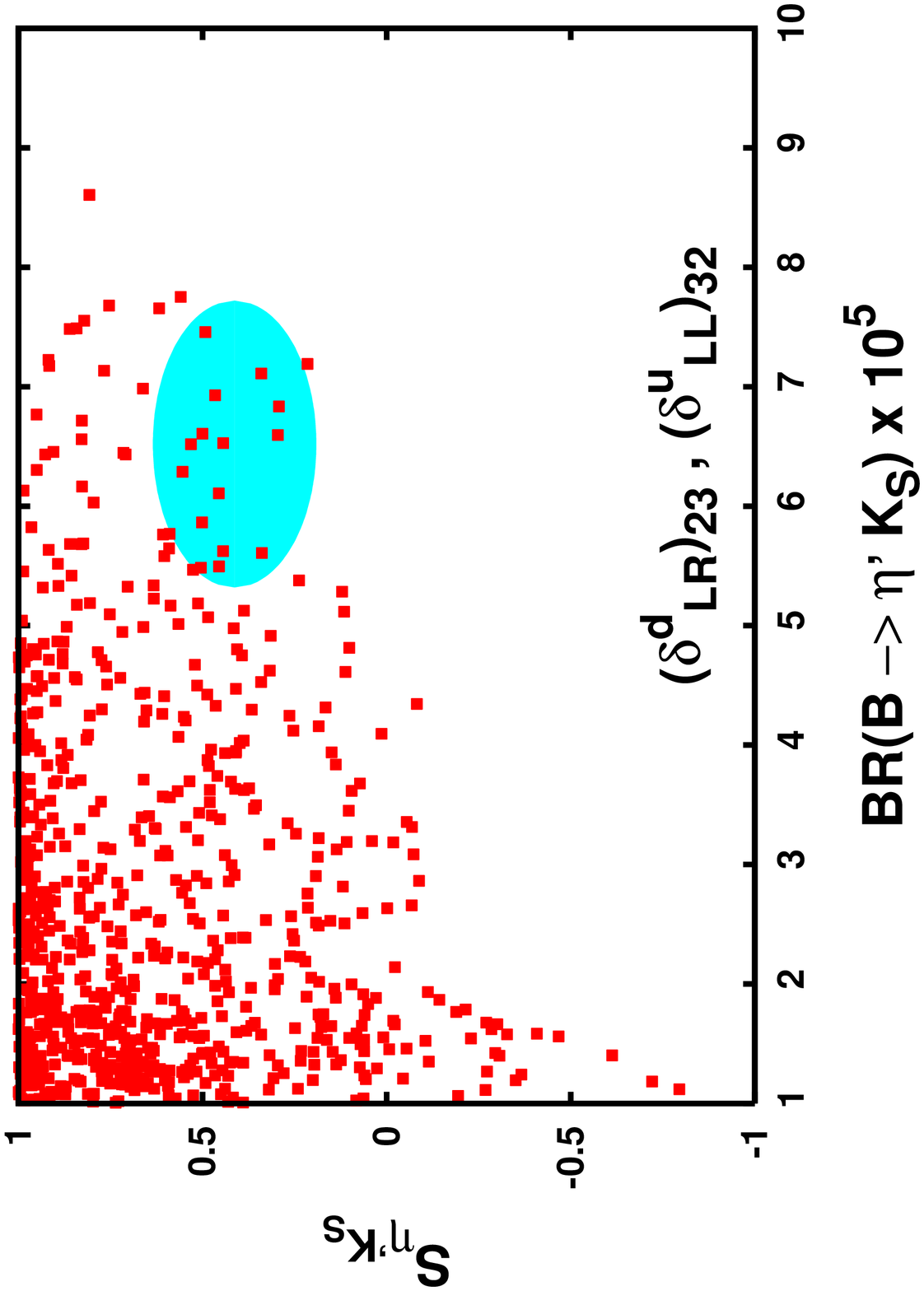}
\end{center}
\caption{\small As in Fig.~\ref{BRS_DLR}, but
for gluino and chargino contributions with
mass insertions $\dd{LR}{23}$ and $\du{LL}{32}$,
with the assumption of 
${\rm arg}[\dd{LR}{23}]= {\rm arg}[\du{LL}{32}]$.
}
\label{BRS_DLR_ULL}
\end{figure}
%----------------------------------------------------------

Finally, the combination 
from gluino and chargino exchanges 
on $BR(B\to \phi K)$ and $BR(B\to \eta^{\prime} K)$ 
versus the respective CP asymmetries,
are shown in Fig.~\ref{BRS_DLR_ULL}. 
We scan on the allowed range of the most relevant mass insertions for these 
two contributions: 
$\dd{LR}{23}$ (gluino) and $\du{LL}{32}$ (chargino). 
We also vary the other parameters as in the previous cases. 
The message we can learn from these results is that 
both gluino (LR) and chargino (LL) contributions
can easily accommodate the experimental results 
for the CP asymmetries and branching ratios of
$BR(B\to \phi K)$ and $BR(B\to \eta^{\prime} K)$.
As already stressed in section 4, this phenomenon is related to the fact 
that $b\to s \gamma$ constraints are more
relaxed for chargino contributions at large $\tan\beta$,
due to potentially destructive interferences of gluino and 
chargino amplitudes in $b\to s\gamma$ decay.
The main difference emerging from  results in Figs.
\ref{BRS_DLRRL} and \ref{BRS_DLR_ULL}, is that for the latter
scenario large and negative values of 
asymmetry ($S_{\phi, \eta^{\prime}}\simeq -1$) can also be easily
achieved, even though they are now outside the allowed experimental range.

%------------------------------------------------------------
\begin{figure}[tpb]
\begin{center}
\dofigs{3.1in}{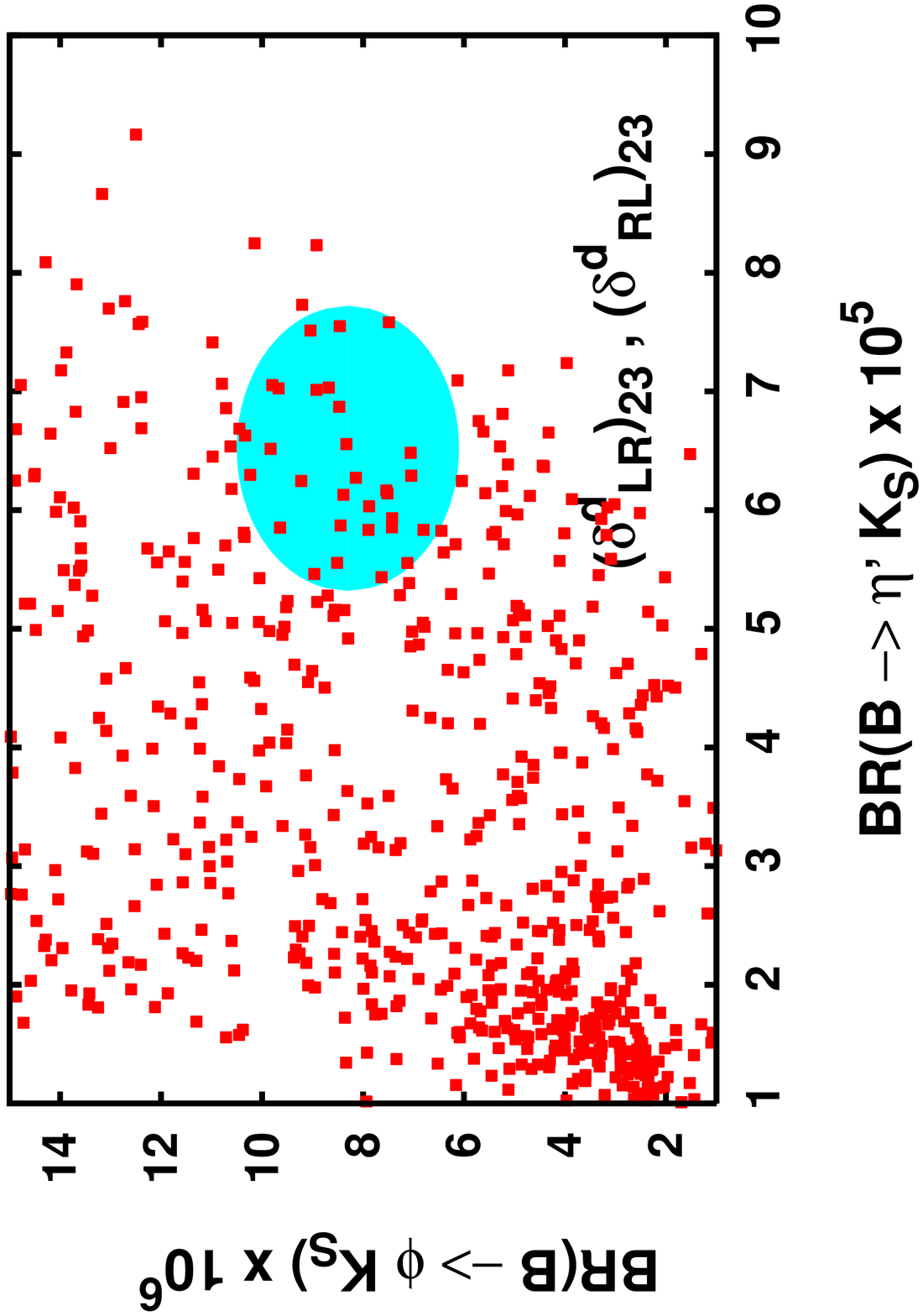}{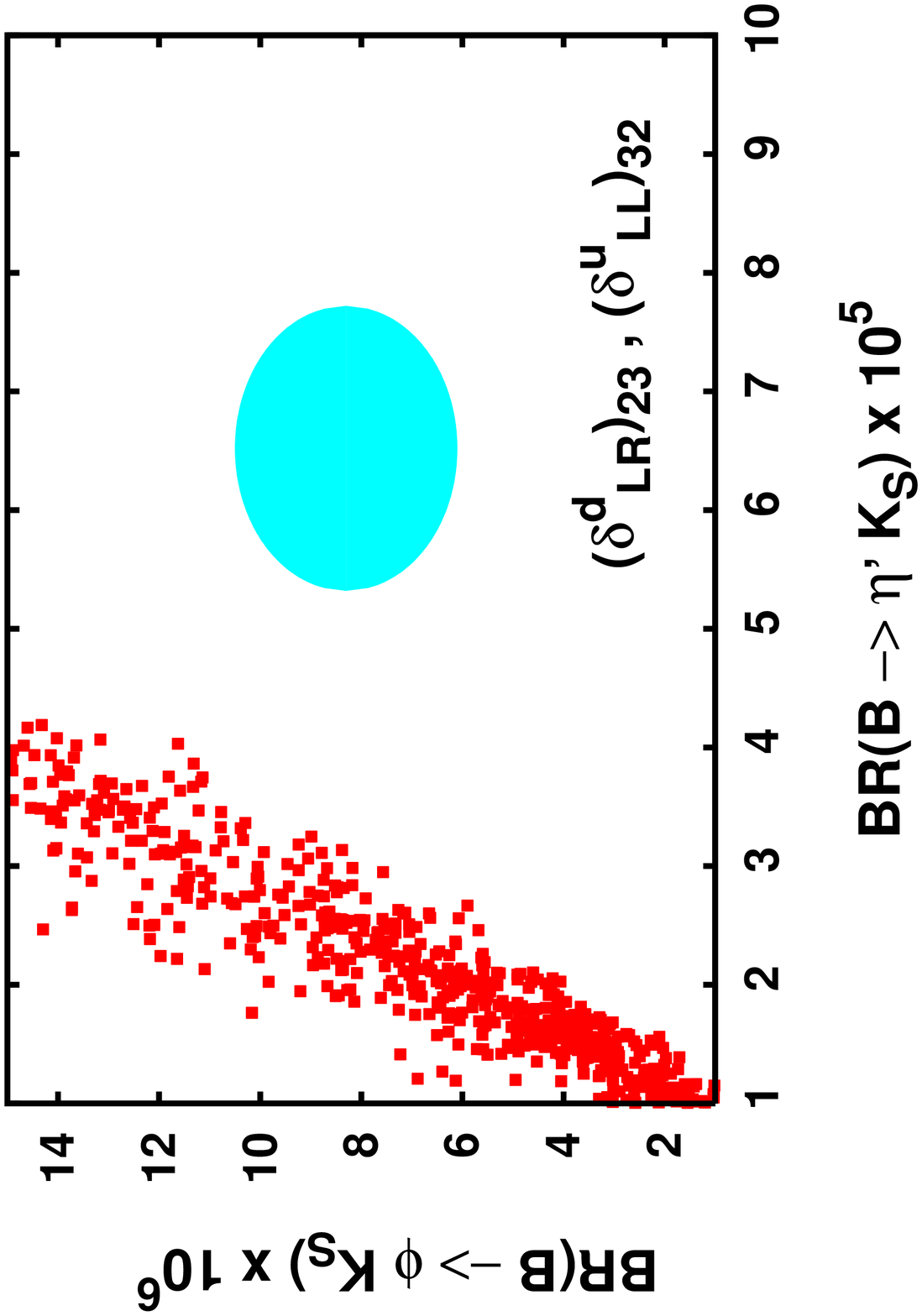}
\end{center}
\caption{\small Correlations of 
$BR(B\to \phi K )$ versus $BR(B\to \eta^{\prime} K )$
for gluino contributions with two mass insertions
$\dd{LR}{23}$ and $\dd{RL}{23}$ (left) and
for gluino and chargino contributions with two mass insertions
$\dd{LR}{23}$ and $\du{LL}{32}$ (right).
The $2\sigma$ constraints on $S_{\phi K_S}$ and $S_{\eta^{\prime} K_S}$ are
applied.
}
\label{BR_CORR}
\end{figure}
%----------------------------------------------------------

It is crucial to see, if regions of points which fit inside 
the two $2\sigma$ ellipses in Figs.~\ref{BRS_DLRRL} and \ref{BRS_DLR_ULL},
actually correspond to the same SUSY configurations.
For this purpose, 
in Fig.~\ref{BR_CORR} we plot correlations of $BR(B\to \phi K )$ 
versus $BR(B\to \eta^{\prime} K )$ for the same scenarios considered in 
Figs.~\ref{BRS_DLRRL} and \ref{BRS_DLR_ULL}, where 
all points satisfy the constraints 
on $S_{\phi K_S}$ and $S_{\eta^{\prime} K_S}$
at $2\sigma$ level.
As we can see from results in Fig.~\ref{BR_CORR}, 
only the scenario in which both LR and RL mass insertions
in gluino exchanges contribute, 
can naturally enhance  $BR(B\to \eta^{\prime} K )$ inside
the allowed experimental range, while 
respecting all the other constraints on CP asymmetries.
On the contrary, in the scenario in which both LR(gluino) 
and LL(chargino) are contributing, this enhancement is ruled out
by simultaneous constraints on CP asymmetries.

As already mentioned, $BR(B\to \eta^{\prime} K )$ suffers from
large theoretical uncertainties due to the
peculiar structure of $\eta^{\prime}$ meson. 
Therefore, it is not a conservative approach 
to constrain NP models from the measurements of $BR(B\to \eta^{\prime} K )$,
until the role of potential new mechanisms responsible of the enhancement of 
$BR(B\to \eta^{\prime} K )$ in the SM will be not completely clarified.

%%%%%%%%%%%%%%%%%%%%%%%%%%%%%%%%%%%%%%%%%%%%%%%%%%%%%%%%%%%%%%%%%%%%%%%%%%%%%%
%
\section{Direct CP asymmetry in $b \to s \gamma$ versus 
$S_{\phi(\eta^{\prime}) K_S}$}
%
%%%%%%%%%%%%%%%%%%%%%%%%%%%%%%%%%%%%%%%%%%%%%%%%%%%%%%%%%%%%%%%%%%%%%%%%%%%%%%
In this section we analyze the correlation for SUSY predictions
between the direct CP asymmetry $A_{CP}(b\to s \gamma)$ in \bsg decay
and the other ones in $B\to \phi (\eta^{\prime}) K_S$.
The CP asymmetry in \bsg is measured 
in the inclusive radiative decay of $B \to X_s \gamma$ by the quantity 
\bea
A_{CP}(b\to s \gamma) = \frac{\Gamma(\bar{B} \to X_s \gamma) - \Gamma(B \to
X_{\bar{s}}
\gamma)}{\Gamma(\bar{B} \to X_s \gamma) + \Gamma(B \to X_{\bar{s}}
\gamma)}.
\label{Absg}
\eea
The SM prediction for $A_{CP}(b\to s \gamma)$ is very small, less than
$1\%$ in magnitude, but known with high precision \cite{ACPbsg}.
Indeed, inclusive decay rates of B mesons are free from large
theoretical uncertainties since they can be reliably calculated in QCD using 
the OPE.
Thus, the observation of sizeable effects in 
$A_{CP}(b\to s \gamma)$ would be a clean signal of new physics.
In particular, large asymmetries are expected in models with enhanced 
chromo-magnetic dipole operators, like for instance supersymmetric models
\cite{ACPbsg}. 

The most recent result reported by BaBar
collaboration for $A_{CP}(b\to s \gamma)$ is \cite{cleo}
\begin{equation}
A_{CP}^{b\to s \gamma} = 0.025 \pm 0.050~ (stat.)~ \pm 0.015~ (syst.),
\end{equation}
which corresponds to the following allowed range at $90\%$ confidence level:
\begin{equation}
-6\% < A_{CP}^{b\to s \gamma}  < 11 \%~.
\end{equation}
Clearly, the present experimental sensitivity is not accurate enough
to strongly constrain the SM predictions at percent level.

Recently it has been shown that the SUSY
contribution to the CP asymmetry in the \bsg decay,
even with the inclusion of 
experimental constraints on electric dipole moments and
branching ratio of $B \to X_s \gamma$,
could be much larger than the SM expectation \cite{bailin} .
Therefore, in the light of present experimental results,
it would be challenging to analyze the SUSY predictions for the 
correlation between $A_{CP}(b\to s \gamma)$ and
$S_{\phi,(\eta^{\prime}) K_S}$, since in SUSY models these asymmetries are 
strongly correlated.

The relevant operators of the effective Hamiltonian in Eq.(\ref{Heff})
that play a role in $A_{CP}(b\to s \gamma)$, 
are given by $Q_2$, $Q_{7\gamma}, Q_{8g}$. We remind here that
$Q_2$, defined in Eqs.(\ref{Qbasis}),
is the usual current-current operator induced at tree-level.
Then, the expression for $A_{CP}(b\to s \gamma)$ 
at the NLO accuracy, is given by \cite{ACPbsg}
\bea
A_{CP}(b\to s \gamma) &=& \frac{\alpha_s(m_b)}{\vert C_{7\gamma} \vert^2} 
\Bigg\{
\frac{40}{81}{\rm Im}\left[C_2C_{7\gamma}^{\star}\right]-
\frac{8z}{9}\left[v(z)+b(z,\delta)\right] \, 
{\rm Im}\left[\left(1+\epsilon_s\right)C_2C_{7\gamma}^{\star}\right]
\nonumber\\
&-&\frac{4}{9}{\rm Im}\left[C_{8g}C_{7\gamma}^{\star}\right]
+\frac{8z}{27}b(z,\delta){\rm Im}\left[\left(1+\epsilon_s\right)
C_2C_{8g}^{\star}
\right]\Bigg\},
\label{ACP}
\eea
where 
$z=(m_c/m_b)^2$ and $v(z)$ is \cite{ACPbsg}
\begin{equation}
v(z) = \left( 5 + \ln z + \ln^2 z - \frac{\pi^2}{3}\right) + \left(\ln^2 z -
 \frac{\pi^2}
{3}\right) z + \left( \frac{28}{9} - \frac{4}{3} \ln z \right) z^2 +
 \mathcal{O}(z^3).
\end{equation}
The function $b(z,\delta)$ is defined as $b(z,\delta) = g(z,1) -
 g(z,1-\delta)$, where 
the parameter $\delta$ is related to the experimental cut on the 
photon energy, $E_{\gamma} > (1- \delta) m_b/2$, and finally 
$g(z,y)$ is given by 
\begin{equation}
g(z,y)= \theta(y-4z)\left\{\left(y^2 - 4y z+6z^2\right)\ln
\left(\sqrt{\frac{y}{4 z}}+
\sqrt{\frac{y}{4z} -1}\right) - \frac{3 y(y-2z)}{4} \sqrt{1- \frac{4 z}{y}}
\right\}.
\end{equation}

In the SM, the Wilson coefficients $C_{2}$, $C_{7\gamma}$, and $C_{8g}$
are real, in particular $C_2^{SM} \simeq 1.08$, $C_{7\gamma}=
-0.318$, and $C_{8g}=-0.151$, so that the only source of CP violation 
comes from the $\epsilon_s$ parameter inside Eq.(\ref{ACP}) which is 
defined in terms of the CKM matrix elements as
\begin{equation}
\epsilon_s=\frac{V_{us}^{\star}V_{ub}}{V_{ts}^{\star}V_{tb}}
\approx \lambda^2(i \eta -\rho)\sim {\cal O}(10^{-2})\, .
\end{equation}

In general SUSY models, $C_{7\gamma}$ and $C_{8g}$ may be complex, and the
corresponding phase can provide the dominant CP violating 
source to $A_{CP}^{b\to s \gamma}$.
In MIA framework, the SUSY contributions to the Wilson coefficients
$C_{7\gamma}$ and $C_{8g}$ are given in terms of $(\delta^d_{AB})_{23}$ 
(for gluino)
and $(\delta^u_{AB})_{32}$ (for chargino). Therefore, the SUSY phase in 
these mass 
insertions is the only source of CP violation in this process.  
As pointed out in previous sections, in general SUSY models, effects induced
by the dipole operators $\tilde{Q}_{7\gamma}$ and $\tilde{Q}_{8g}$, which
have opposite chirality to $Q_{7\gamma}$ and $Q_{8g}$, are non-negligible. 
In the MSSM these contributions are suppressed by terms of order 
${\cal O}(m_s/m_b)$
due to the universality of the $A$--terms. However, in general models one 
should take 
them into account, since they might be of the same order than $C_{7\gamma}$ 
and $C_{8g}$. 
Denoting by $\tilde{C}_{7\gamma}$ and $\tilde{C}_{8g}$
the Wilson 
coefficients multiplying the new operators $\tilde{Q}_{7\gamma}$ and 
$\tilde{Q}_{8g}$,
the expression for the asymmetry in Eq.(\ref{ACP})
will be modified by making the replacement 
\begin{equation}
C_i C_j^* \to C_i C_j^* + \tilde{C}_i \tilde{C}_j^*.
\label{chirality}
\end{equation}

Notice that, since\footnote{In principle,
$\tilde{C}_2$ might receive some contributions
from RGE due to the mixing of the other $\tilde{Q}_i$ operators with
$\tilde{Q}_2$.
However, the radiative effects induced by $\tilde{C}_i$ on $\tilde{C}_2$ 
are quite small and so we will neglect them in our analysis.}
$\tilde{C}_2=0$, the only modification in the numerator
of Eq.(\ref{ACP}) is due to the new term $\tilde{C}_{7\gamma}\tilde{C}_{8g}$.
However, if only one single mass insertion is taken into account,
both $\tilde{C}_{7\gamma}$ and $\tilde{C}_{8g}$ will be 
proportional to the same mass insertion, and therefore 
${\rm Im}\left(\tilde{C}_{8g}\tilde{C}^{\star}_{7\gamma}\right)=0$.
Therefore, the effects of these new operators will enter only through
$\tilde{C}_{7\gamma}$ in the denominator, by means of the shift
$|C_{7\gamma}|^2\to |C_{7\gamma}|^2+|\tilde{C}_{7\gamma}|^2$.

It is worth mentioning that in the scenarios dominated by one single mass 
insertion, the leading SUSY contribution to the CP asymmetry 
$A_{CP}(b\to s \gamma)$ is due to the $C_{7\gamma}$. 
This can be understood as follows.
Using our previous inputs for charm
and bottom masses and also assuming, for instance, an energy resolution 
$\delta \simeq 0.3$ so that $b(z,\delta)\simeq 0.17$,
Eq.(\ref{ACP}) can be written as
\bea
A_{CP}(b\to s \gamma) \simeq 
\frac{\alpha_s(m_b)}{\vert C_{7\gamma} \vert^2}\left( 
0.045~{\rm Im}[C_{7\gamma}^*] - 0.44~ {\rm Im}[C_{7\gamma} C^*_{8g}] + 0.006~ 
{\rm Im}[C_{8g}^*]\right).
\label{asymmetry2}
\eea
We remind here that the low energy coefficients 
$C_{7\gamma}$ and $C^*_{8g}$ at $\mu_b$ scale 
are related to the high energy ones by
\bea
&&C_{7\gamma}^{SUSY}(m_b) \simeq 0.66~ C_{7\gamma}^{SUSY}(\mu_W) +
 0.09~ C_{8g}^{SUSY}(\mu_W)
,\\
&& C_{8g}^{SUSY}(m_b) \simeq 0.69~ C_{8g}^{SUSY}(\mu_W).
\eea
As shown in section 2, the kind of mass insertions that will appear in 
$C_{7\gamma}(\mu_W)$ are the same that determine $C_{8g}(\mu_W)$. 
In the case of SUSY contribution from one single mass
insertion, both $C_{7\gamma}(\mu_b)$ and $C^*_{8g}(\mu_b)$ 
will acquire the same phase 
and so the second term in Eq.(\ref{ACP}) identically vanishes.

In conclusion, the relevant 
mass insertions contributing to $A_{CP}(b\to s \gamma)$
give also the dominant effects in $S_{\phi K_S}$ 
and $S_{\eta^{\prime} K_S}$. 
Therefore, it is expected that these three CP 
asymmetries should be highly correlated, since
all of them depend on the same source of SUSY CP violation.   

\begin{figure}[tpb]
\begin{center}
\dofigs{3.1in}{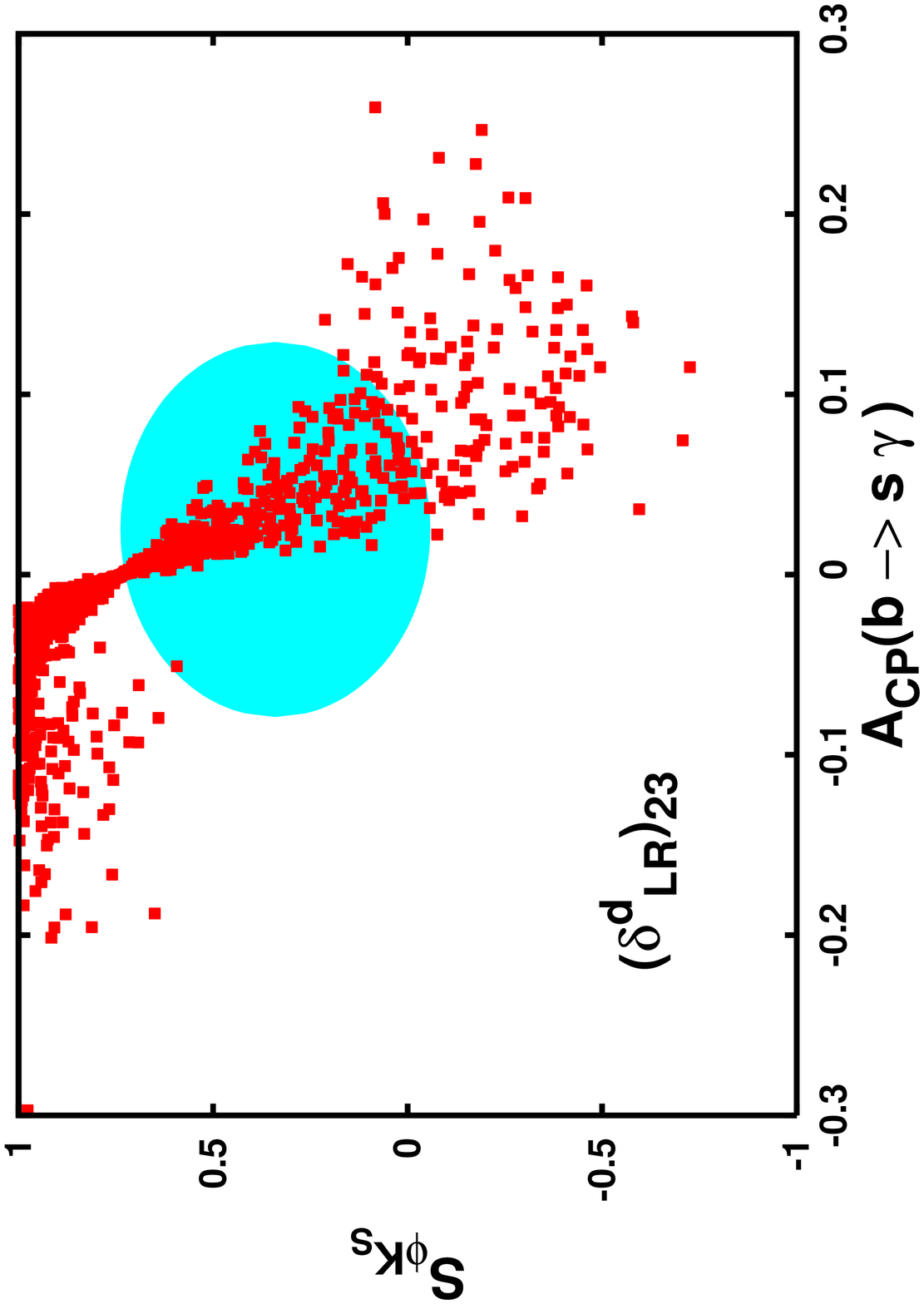}{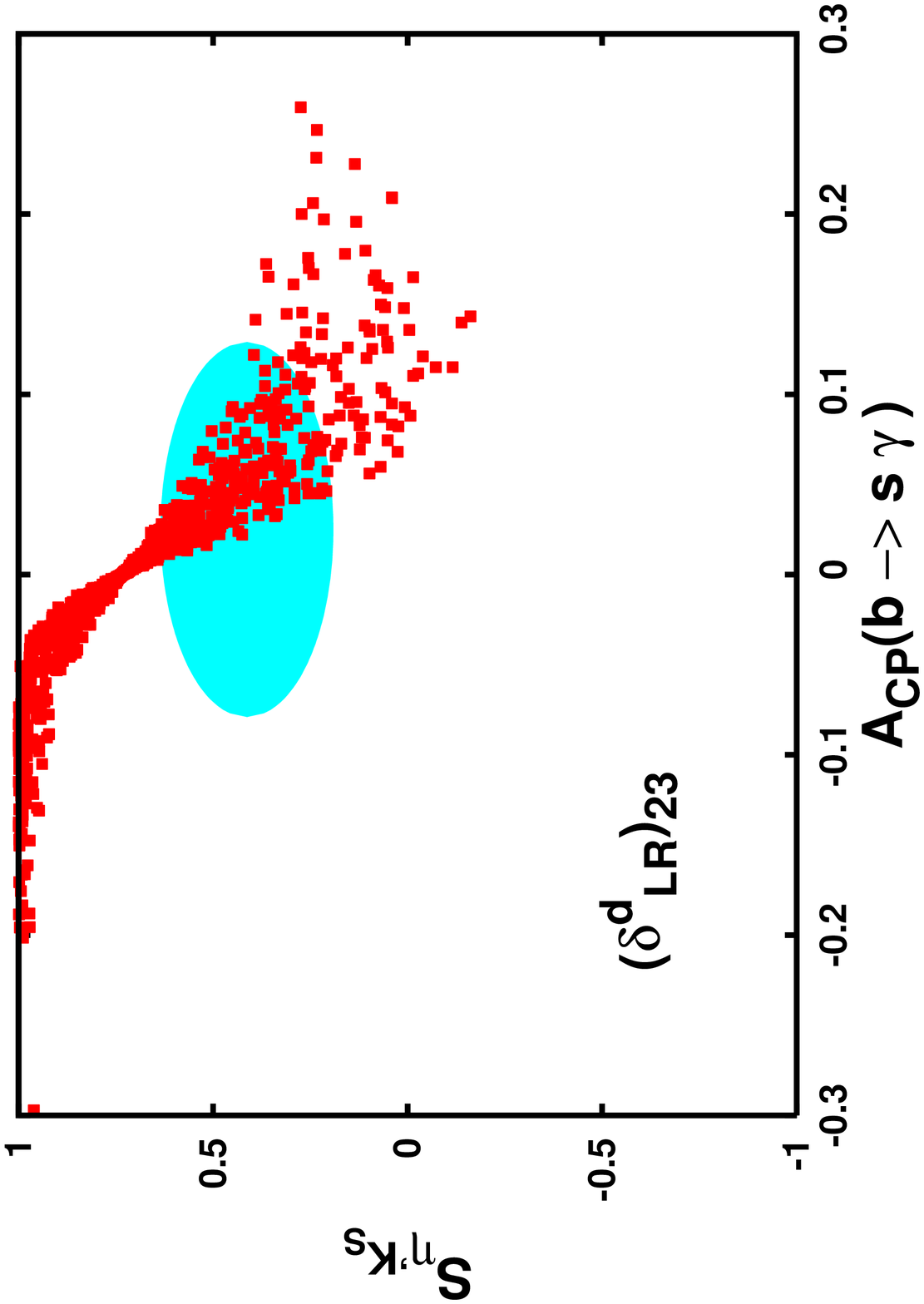}
\end{center}
\caption{\small 
Correlations of $S_{\phi K_S}$ versus $A_{CP}(b\to s \gamma )$ (left) and
$S_{\eta^{\prime} K_S}$ versus $A_{CP}(b\to s \gamma )$ 
(right), for gluino contributions with one single 
mass insertion $\dd{LR}{23}$.}
\label{ABSG_DLR}
\end{figure}

\begin{figure}[tpb]
\begin{center}
\dofigs{3.1in}{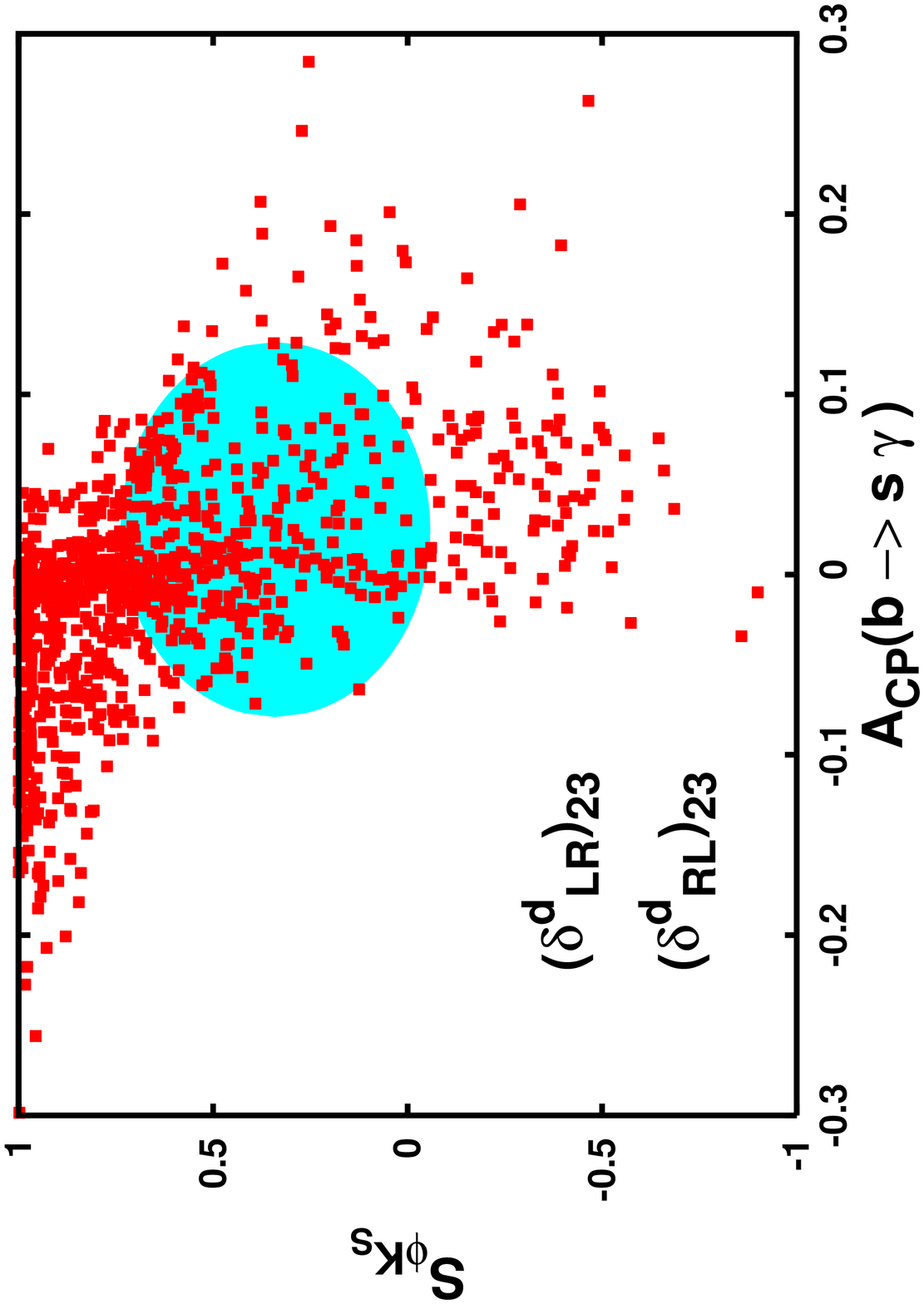}{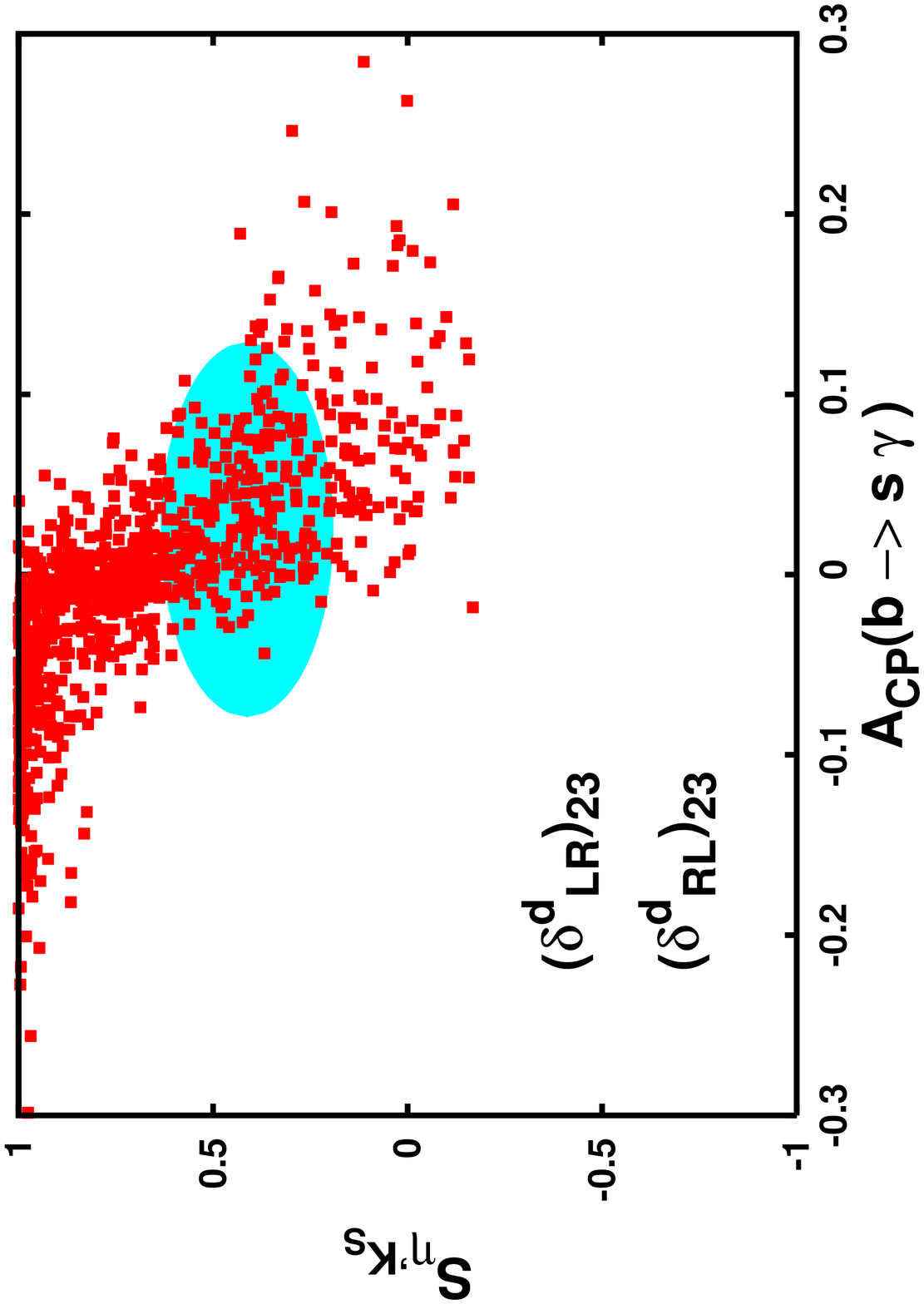}
\end{center}
\caption{\small As in Fig.~\ref{ABSG_DLR}, but
for gluino contributions with two 
mass insertions $\dd{LR}{23}$ and $\dd{RL}{23}$.
}
\label{ABSG_DLRRL}
\end{figure}

In Fig.~\ref{ABSG_DLR} we plot the correlations between $S_{K_S \phi}$ and 
$A_{CP}(b\to s\gamma)$ and also 
$S_{\eta^{\prime} K_S}$ and  $A_{CP}(b\to s\gamma)$ for gluino exchanges 
with one mass insertion, namely $\dd{LR}{23}$.
As we can see from  results in Fig.~\ref{ABSG_DLR}, a specific trend emerges
from this scenario. The simultaneous combination of 
$S_{\phi K_S}$ and $S_{\eta^{\prime} K_S}$ constraints, favors
the SUSY predictions to be 
in the region of positive values of $A_{CP}(b\to s\gamma)$ 
asymmetry.\footnote{
This result is also in agreement with the corresponding predictions
for the correlation of $S_{\phi K_S}$ versus $A_{CP}(b\to s\gamma)$ in the
article of Kane {\it et al.} in Ref.\cite{PhiKs_gluino}.}
As shown in Fig.~\ref{ABSG_DLRRL},
analogous results are also obtained when two mass 
insertions in gluino sector are taken simultaneously,
namely  $\dd{LR}{23}$ and $\dd{LR}{23}$. 
However, in this case also negative values of $A_{CP}^{b\to s\gamma}$
are allowed.
Although these points are favored by $S_{\phi K_S}$, they are
disfavored by $S_{\eta^{\prime} K_S}$.

\begin{figure}[tpb]
\begin{center}
\dofigs{3.1in}{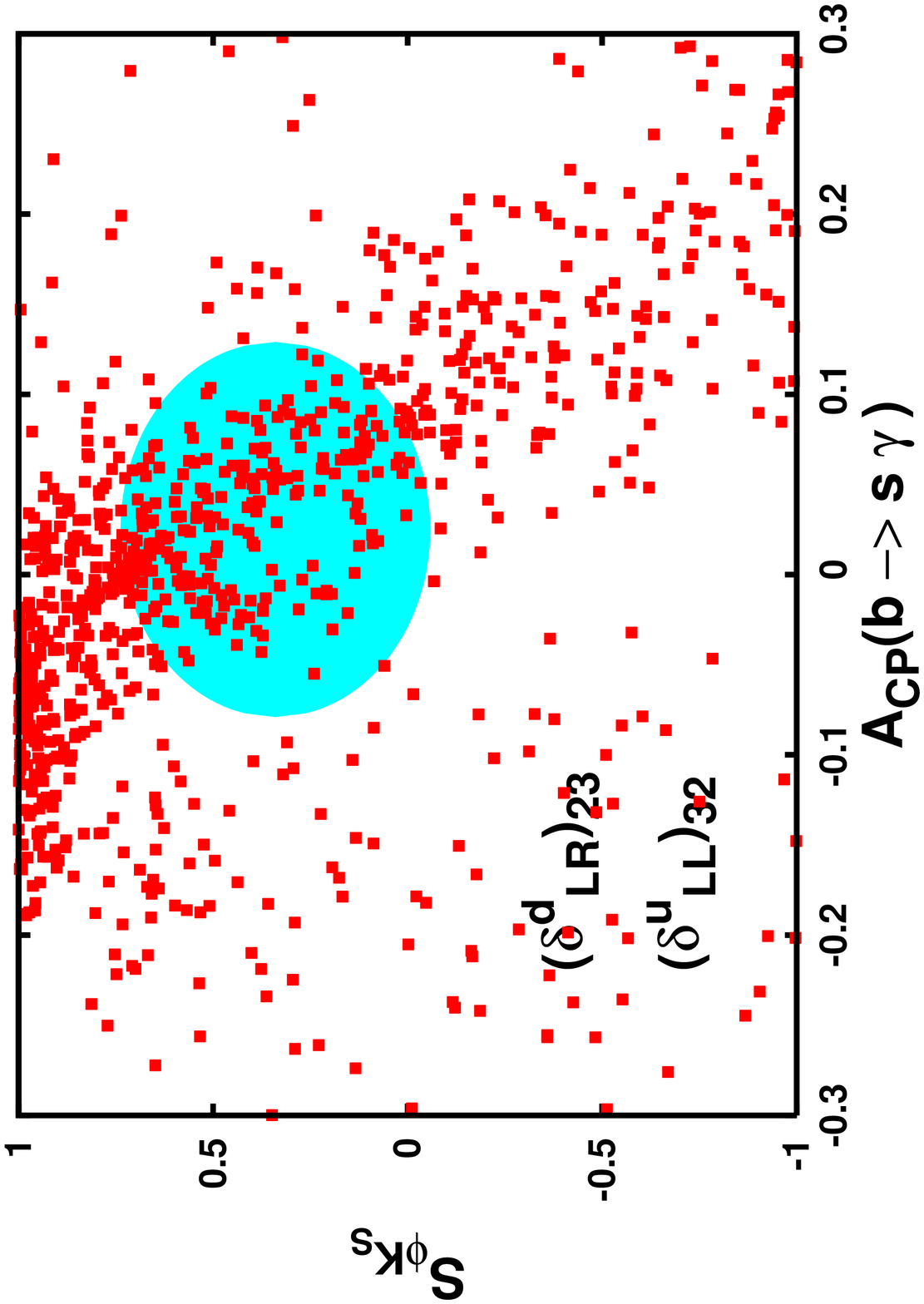}{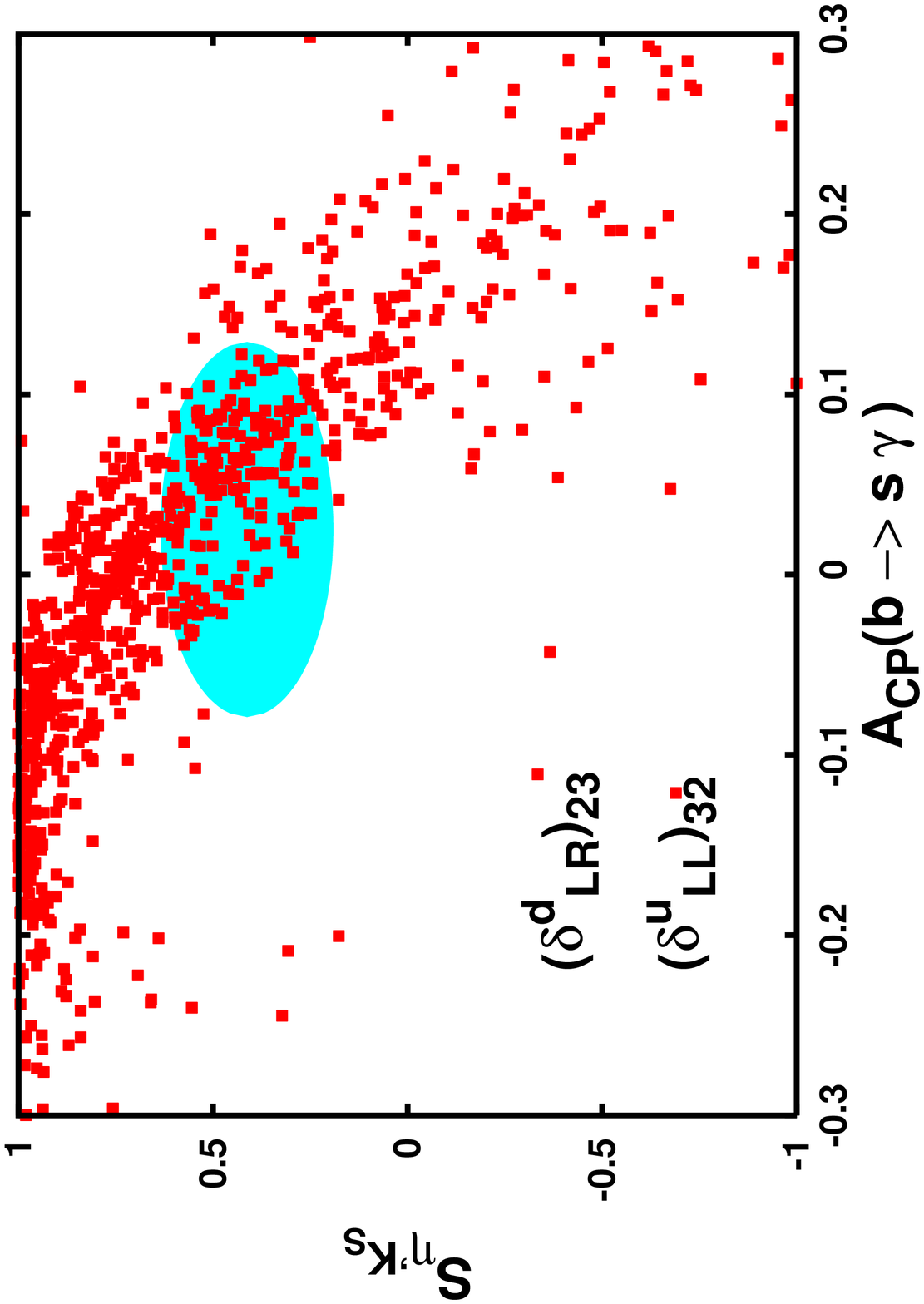}
\end{center}
\caption{\small As in Fig.~\ref{ABSG_DLR}, but
for gluino and chargino contributions with
mass insertions $\dd{LR}{23}$ and $\du{LL}{32}$ respectively.
}
\label{ABSG_DLR_ULL}
\end{figure}

Finally, in  Fig.~\ref{ABSG_DLR_ULL} we show our results for two
mass insertions $\dd{LR}{23}$ and $\du{LL}{32}$
with both gluino and chargino exchanges.
In this case we see that $S_{\phi K_S}$ constraints 
do not set any restriction on $A_{CP}(b\to s\gamma)$, and 
also large and positive values of $A_{CP}(b\to s\gamma)$ asymmetry 
can be achieved.
However, by imposing the constraints on
$S_{\eta^{\prime} K_S}$, see plot on the right side of Fig.~\ref{ABSG_DLR_ULL},
the region of negative $A_{CP}(b\to s\gamma)$ is disfavored in this scenario 
as well.

%------------------------------------------------------------
%----------------------------------------------------------

%------------------------------------------------------------
%----------------------------------------------------------

%------------------------------------------------------------
%----------------------------------------------------------

%%%%%%%%%%%%%%%%%%%%%%%%%%%%%%%%%%%%%%%%%%%%%%%%%%%%%%%%%%%%%%%%%%%%%
% 
\section{Conclusions}
%
%%%%%%%%%%%%%%%%%%%%%%%%%%%%%%%%%%%%%%%%%%%%%%%%%%%%%%%%%%%%%%%%%%%%%
The $B$-meson decays to $\phi K$, $\eta^{\prime} K$, and to
$X_S\gamma$ provide a clean window to the physics beyond the SM.
In this paper we have systematically analyzed the supersymmetric 
contributions to the CP asymmetries and the branching ratios of 
these processes in a model independent way.
The relevant SUSY contributions
in the $b\to s$ transitions, namely chargino and gluino exchanges 
in box and penguin diagrams, have been considered 
by using the mass insertion method.

We have provided analytical expressions for all the relevant Wilson 
coefficients.
Both naive factorization and QCD factorization approximation
to determine the hadronic matrix elements have been employed.
We showed that the 
SUSY predictions for the mixing CP asymmetry of $S_{\phi K_S}$ and 
$S_{\eta^{\prime} K_S}$ are not very sensitive to the kind of approximation
adopted for evaluating hadronic matrix elements.

We found that due to the stringent constraints from the experimental 
measurements of $BR(b\to s \gamma)$, the scenario with 
pure chargino exchanges cannot give large and negative 
values for CP asymmetry $S_{\phi K_S}$. Indeed, by combining 
present experimental constraints on  $S_{\phi K_S}$ and 
$S_{\eta^{\prime} K_S}$ asymmetries at $2\sigma$ level, pure chargino 
contributions can be almost ruled out. 
On the other hand, it is quite possible for gluino exchanges to account for
$S_{\phi K_S}$ and $S_{\eta^{\prime} K_S}$ at the same time.
Interestingly, we have shown that the charged Higgs, if not too heavy,
may change the chargino and gluino contributions to enhance the CP 
asymmetries considerably.

The branching ratios of these decays have also been considered.
It has been noticed that the $\rho $ parameter is strictly
bounded by the $B\to \phi K$ branching ratio, and this influences
strongly the possible SUSY contributions to the asymmetries.
Investigating the correlations between CP asymmetries and branching 
ratios, pointed out that the most favored SUSY scenarios, which 
easily satisfy all 
the experimental results, are the ones with inclusion of 
two mass insertions. In particular, a pure gluino dominated scenario, 
in which both $LR$  and $RL$ mass insertions are contributing 
and the one in which both gluino and chargino 
contribute with $LR$ and $LL$ mass insertions 
in down- and up-squark sectors, respectively.
In the latter scenario we show that chargino exchanges provide sizeable 
effects. This is due to the fact that $b\to s \gamma$ 
constraints could be relaxed by potentially destructive 
interferences between gluino and chargino amplitudes.
Finally, 
it is remarkable to notice that in the scenario in which 
both LR and RL down squark mass insertions dominate,
the observed large enhancement 
of  $BR(B\to \eta^{\prime} K )$ could be naturally explained, while 
respecting all the other experimental constraints on CP asymmetries.

We also discussed the correlations between the CP asymmetries of these 
processes and the direct CP asymmetry in $b\to s \gamma$ decay. 
In this case, we found that the general trend of 
SUSY models, satisfying all the experimental constraints, 
strongly favors large and positive contributions 
to $b\to s \gamma$ asymmetry.

More precise measurements would certainly allow us to draw more definite
conclusions on the SUSY models that can accommodate these data.

\vspace{0.5in}

\noindent
{\bf Acknowledgments} 

\vspace{0.2in} 
\noindent
The authors thank Debrupa Chakraverty for her efforts in the initial
stages of this work. We would like to thank G. D'Agostini,
E. Kou, and K. \"Osterberg for very useful discussions. 
E.G. acknowledges the CERN PH-TH division and IPPP, CPT 
of Durham University for their kind hospitality during the
preparation of this work.
The authors appreciate the financial support from the Academy of Finland
(project numbers 104368 and 54023) and from the Associate Scheme of ICTP.

%%%%%%%%%%%%%%%%%%%%%%%%%%%%%%%%%%%%%%%%%%%%%%%%%%%%%%%%%%%%%%%%%%%

\newpage

\noindent
{\bf\Large{Appendix}}

\appendix
%%%%%%%%%%%%%%%%%%%%%%%%%%%%%%%%%%%%%%%%%%%%%%%%%%%%%%%%%%%

%%%%%%%%%%%%%%%%%%%%%%%%%%%%%%%%%%%%%%%%%%%%%%%%%%%%%%%%%%
\section{Chargino contributions in MIA }
Here we provide the expressions for chargino contributions, at leading
order in MIA, to the $R_F^{AB}$ quantities in Eq.(\ref{RFch}), where $F$ refers
to $D,E,C,B_{(u,d)}$, and $M_{\gamma,g}$ \cite{our}. 
\bea
R_D^{LL}&=&\sum_{i=1,2} \,
|V_{i1}|^2 \, x_{\W i}\, P_D(x_i)
\nonumber
\\
R_D^{RL}&=&-\sum_{i=1,2} V_{i2}^{\star}V_{i1} \, x_{\W i}\, P_D(x_i)
\nonumber
\\
R_D^{RR}&=&\sum_{i=1,2} \, |V_{i2}|^2 \, x_{\W i}\, P_D(x_i)
\nonumber
\\
R_D^{LR}&=&\left(R_D^{RL}\right)^{\star}
\nonumber
\\
%%%%%%%%%%%%%%%%%%%%%%%%%%%%%%
R_E^{LL}&=&\sum_{i=1,2} \,
|V_{i1}|^2 \, x_{\W i}\, P_E(x_i)
\nonumber
\\
R_E^{RL}&=&-\sum_{i=1,2} V_{i2}^{\star}V_{i1} \, x_{\W i}\, P_E(x_i)
\nonumber
\\
R_E^{RR}&=&\sum_{i=1,2} \, |V_{i2}|^2 \, x_{\W i}\, P_E(x_i)
\nonumber
\\
R_E^{LR}&=&\left(R_E^{RL}\right)^{\star}
\nonumber
\\
%%%%%%%%%%%%%%%%%%%%%%%%%%%%%
R_C^{LL}&=&
\sum_{i=1,2}|V_{i1}|^2 \, P_C^{(0)}(\bar x_i)
+\sum_{i,j=1,2} \left[ U_{i1}V_{i1}U_{j1}^{\star}V_{j1}^{\star}\,
P_C^{(2)}(x_i,x_j) \right.
\nonumber\\
&+& \left. |V_{i1}|^2 |V_{j1}|^2
\left(\frac{1}{8}-P_C^{(1)}(x_i, x_j)\right)\right]
\nonumber
\\
R_C^{RL}&=&-\frac{1}{2}
\sum_{i=1,2}\, V_{i2}^{\star}V_{i1}\,  P_C^{(0)}(\bar x_i)
- \sum_{i,j=1,2}\, V_{j2}^{\star}V_{i1}\left(
U_{i1}U_{j1}^{\star}\, P_C^{(2)}(x_i, x_j)
\right.\nonumber \\
&+&
\left.
V_{i1}^{\star} V_{j1}\, P_C^{(1)}(x_i, x_j)\right)
\nonumber
\\
R_C^{LR}&=&\left(R_C^{RL}\right)^{\star},
\nonumber
\\
R_C^{RR}&=&
\sum_{i,j=1,2}\, V_{j2}^{\star}V_{i2}\left(
U_{i1}U_{j1}^{\star}\, P_C^{(2)}(x_i, x_j) +
V_{i1}^{\star} V_{j1}\, P_C^{(1)}(x_i, x_j)\right)
\nonumber
\\
%%%%%%%%%%%%%%%%%%%%%%%%%%%%%%%
R_{B_{u}}^{LL}&=&
2\sum_{i,j=1,2}\, V_{i1} V_{j1}^{\star}
U_{i1}U_{j1}^{\star}\,x_{Wj}\sqrt{x_{ij}}~
P_{B}^{u}(\bar x_j, x_{ij})
\nonumber
\\
R_{B_{u}}^{RL}&=&-
2\sum_{i,j=1,2}\, V_{i1} V_{j2}^{\star}
U_{i1}U_{j1}^{\star}\,x_{Wj}\sqrt{x_{ij}}~
P_{B}^{u}(\bar x_j, x_{ij})
\nonumber
\\
R_{B_{u}}^{LR}&=&\left(R_{B_{u}}^{RL}\right)^{\star}
\nonumber
\\
R_{B_{u}}^{RR}&=&
2\sum_{i,j=1,2}\, V_{i2} V_{j2}^{\star}
U_{i1}U_{j1}^{\star}\,x_{Wj}\sqrt{x_{ij}}~
P_{B}^{u}(\bar x_j, x_{ij})
\nonumber
\\
%%%%%%%%%%%%%%%%%%%%
R_{B_{d}}^{LL}&=&
\sum_{i,j=1,2}\, |V_{i1}|^2 |V_{j1}|^2 \,x_{Wj}\,
P_{B}^{d}(\bar x_j, x_{ij})
\nonumber
\\
R_{B_{d}}^{RL}&=&-
\sum_{i,j=1,2}\, V_{i2}^{\star} V_{i1}
|V_{j1}|^2\,x_{Wj}\, P_{B}^{d}(\bar x_j, x_{ij})
\nonumber
\\
R_{B_{d}}^{LR}&=&\left(R_{B_{d}}^{RL}\right)^{\star}
\nonumber
\\
R_{B_{d}}^{RR}&=&
\sum_{i,j=1,2}\, V_{i2}^{\star} V_{i1}
V_{j1}^{\star}V_{j2}\, x_{Wj}\, P_{B}^{d}(\bar x_j, x_{ij})
%%%%%%%%%%%%%%%%%%%%
\nonumber
\\
R_{M_{\gamma,g}}^{LL}&=&\sum_i |V_{i1}|^2\,
x_{Wi}\, P_{M_{\gamma,g}}^{LL}(x_i)
-Y_b\sum_i V_{i1} U_{i2}\,
x_{Wi}\, \frac{m_{\chi_i}}{m_b} P_{M_{\gamma,g}}^{LR}(x_i)
\nonumber
\\
R_{M_{\gamma,g}}^{LR}&=&-\sum_i V_{i1}^{\star}V_{i2}\,
x_{Wi}\, P_{M_{\gamma,g}}^{LL}(x_i)
+Y_b\sum_i  V_{i2} U_{i2}\,
x_{Wi}\, \frac{m_{\chi_i}}{m_b} P_{M_{\gamma,g}}^{LR}(x_i)
\nonumber
\\
R_{M_{\gamma,g}}^{RL}&=&
-\sum_i V_{i1}V_{i2}^{\star}\,
x_{Wi}\, P_{M_{\gamma,g}}^{LL}(x_i)
\nonumber
\\
R_{M_{\gamma,g}}^{RR}&=&
\sum_i |V_{i2}|^2\,
x_{Wi}\, P_{M_{\gamma,g}}^{LL}(x_i)\, .
\eea
Regarding the $R_F^0$ terms at the zero order in mass insertion, we have
\bea
R_D^0 &=& 2x_{W}\sum_{i=1,2} |V_{i2}|^2 D_{\chi}(x_i) 
\nonumber
\\
R_E^0 &=& 2x_{W}\sum_{i=1,2} |V_{i2}|^2 E_{\chi}(x_i) 
\nonumber
\\
R_C^0 &=& 2 \sum_{i,j=1,2} U_{i1} U^{\star}_{j1} 
V_{i2} V^{\star}_{j2} C_{\chi}^{(2)}(x_i,x_j)
\nonumber
\\
R_{B_d}^0&=&-\frac{1}{2}\sum_{i,j=1,2} x_{Wj} V_{i2}^{\star} V_{i1}
V_{j1}^{\star} V_{j2} B^{(d)}_{\chi}(\bar{x}_j,\bar{x_j},x_{ij})
\nonumber
\\
R_{B_u}^0&=&\sum_{i,j=1,2} x_{Wj} V_{i2} V_{j2}^{\star}
U_{i1}^{\star} U_{j1}^{\star} B^{(u)}_{\chi}(\bar{x}_j,\bar{x_j},x_{ij})
\nonumber
\\
R_{M_{\gamma}}^0&=&-x_{W}\sum_{i=1,2} |V_{i2}|^2
\left(F_1(x_{i})+\frac{2}{3}F_2(x_i)\right)
\nonumber
\\
R_{M_{g}}^0&=&-x_{W}\sum_{i=1,2} |V_{i2}|^2 F_2(x_{i})
\eea
where $Y_b$ is the Yukawa coupling of bottom quark,
$x_{\W i}\equiv m_W^2/m_{\chi_i}^2$, $x_{\W}\equiv m_W^2/\tilde{m}^2$,
$x_{i}\equiv m_{\chi_i}^2/\tilde{m}^2$, 
$\bar x_i \equiv \tilde{m}^2/m_{\chi_i}^2$,
and $x_{ij}\equiv m_{\chi_i}^2/m_{\chi_j}^2$. The chargino mixing matrices 
$U$ and $V$, are defined in Eq.(\ref{chmatrix}).
The loop functions of penguin $P_{D,E,C}$, 
$D_{\chi},E_{\chi},C_{\chi}$, 
box $B^{(u,d,\tilde{g})}$, $P_B^{(u,d,\tilde{g})}$, and
magnetic and chromo-magnetic penguin diagrams
$F_{1,2}$, $P_{M_{\gamma,g}}^{LL}$, $P_{M_{\gamma,g}}^{LR}$
are reported in appendix 
C.\footnote{The minus sign appearing in front of the right-hand-side of
equation above for $R_{B_d}^0$, takes into account the correction 
of a sign mistake in the first reference of \cite{GG}
for the chargino contributions to 
down-type box diagrams. This mistake 
was also pointed out in the second reference of \cite{GG}. This sign
correction has also been re-absorbed in the $P_{B}^d$ function given 
in appendix C.}
%%%%%%%%%%%%%%%%%%%%%%%%%%%%%%%%%%%%%%%%%%%%%%%%%%%%%%%%%%%
\section{Chargino contributions in MIA for the light $\tilde t_R$ case}
In this appendix we provide the relevant formulas for the case in which
the mass of the stop-right ($m^2_{{\tilde t}_R}$)
is lighter than the average squark mass $\tilde{m}^2$.
We recall here
that this will modify only the expressions of $R_F^{RL}$, $R_F^{RR}$,
and $R_F^{0}$, since the light stop right does not affect $R_F^{LL}$.
In the case of $R_{D,E,C}^{RR}$ and $R_{D,E,C}^{0}$
the functional forms of $R_{D,E,C}^{RR}$ and $R_{D,E,C}^{0}$
remain unchanged, while the arguments of the functions involved are changed as 
$x_W   \to   x_{Wt}$,
$x_i   \to   x_{it}$,  and ${\bar x}_i \to {\bar x}_{it}$,
where $x_{Wt}\equiv m_W^2/m^2_{\tilde{t}_R}$,
$x_{it}\equiv m_{\chi_i}^2/{m^2_{{\tilde t}_R}}$,
${\bar x_{it}} \equiv {m^2_{{\tilde t}_R}}/m_{\chi_i}^2$.
Regarding the box-type contributions\footnote{Here
we have corrected the formula in \cite{our}.
The effects are ${\cal{O}}(\lambda^2)$ and do not affect the numerical 
results.}
 $R_{B_{u,d}}^{RR}$ and 
$R_{B_{u,d}}^{0}$,
the functions $P_B^{(u,d)}$ and $B_{\chi}^{(u,d)}$
appearing there should be replaced as follows
\bea
P_B^{(u,d)}(\bar{x}_j,x_{ij})&\to&
\bar{P}_B^{(u,d)}(\bar{x}_{jt},\bar{x}_j,x_{ij})
\nonumber
\\
B^{(u,d)}_{\chi}(\bar{x}_j,\bar{x}_j,x_{ij})&\to&
B^{(u,d)}_{\chi}(\bar{x}_{jt},\bar{x}_j,x_{ij})\, ,
\eea
where
\begin{equation}
\bar{P}_B^{(u,d)}(x,y,z)=\frac{1}{2}
x\frac{\partial}{\partial x}B^{(u,d)}_{\chi}(x,y,z)\, .
\label{Pbarbox}
\end{equation}
In the case of $R_F^{LR}$ and $R_F^{RL}$
the analytical expression of loop functions of penguin
$P_{D,E,C}$, box $P_B^{(u,d)}$, and
magnetic and chromo-magnetic penguin diagrams
$P_{M_{\gamma,g}}^{LL}$ and $P_{M_{\gamma,g}}^{LR}$ respectively,
should be replaced with the following ones
\bea
P_{D,E}(x_i)&\to&P_{D,E}(x_i,x_{it})
\nonumber \\
P_C^0(\bar{x}_i)&\to&P_C^0(\bar{x}_i,\bar{x}_{it})
\nonumber \\
P_C^{(1,2)}(x_i,x_j)&\to&P_C^{(1,2)}(x_i,x_{it}, x_j,x_{jt})
\nonumber \\
P_B^{(u,d)}({\bar x}_j, x_{ij})&\to&
P_B^{(u,d)}({\bar x}_j,{\bar x}_{jt}, x_{ij})
\eea
where
\bea
P_D(x_i,x_{it})&=&{2\over{(x_t -1)}}
\left[x_{it} D_\chi(x_{it}) - x_i D_\chi(x_i)\right]
\nonumber \\
P_E(x_i,x_{it})&=&{2\over{(x_t -1)}}
\left[x_{it} E_\chi(x_{it}) - x_i E_\chi(x_i)\right]
\nonumber \\
P_C^{(1,2)}(x_i,x_{it}, x_j,x_{jt})&=& {2\over{(x_t -1)}}
  \left[ C_{\chi}^{(1,2)}(x_{jt},x_{it}) -
C_{\chi}^{(1,2)}(x_{j},x_{i})\right]
\nonumber \\
P_C^{(0)}({\bar x}_i,{\bar x}_{it})&=&{4\over{(x_t -1)}}
  \left[ C_{\chi}^{(1)}({\bar x}_{it},{\bar x}_{i}) -
C_{\chi}^{(1)}({\bar x}_{i},{\bar x}_{i}) \right]
\nonumber \\
P_B^{(u)}({\bar x}_j,{\bar x}_{jt}, x_{ij})&=&{1\over{2(x_t -1)}}
\left[ B_{\chi}^{(u)}({\bar x}_{jt},{\bar x}_j, x_{ij})
-B_{\chi}^{(u)}({\bar x}_{j},{\bar x}_j, x_{ij}) \right]
\nonumber \\
P_B^{(d)}({\bar x}_j,{\bar x}_{jt}, x_{ij})&=&-{1\over{2(x_t -1)}}
\left[B_{\chi}^{(d)}({\bar x}_{jt},{\bar x}_j, x_{ij})
-B_{\chi}^{(d)}({\bar x}_{j},{\bar x}_j, x_{ij}) \right]
\nonumber \\
P_{M_{\gamma}}^{LL}(x_i,x_{it})&=&{1\over{(x_t -1)}}
\left[x_{it}\left(  F_{1}(x_{it})+\frac{2}{3} F_2(x_{it})\right)
-x_{i}\left(  F_{1}(x_{i})+\frac{2}{3} F_2(x_{i})\right)\right]
\nonumber \\
P_{M_{\gamma}}^{LR}(x_i,x_{it})&=&{1\over{(x_t -1)}}
\left[x_{it}\left(  F_{3}(x_{it})+\frac{2}{3} F_4(x_{it})\right)
-x_{i}\left(F_{3}(x_{i})+\frac{2}{3} F_4(x_{i})\right)\right]
\nonumber \\
P_{M_{g}}^{LL}(x_i,x_{it})&=&{1\over{(x_t -1)}}
\left[x_{it} F_2(x_{it}) -x_{i} F_2(x_{i})\right]
\nonumber \\
P_{M_{g}}^{LR}(x_i,x_{it})&=&{1\over{(x_t -1)}}
\left[x_{it} F_4(x_{it}) -x_{i} F_4(x_{i})\right]\, .
\label{Frightstop}
\eea
The expressions for the loop functions involved in
Eqs.(\ref{Pbarbox}) and (\ref{Frightstop}) can be found in Appendix C.
%%%%%%%%%%%%%%%%%%%%%%%%%%%%%%%%%%%%%%%%%%%%%%%%%%%%%%%%%
\section{Basic SM and SUSY Loop functions}
Here we report the loop functions entering in the SM and SUSY diagrams at 
1-loop.
\vskip 1truecm
\begin{itemize}
\item  {\bf \large SM functions}

\bea
{\rm D}(x)&=&\frac{x^2\left(25-19x\right)}{36\left(x-1\right)^3}+
\frac{\left(-3x^4+30x^3-54x^2+32x-8\right)}{18\left(x-1\right)^4}
\log{x}
\nonumber\\
{\rm C}(x)&=&\frac{x\left(x-6\right)}{8\left(x-1\right)}+
\frac{x\left(3x+2\right)}{8\left(x-1\right)^2}
\log{x}
\nonumber\\
{\rm E}(x)&=&\frac{x\left(x^2+11x-18\right)}{12\left(x-1\right)^3}+
\frac{\left(-9x^2+16x-4\right)}{6\left(x-1\right)^4}
\log{x}
\nonumber\\
{\rm B}(x)&=&-\frac{x}{4(x-1)}+\frac{x}{4(x-1)^2}\log{x}
\eea
\vskip 1truecm

\item  {\bf \large Charged Higgs functions}

\bea
{\rm D_H}(x)&=&
\frac{x\left(47x^2-79x+38\right)}{108\left(x-1\right)^3}
+\frac{x\left(-3x^2+6x-4\right)}{18\left(x-1\right)^4}
\log{x}
\nonumber\\
{\rm C_H}(x)&=&-\frac{1}{2} {\rm B}(x)
\nonumber\\
{\rm E_H}(x)&=&\frac{x\left(7x^2-29x+16\right)}{36\left(x-1\right)^3}+
\frac{x\left(3x-2\right)}{6\left(x-1\right)^4}
\log{x}
\eea
\vskip 1truecm

\item {\bf \large Chargino functions in MIA }

\bea
P_D(x)&=&
{\frac{2\,x\,\left( -22 + 60\,x - 45\,{x^2} + 4\,{x^3} +
       3\,{x^4} - 3\,\left( 3 - 9\,{x^2} + 4\,{x^3} \right) \,
        \log{x} \right) }{27\,{{\left(1- x  \right) }^5}}}
\nonumber\\
P_E(x)&=&{\frac{x\,\left( -1 + 6\,x - 18\,{x^2} + 10\,{x^3} +
         3\,{x^4} - 12\,{x^3}\,\log{x} \right) }{9\,
     {{\left(1- x \right) }^5}}}
\nonumber\\
P_C^{(0)}(x)&=&
{\frac{x\,\left( 3 - 4\,x + {x^2} + 2\,\log{x} \right) }
   {8\,{{\left(1- x  \right) }^3}}}
\nonumber\\
P_C^{(1)}(x,y)&=&\frac{1}{8\left(x-y\right)}\left[
\frac{x^2\left(x-1-\log{x}\right)}{(x-1)^2}
-\frac{y^2\left(y-1-\log{y}\right)}{(y-1)^2}\right]
\nonumber\\
P_C^{(2)}(x,y)&=&\frac{\sqrt{xy}}{4\left(x-y\right)}\left[
\frac{x\left(x-1-\log{x}\right)}{(x-1)^2}
-\frac{y\left(y-1-\log{y}\right)}{(y-1)^2}\right]
\nonumber\\
P_B^u(x,y)&=&
{\frac{-y - x\,\left( 1 - 3\,x + y \right) }
    {4\,{{\left( x -1  \right) }^2}\,
      {{\left( x - y \right) }^2}}} -
  {\frac{x\,\left( {x^3} + y - 3\,x\,y + {y^2} \right) \,
      \log{x}}{2\,{{\left(x -1  \right) }^3}\,
      {{\left( x - y \right) }^3}}}
\nonumber \\
&+&  {\frac{x\,y\,\log{y}}
    {2\,{{\left( x - y \right) }^3}\,\left(y -1  \right) }}
\nonumber\\
P_B^d(x,y)&=&-
{\frac{x\,\left( 3\,y - x\,\left( 1 + x + y \right)  \right) }
    {4\,{{\left(x -1  \right) }^2}\,
      {{\left( x - y \right) }^2}}} -
  {\frac{x\,\left( {x^3} + \left( x -3 \right) \,{x^2}\,y +
        {y^2} \right) \,\log{x}}{2\,
      {{\left(x -1  \right) }^3}\,{{\left( x - y \right) }^3}}
    }
\nonumber\\
&+& {\frac{x\,{y^2}\,\log{y}}
    {2\,{{\left( x - y \right) }^3}\,\left(y -1 \right) }}
\nonumber\\
P_{M_{\gamma}}^{LL}(x)&=&
\frac{x\left(-2-9x+18x^2-7 x^3+3 x\left(x^2-3\right)\log{x}\right)}
{9\left(1-x\right)^5}
\nonumber\\
P_{M_{\gamma}}^{LR}(x)&=&
\frac{x\left(13-20x+7x^2+\left(6+4x-4x^2\right)\log{x}\right)}
{6\left(1-x\right)^4}
\nonumber\\
P_{M_{g}}^{LL}(x)&=&
\frac{x\left(-1+9x+9x^2-17 x^3+6 x^2\left(3+x\right)\log{x}\right)}
{12\left(1-x\right)^5}
\nonumber\\
P_{M_{g}}^{LR}(x)&=&
\frac{x\left(-1 -4x+5x^2-2x\left(2+x\right)\log{x}\right)}
{2\left(1-x\right)^4}
\eea
\vskip 1truecm

\item {\bf \large Chargino functions in MIA with a light $\tilde{t}_R$}

\bea
D_{\chi}(x)&=&\frac{\left(-43x^2+101x-52\right)}{108\left(x-1\right)^3}+
\frac{\left(2x^3-9x+6\right)}{18\left(x-1\right)^4}
\log{x}
\nonumber\\
C_{\chi}^{(1)}(x,y)&=&\frac{1}{16\left(y-x\right)}\left[\frac{x^2}{x-1}\log{x}
-\frac{y^2}{y-1}\log{y}\right]
\nonumber\\
C_{\chi}^{(2)}(x,y)&=&\frac{\sqrt{xy}}{8\left(y-x\right)}\left[
\frac{x}{x-1}\log{x}-\frac{y}{y-1}\log{y}\right]
\nonumber\\
E_{\chi}(x)&=&\frac{\left(-11x^2+7x-2\right)}{36\left(x-1\right)^3}+
\frac{x^3}{6\left(x-1\right)^4}
\log{x}
\nonumber\\
B^{(u)}_{\chi}(x,y,z)&=&-\Big[F(x,y,z)+F(y,z,x)+F(z,x,y)\Big]
\nonumber\\
B^{(d)}_{\chi}(x,y,z)&=&xF(x,y,z)+yF(y,z,x)+zF(z,x,y)
\nonumber\\
F(x,y,z)&=&\frac{x\log{x}}{(x-1)(x-y)(x-z)}
\nonumber\\
F_1(x)&=&\frac{\left(x^3-6x^2+3x+2+6x\log{x}\right)}{12\left(x-1\right)^4}
\nonumber\\
F_2(x)&=&\frac{\left(2x^3+3x^2-6x+1-6x^2\log{x}\right)}{12\left(x-1\right)^4}
\nonumber\\
F_3(x)&=&\frac{\left(x^2-4x+3+2\log{x}\right)}{2\left(x-1\right)^3}
\nonumber\\
F_4(x)&=&\frac{\left(x^2-1+2x\log{x}\right)}{2\left(x-1\right)^3}
\eea
\vskip 1truecm

\item {\bf \large Gluino functions in MIA}

\bea
{\mbox P}_{1}(x) &=& {{1 - 6\,x + 18\,{x^2} - 10\,{x^3} - 3\,{x^4} + 
      12\,{x^3}\,\ln (x)}\over {18\,{{\left( x-1 \right) }^5}}}\;
\nonumber \\ 
{\mbox P}_{2}(x) &=& {{7 - 18\,x + 9\,{x^2} + 2\,{x^3} + 3\,\ln (x) - 
      9\,{x^2}\,\ln (x)}\over {9\,{{\left( x-1 \right) }^5}}}\;
 \nonumber \\
{\mbox M}_{1} (x)&=& \frac{1 + 4x - 5x^2 + 4x\ln (x) + 2x^2\ln (x)}
{2(1 - x)^4}\;
 \nonumber \\
{\mbox M}_{2} (x)&=& \frac{-5 + 4x + x^2 - 2\ln(x) - 4x\ln(x)}
{2(1 - x)^4}   \;
 \nonumber \\
{\mbox M}_{3} (x)&=& {{-1 + 9\,x + 9\,{x^2} - 17\,{x^3} + 18\,{x^2}\,\ln (x) + 
      6\,{x^3}\,\ln (x)}\over {12\,{{\left( x-1 \right) }^5}}}\;
 \nonumber \\
{\mbox M}_{4} (x)&=& {{-1 - 9\,x + 9\,{x^2} + {x^3} - 6\,x\,\ln (x) - 
      6\,{x^2}\,\ln (x)}\over {6\,{{\left( x-1 \right) }^5}}}\;
 \nonumber \\
{\mbox B}_{1}(x) &=& \frac{1}{4} {\mbox M}_1(x)
\nonumber \\   
{\mbox B}_{2}(x) &=& -x {\mbox M}_2(x)\, .
\eea
\end{itemize}
%%%%%%%%%%%%%%%%%%%%%%%%%%%%%%%%%%%%%%%%%%%%%%%%%%%
\newpage

%%%%%%%%%%%%%%%%%%%%%%%%%%%%%%%%%%%%%%%%%%%%%%%%%%%

\end{document}